%% file: hr8799.tex
\documentclass[emulateapj]{emulateapj}
\usepackage{natbib}
\usepackage{graphicx}

\begin{document}
\title{A Combined Subaru/VLT/MMT 1--5 $\mu m$ Study of Planets Orbiting HR 8799: 
Implications for Atmospheric Properties, Masses, and Formation}
\author{Thayne Currie\altaffilmark{1}, Adam Burrows\altaffilmark{2}, 
Yoichi Itoh\altaffilmark{3}, Soko Matsumura\altaffilmark{4}, Misato Fukagawa\altaffilmark{5}, Daniel 
Apai \altaffilmark{6}, 
Nikku Madhusudhan\altaffilmark{2},
Philip M. Hinz\altaffilmark{7}, T. J. Rodigas\altaffilmark{7}, 
Markus Kasper\altaffilmark{8},
T-S. Pyo\altaffilmark{9}, Satoshi Ogino\altaffilmark{3}
}
\altaffiltext{1}{NASA-Goddard Space Flight Center}
\altaffiltext{2}{Department of Astrophysical Sciences, Princeton University}
\altaffiltext{3}{Graduate School of Science, Kobe University}
\altaffiltext{4}{Department of Astronomy, University of Maryland-College Park}
\altaffiltext{5}{Department of Earth and Space Science, Graduate School of Science, Osaka University}
\altaffiltext{6}{Space Telescope Science Institute}
\altaffiltext{7}{Steward Observatory, Department of Astronomy, University of Arizona}
\altaffiltext{8}{European Southern Observatory}
\altaffiltext{9}{National Astronomical Observatories of Japan}
\begin{abstract}
We present new 1--1.25 $\mu m$ (z and J band) Subaru/IRCS and 2 $\mu m$ (K band) VLT/NaCo data for HR 8799 and a rereduction 
of the 3--5 $\mu m$ MMT/Clio data first presented by \citet{Hinz2010}.  
Our VLT/NaCo data yields a detection of a fourth planet at a projected separation of $\sim$ 15 AU -- ``HR 8799e".
We also report new, albeit weak detections of HR 8799b at 1.03 $\mu m$ and 3.3 $\mu m$.  
Empirical comparisons to field brown dwarfs show that at least HR 8799b and HR8799c,
 and possibly HR 8799d, have near-to-mid IR colors/magnitudes significantly discrepant from the L/T dwarf sequence.
Standard cloud deck atmosphere models appropriate for brown dwarfs provide only (marginally) 
statistically meaningful fits to HR 8799b and c for unphysically small radii.  
   Models with thicker cloud layers not present in brown dwarfs reproduce 
the planets' SEDs far more accurately and without the need for rescaling the planets' radii.  
Our preliminary modeling suggests that HR 8799b has log(g) = 4--4.5, T$_{eff}$ = 900K, 
while HR 8799c, d, and (by inference) e have log(g) = 4--4.5, T$_{eff}$ = 1000--1200K.
Combining results from planet evolution models and new dynamical stability limits implies
that the masses of HR 8799b, c, d, and e are 6--7 M$_{J}$, 7--10 M$_{J}$, 7--10 M$_{J}$ 
and 7--10 M$_{J}$.  ''Patchy" cloud prescriptions may provide even better fits to the data and may 
lower the estimated surface gravities and masses.
Finally, contrary to some recent claims, forming the HR 8799 planets by core accretion
is still plausible, although such systems are likely rare.
\end{abstract}
\section{Introduction}
The HR 8799 planetary system is the first directly imaged multiplanetary system \citep{Marois2008}. 
Along with Fomalhaut and $\beta$ Pic, it is also the only imaged system with companion mass ratios 
and separations reasonably close to the giant planets in the Solar System
\citep{Kalas2008,Lagrange2009,Lagrange2010}\footnote{Here, we consider 1RXJ1609.1-210524b discovered 
by \citet{Lafreniere2008a} to be a more complicated case as its mass ratio and separation are continuous 
with brown dwarf companions (see Discussion Section)}.  After the initial 
detection of HR 8799bcd, one or more planets were recovered in prior datasets \citep{Lafreniere2009, 
Fukagawa2009, Metchev2009}.  
Recently, \citet{Marois2011} imaged a fourth planet -- HR 8799e -- which we 
independently detected (see \S 2).  

Mass estimates based on cooling models yield 5--11 M$_{J}$ for HR 8799b and 
7--13 M$_{J}$ for the other planets \citep{Marois2008,Marois2011}.  Dynamical constraints 
placed by HR 8799bcd imply that the companions likely have masses below the 
deuterium-burning limit \citep[][]{Spiegel2010} and are kept stable by resonant interactions 
\citep{Fabrycky2009,MoroMartin2010}.  Including 
the fourth planet, \citet{Marois2011} argue that the planets most likely have 
masses at the low end of the range allowed by cooling models.  
With masses of $\approx$ 5--13 M$_{J}$, the HR 8799 planets then bridge the gap between the solar 
system's gas giants/Jupiter-mass planets detected by radial velocity surveys 
\citep[e.g.][]{Howard2010} and low-mass brown dwarf companions to nearby stars 
such as GJ 758B and PZ Tel \citep{Thalmann2009,Currie2010a,Biller2010}.  

Recent studies complicate our understanding of the relationship between
brown dwarfs, the gas giants detected in RV surveys, and the HR 8799 
planets.  The planets' masses 
are significantly larger than most planets detected by radial velocity and transit methods.  
\citet{Marois2008} noted that the planets appear 
slightly redder than the distribution of H/H-K$_{s}$ colors for old field brown dwarfs.  The 
K-band spectrum of HR 8799b is not well matched by typical L and T-type brown dwarf spectra 
\citep{Bowler2010}.

Comparisons between the HR 8799 planet photometry/spectroscopy and atmosphere models 
reveal additional difficulties in understanding their properties within the theoretical framework of 
standard, cloud deck models that track the field L/T dwarf sequence.
In the discovery paper, \citet{Marois2008} briefly mention a discrepancy between temperatures derived from 
atmosphere models and those estimated from more simple, and presumably most accurate, 
cooling model estimates.  
More recently, \citet{Bowler2010} provide a detailed comparison 
between the HR 8799b spectra and 1.1--4.1 $\mu m$ photometry and predictions 
from standard atmosphere models.  They show that the `best-fit' temperatures derived 
from modeling are inconsistent with cooling model estimates.
  They also explicitly show that the 
implied radii for best-fit models are well below the 1.1--1.3 R$_{J}$ range allowed 
by standard cooling models (e.g. 0.3--0.5 R$_{J}$).

To interpret these modeling difficulties, \citet{Bowler2010} argue that a different atmospheric 
structure, namely atmospheres with stronger cloud coverage, may better explain 
the HR 8799b SED.  Since atmospheric dust entrained in clouds absorbs more efficiently at 
shorter wavelengths, photometry for HR 8799b at wavelengths shortward of J band 
would provide a crucial test of the planet's level of cloud coverage \citep[cf.][]{Burrows2006}.
The \citet{Bowler2010} study also found difficulty in reconciling their model fitting of 
detections from \citet{Marois2008} with 3--5 $\mu m$ upper limits from \citet{Hinz2010}.  
More sensitive photometry at these wavelengths would then provide better modeling constraints. 

In this study, we investigate the atmospheres and dynamics of the HR 8799 planets using
 new observations obtained at the Subaru Telescope and VLT and a rereduction of MMT data 
presented by \citet{Hinz2010}.  Combined with photometry presented by \citet{Marois2008}, 
our data yield nine photometric points spanning 1--5 $\mu m$ for a detailed 
comparison to the properties of field brown dwarfs.  This wavelength range also provides
 a sensitive probe of the effects of surface gravity, temperature, (non)equilibrium chemistry, metallicity, 
and cloud coverage.  

We compare the planets' SEDs to atmosphere models exploring a phase space 
defined by these effects.  By quantifying the model fits, we determine the range of parameter 
space that fails to characterize the planets' SEDs and identify the subset of models that 
more accurately reproduce the data and may better represent their atmospheres' physical properties.
These results will then be used to more thorougly and accurately probe the planets' 
atmospheric properties in a companion paper (Madhusudhan et al. 2011, in prep.).

Our study is structured as follows.  \S 2 describes our observations, image processing, and 
detections for each dataset.  
The first part of \S 3 compares the HR 8799 planet photometry 
to the L/T dwarf sequence and the IR properties of other very low-mass objects 
(M $<$ 25 M$_{J}$).  The rest of \S 3 presents preliminary comparisons between the HR 8799 
planet SEDs and planetary atmosphere models.  
\S 4 describes simple dynamical modeling of the system to 
identify the range of masses for dynamically stable orbits.  
\S 5 summarizes our results, discusses our 
work within the context of previous studies of HR 8799 and planet imaging in general, discusses 
how our results fit within the context of planet evolution models, and comments on the plausible 
formation mechanism(s) for the planets.
\section{Data}
\subsection{Observations}
Our study combines data from three facilities -- VLT/NaCo, Subaru/IRCS, 
and MMT/Clio -- at six broadband filters centered on 1.03 $\mu m$ to 4.8 $\mu m$.
The VLT data are the most sensitive and were obtained to 
place limits on the presence of other candidate planets in the system.  
The Subaru data at 1.03--1.25 $\mu m$ were taken to probe the effect of 
clouds on the planets' atmospheres.  
Finally, we rereduced the 3--5 $\mu m$ 
MMT/Clio data first presented by \citet{Hinz2010} using our reduction 
pipline, which utilizes advanced image registration, 
PSF removal, speckle suppression, and flux calibration 
routines \citep[e.g. LOCI][]{Lafreniere2007a} also used in \citet{Marois2008}. 

All of our data were taken in angular differential imaging mode \citep{Marois2006}, 
where the instrument rotator is adjusted to stay at a fixed angle with 
respect to the (changing) parallactic angle, resulting in the field of 
view rotating with time.  Combined with the \citeauthor{Marois2008} 
data, we thus have data spanning nine photometric filters that 
is largely reduced self consistently. 
Table \ref{obstable} summarize basic properties of our observations.
\subsubsection{VLT/NaCo K$_{s}$ band Data}
HR 8799 was imaged with VLT/NaCo on six separate nights 
in October 2009 as a part of a separate study
of the HR 8799 planets (P.I. Daniel Apai; 
Apai et al. 2011, in prep.).  Once publically available, the 
science and calibration data were downloaded from the ESO VLT archive 
for October 8--11, nights over which the field rotation 
for the HR 8799 data was $>$ 30--45 degrees, resulting in a 
small (r $\sim$ 0.22") inner working angle.  

The data were taken with the 13.27 mas pixel scale, 
without coronographic masks, and in pupil tracking mode allowing 
for angular differential imaging.  All data consist of coadded 0.345s exposures  
totaling $\sim$ 43s a piece and are stored in the standard 
NaCo datacube format.  In this paper, 
we focus specifically on the October 8 data, which 
had the highest quality and greatest amount of field rotation.
Apai et al. (2011, in prep.) will later present a larger study combining 
all October 2009 data and Fall 2010 data.
\subsubsection{Subaru/IRCS z(Y) and J band data}
HR 8799 was targeted for direct imaging on August 15, 2009 
with the Subaru Telescope using the Infrared 
Camera and Spectrograph \citep[IRCS;][]{Tokunaga1998} and AO-188 
adaptive optics system in natural guide star mode.  
The data were taken in the Mauna Kea J band filter ($\sim$ 
1.25 $\mu m$) and the z filter centered on 1.033 $\mu m$, 
analogous to the better-known Y band filter \citep[e.g.][]{Hillenbrand2002}\footnote{The 
zeropoint wavelength for the z filter listed on the IRCS webpage is 1.033 $\mu m$ with a 
width of 0.073 $\mu m$.  The Y-band filters for comparable cameras are slightly 
wider but otherwise quite similar: filters for Keck/NIRC (there called "Z"), UKIRT/WFCAM and Gemini/NIRI have zeropoint 
wavelengths of 1.032, 1.031 and 1.02 $\mu m$ and widths of 0.156, 0.1 and 0.1 $\mu m$, respectively.  
The IRCS z filter should not be confused with the Sloan z' filter, which covers shorter wavelengths.}   
During our observations, conditions were photometric with fair natural 
seeing ($\sim$ 0.65--0.75").  AO-188 yielded a corrected image 
with FWHM(PSF) $\sim$ 0.06" in z and 0.064" in J.  For all observations, 
the native pixel scale is 20.57 mas/pixel;
we used the 0.8" diameter, non-transmissive coronographic mask to block most 
of the primary starlight.

The z data were taken using 30 second exposures consisting of 6 coadded 
frames to avoid saturation at separations corresponding to HR8799bcd 
for a total integration time of 4500s.  The J band data consist of 
25 second coadded exposures for a total integration time of 1080s.  
The z data were observed through transit 
yielding a total field rotation of 172 degrees.  The J data were taken about an 
hour after transit resulting in very poor field rotation ($\sim$ 6.4 degrees).

\subsubsection{MMT/Clio 3--5 $\mu m$ Data}
MMT/Clio observations were previously described by \citet{Hinz2010}.  Briefly, 
HR 8799 was imaged in three separate observing runs -- November 21, 2008, 
January 10, 2009, and September 12, 2009 -- at the L' (3.8 $\mu m$) and 
Barr M (4.8 $\mu m$) and a shorter wavelength filter centered on 3.3 $\mu m$ 
methane absorption feature and extending from 3.12 $\mu m$ to 3.53 $\mu m$.  
We focus on the November 2008 and September 2009 runs, which had sufficient 
field rotation for angular differential imaging.
The pixel scale for all Clio data is 48.57 mas/pixel.

The [3.3], L', and M data were imaged 
for 6780s, 5690s, and 9600s: the total field rotation for data in these three 
filters was 125.3 degrees, 72 degrees, and 31.8 degrees.
While observing conditions for the L' and M data were clear, 
the [3.3] micron data were taken through variable seeing in 
two sets between which the AO system failed to yield an acceptable correction.   

\subsection{Image Processing/Data Reduction}
\subsubsection{Basic Image Processing and Image Registration}
For our Subaru/IRCS and VLT/NaCo data, we first performed standard dark subtraction, flat fielding, and 
bad pixel masking.  While the NaCo data followed a four-point dither pattern 
which should wash out image distortion errors, the IRCS data were not 
dithered.  Thus, each IRCS frame was corrected for image distortion using polynomial 
fits, resulting in a revised pixel scale of 20.53 mas/pixel.  

For the MMT/Clio data we then performed sky subtraction.  
We constructed Clio sky frames from median-combined images obtained 
for each nod position and subtracted to remove the sky background.  
Final pixel values for each VLT/NaCo image were nominally constructed 
from the average pixel value drawn from each frame in the datacube.  
For regions within 1" of estimated stellar centroid, we determined 
the average pixel value after iteratively clipping 5$\sigma$ outliers.
For all datasets, bad pixels were identified as outliers within a 
moving-box median filter, flagged, and interpolated over.  

Our image registration procedure closely follows that of \citet{Lafreniere2007b} 
and \citet{Marois2008}.  We first copied each image into a larger 
blank one, coarsely registering them using a priori knowledge about the 
center of the coronographic mask (for IRCS) or a gaussian fit 
to a convolved version of the image using the IDL function \textrm{gcntrd.pro}.  
For precise image registration, we center one image using a 
2D cross correlation function relating it to a 180 degree rotation of itself.  
We then identify the fractional pixel offsets between the reference image and 
subsequent images yielding the highest correlation.  The region of interest used 
to register IRCS images is focused on diffracted light from the secondary spider. 
For the Clio and NaCo images we used the non-saturated portions of the stellar PSF, 
since the diffracted light from the spider is highly suppressed.  

\subsubsection{Localized Combination of Images (LOCI) Speckle Supression Processing}
Further data reduction follows the ADI/LOCI reduction procedure 
described by \citet{Lafreniere2007a} and \citet{Marois2008}.  
We first subtract out the time-independent component of the stellar PSF, 
exploring two methods.  In the first method, we median combine 
all images for a reference PSF which we subtract from each image, the simple ADI 
method used by \citet{Hinz2010}.  In the second 
method, we construct a two-dimensional radial profile for each image and subtract 
it to remove the smooth seeing halo.  

 Next, we perform the LOCI speckle suppression algorithm on the residual images, derotate the 
processed images and median combine them for a final science image.   
We compared reductions for a range of LOCI input parameters -- dr, N$_{\delta}$, Na, and geom 
\citep[see ][for definitions]{Lafreniere2007a} -- to identify the set that maximized the 
signal-to-noise of the planets, using the set recommended by \citet{Lafreniere2007a} as a starting 
point.   
Our pipeline also produces the simple ADI reduction as a 
byproduct, useful for a separate, sensitive identification of HR 8799b, whose detection in some filters (e.g. [3.3], M) 
may be more severely limited by photon noise than by speckle noise.

\subsection{Planet Detections and Astrometry}
To identify detected planets in our images, we compute the standard deviation and signal-to-noise ratio 
of pixel values in concentric annuli \citep{Currie2010a,Thalmann2009}.  As a check on our results, 
we compare our astrometry for candidate detections in a given filter with that obtained by 
us in other filters and from \citet{Marois2008,Marois2011} during the Fall 2008 and 2009 epochs.  
We claim a detection of a planet independent of other datasets if SNR $>$ 5.  For SNR = 3--5, 
the centroid of the candidate planet detection must be the same as that reported for the 
planet data where SNR $>$ 5 within astrometric errors (typically 0.5 pixels). 
We centroid the planet using the IDL functions \textrm{gcntrd.pro} and \textrm{cntrd.pro} 
 and adopt a minimum astrometric uncertainty of 0.5 pixels to account for image registration and 
centroiding/orientation errors.  The rightmost column of Table \ref{obstable} summarizes our 
planet detections and Table \ref{astromtable} lists their astrometry.

\subsubsection{VLT/NaCo Detections}

\begin{figure}
\centering
\epsscale{0.75}
\plotone{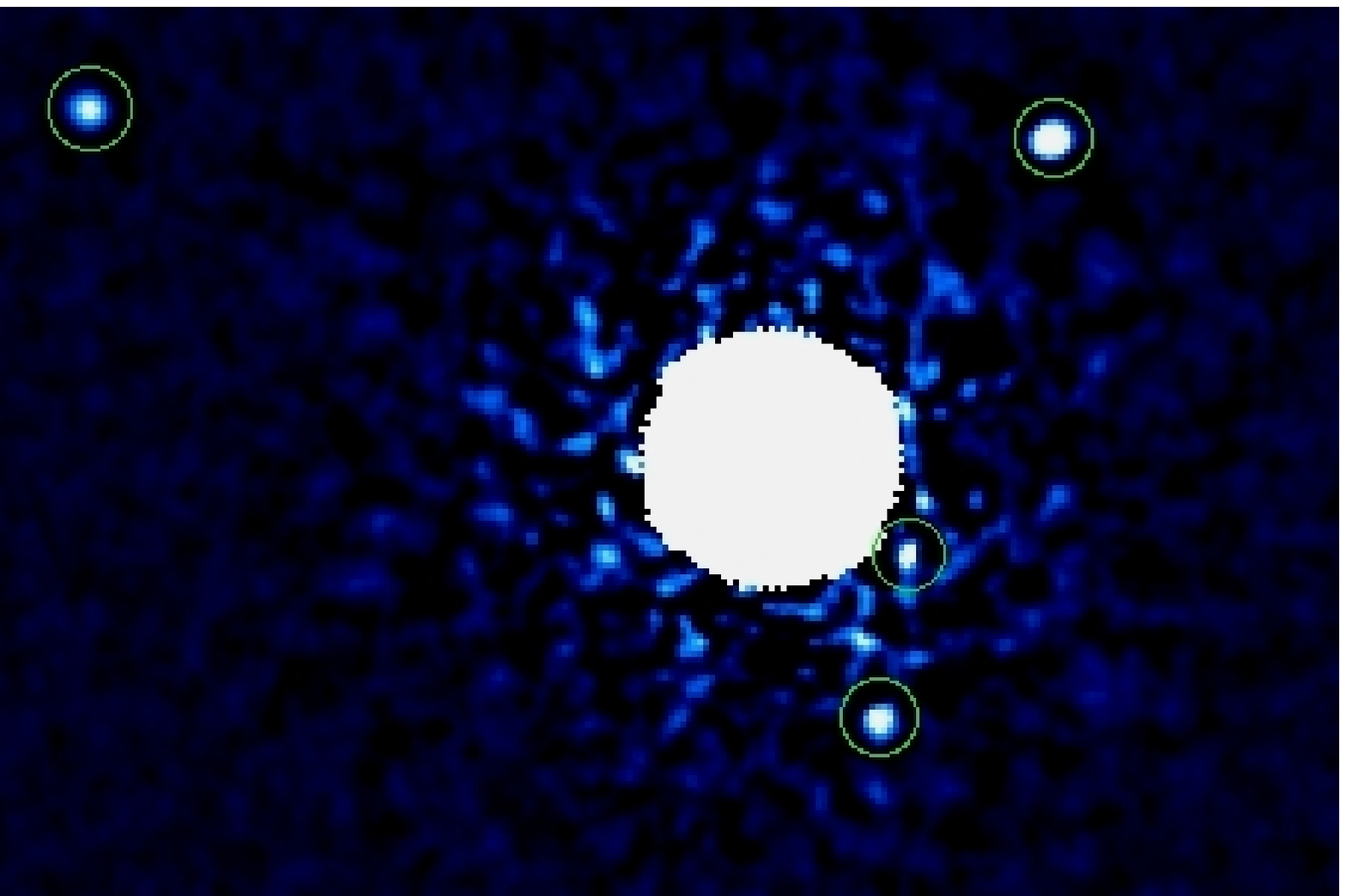}
\caption{VLT/NaCo image of the HR 8799 planetary system.  Previously detected 
planets -- HR 8799b, c, and d -- are easily visible at high signal-to-noise.  
At $\sim$ 0.375" separation, we detect an additional object consistent 
with being a fourth planet orbiting HR 8799 -- HR 8799e.  This same object 
was independently detected by \citet{Marois2011} and confirmed to be a fourth 
planet.  HR 8799e (and d, to a lesser extent) appear slightly smaller than 
b and c because of point source self subtraction inherent in LOCI processing.}
\label{nacoimage}
\end{figure}

Figure \ref{nacoimage} shows our reduced VLT/NaCo K$_{s}$ band image.  HR 8799 b and c are 
detected at better than 25 $\sigma$, while HR 8799d is detected at 10 $\sigma$.
The planets are also free of deep, negative flux troughs at the same separation 
but slightly different position angles that results from LOCI being applied 
to datasets with poor field rotation or those where most exposures are taken well 
before or well after transit \citep[e.g.][]{Marois2010b}.

Additionally, our data show a detection of an another point source 
located interior to and in the same quadrant as HR 8799d consistent with a 
fourth planet -- ''HR 8799e".  
Recently, \citet{Marois2011} announced a multiepoch detection of 
HR 8799e in K-band and L'-band using Keck/NIRC2.
Their detection significance in K-band using Keck/NIRC2 is 
slightly better than ours ($\sim$ 5 $\sigma$ vs. our $\sim$ 4 $\sigma$).
Our photometry using methods described in \S 2.4 yields an absolute 
magnitude of m($K_{s}$) = 12.89 $\pm$ 0.26, consistent with \citeauthor{Marois2011}'s 
estimate of 12.93 $\pm$ 0.22.

\begin{figure}
\centering
\plotone{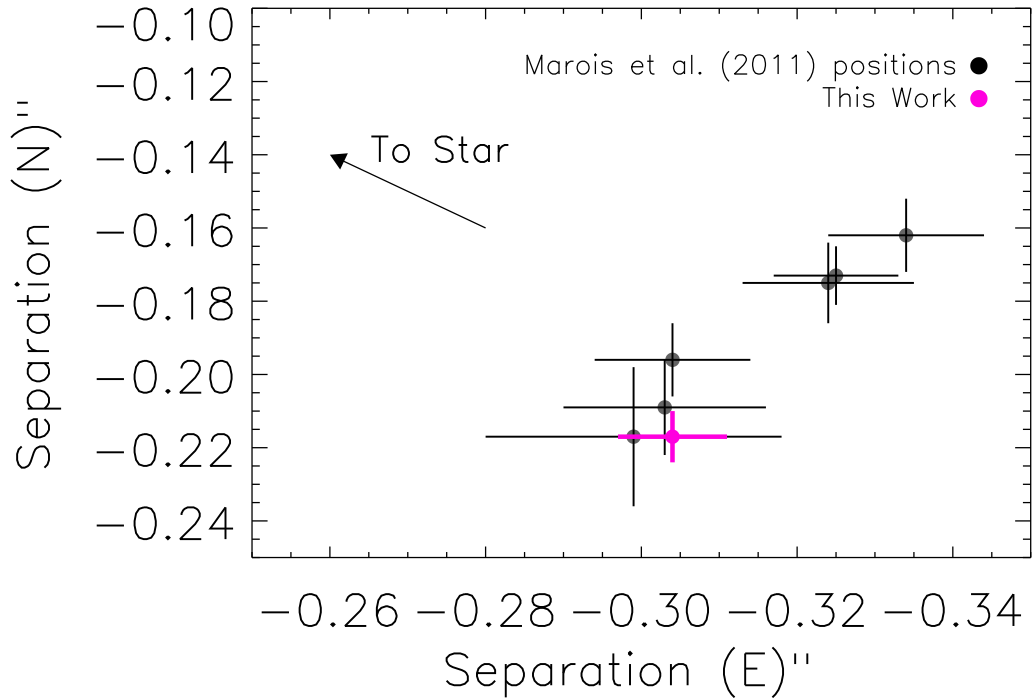}
\caption{Astrometry for HR 8799e comparing positions from \citet{Marois2011} and from our work.  
The arrow identifies the direction to the HR 8799 primary.  The two points from 
\citet{Marois2011} nearest to our October 8, 2009 measurement were taken in 
August 2009 and November 2009, respectively.  Our astrometry are consistent with 
those from \citet{Marois2011} within errors.}
\label{hr8799astrom}
\end{figure}

Figure \ref{hr8799astrom} compares our astrometry.
We measure a centroid position of [E,N] = [-0.306" $\pm$ 0.007", -0.217" $\pm$ 0.007"], 
implying a projected separation of 14.8 AU $\pm$ 0.4 AU.  The average of the August and November 
2009 positions from \citet{Marois2011} is [-0.304,-0.203] with an intrinsic uncertainty 
$\sim$ 0.01".  Our position is then consistent with theirs to within 1.4 $\sigma$.  
Our implied projected physical separation is consistent with
\citeauthor{Marois2011}'s estimates from multiepoch data: 14.5 AU $\pm$ 0.5 AU.

\subsubsection{Subaru/IRCS Detections}
\begin{figure}
\epsscale{0.9}
\centering
\plottwo{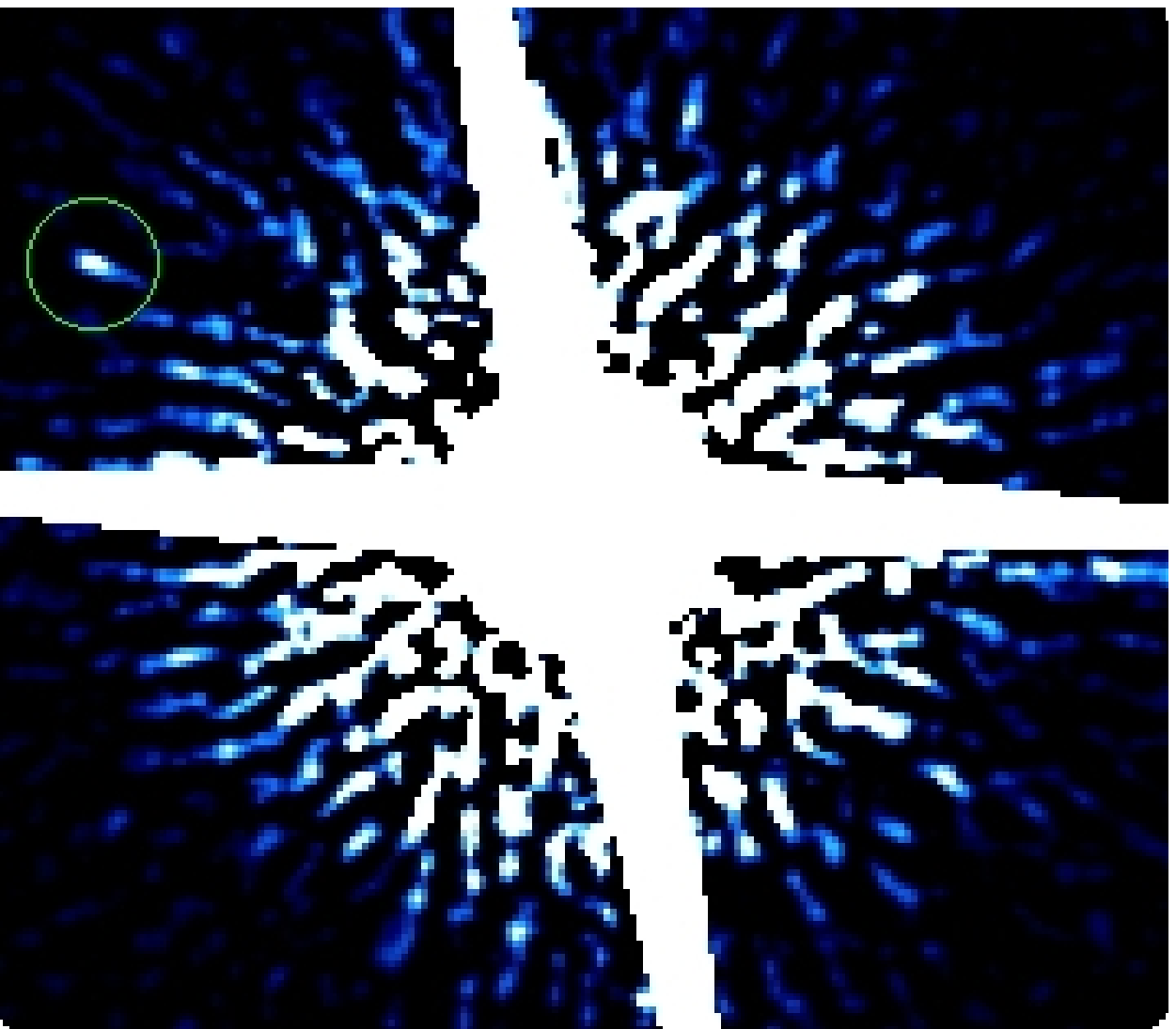}{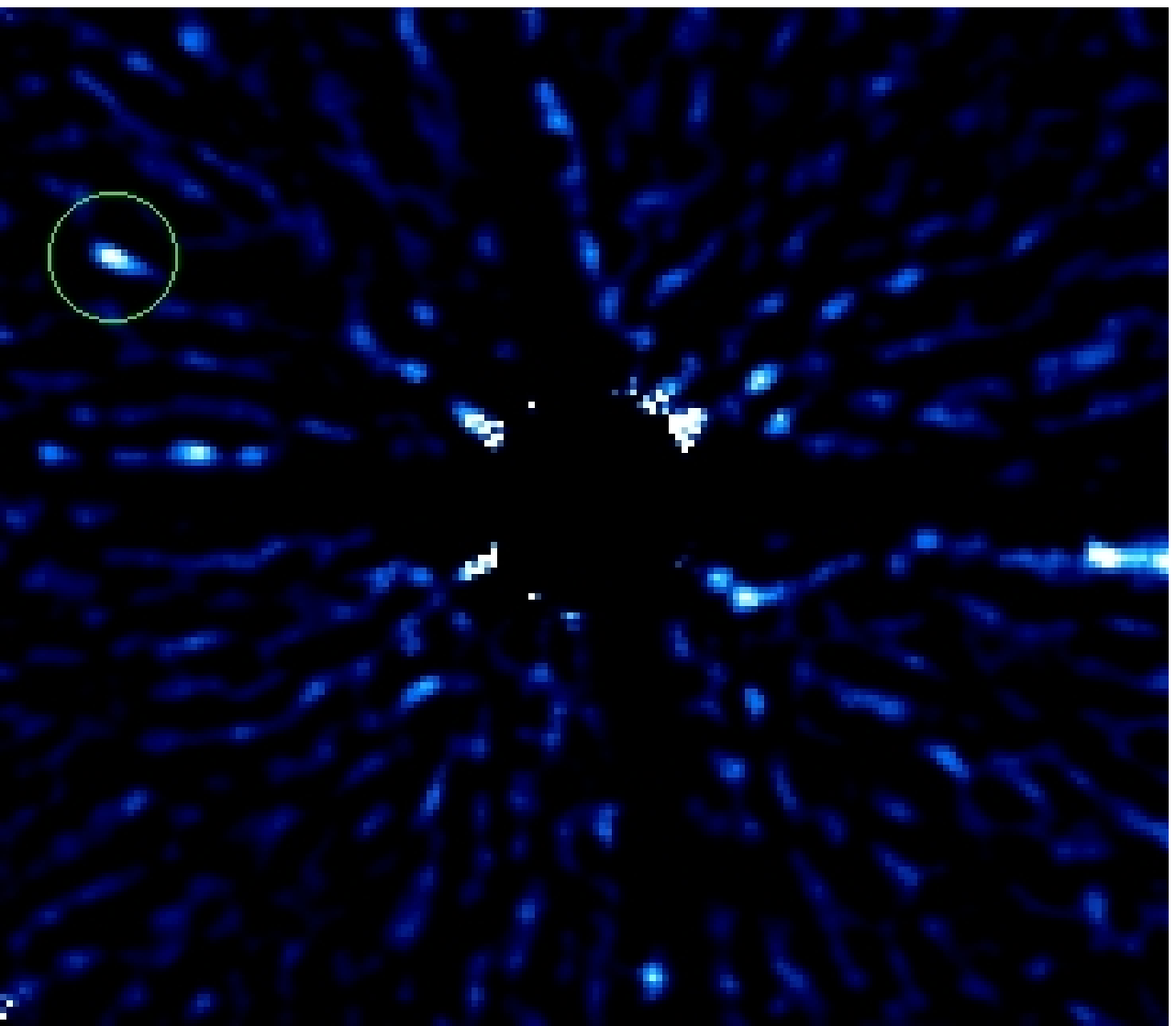}
\plottwo{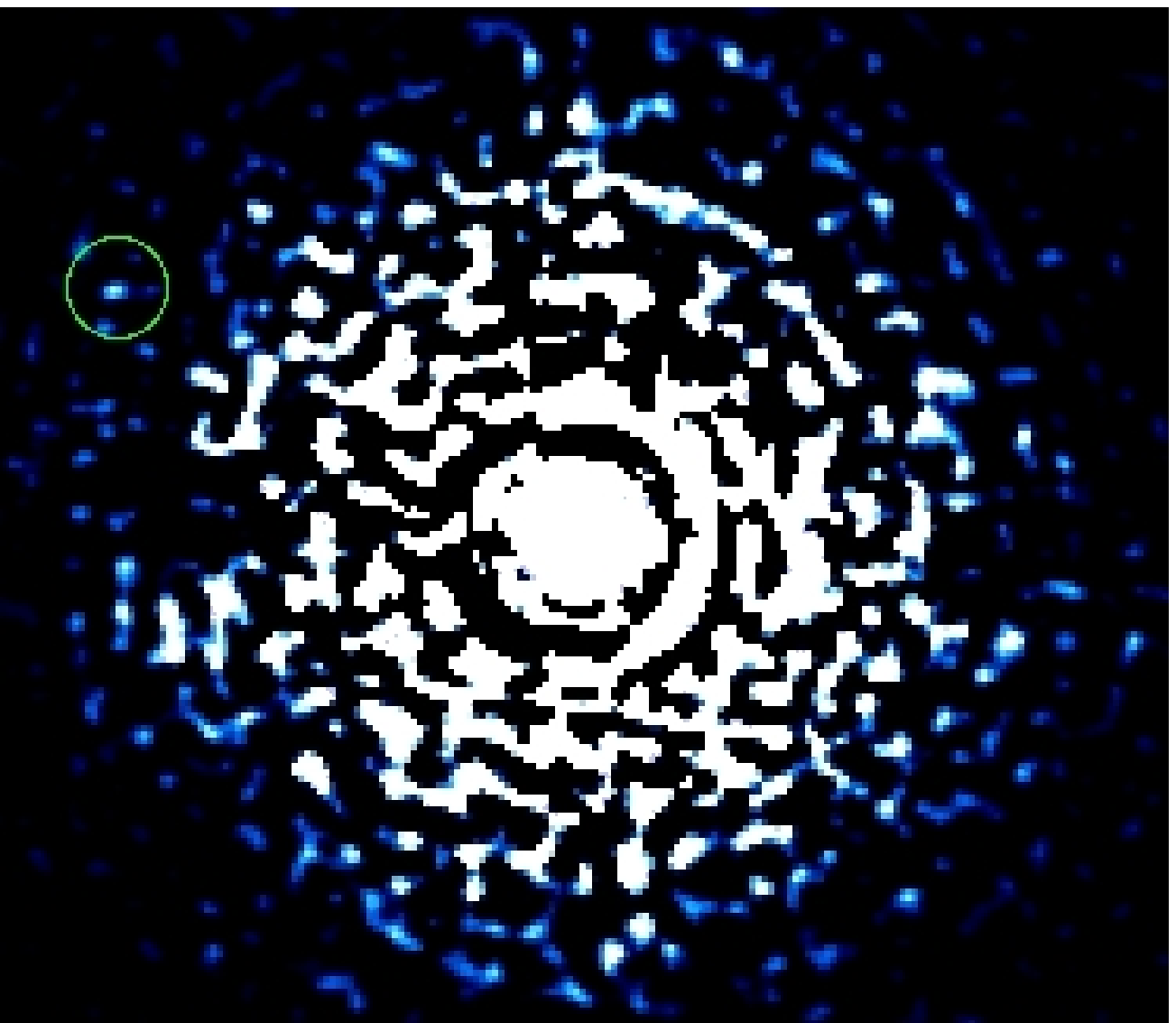}{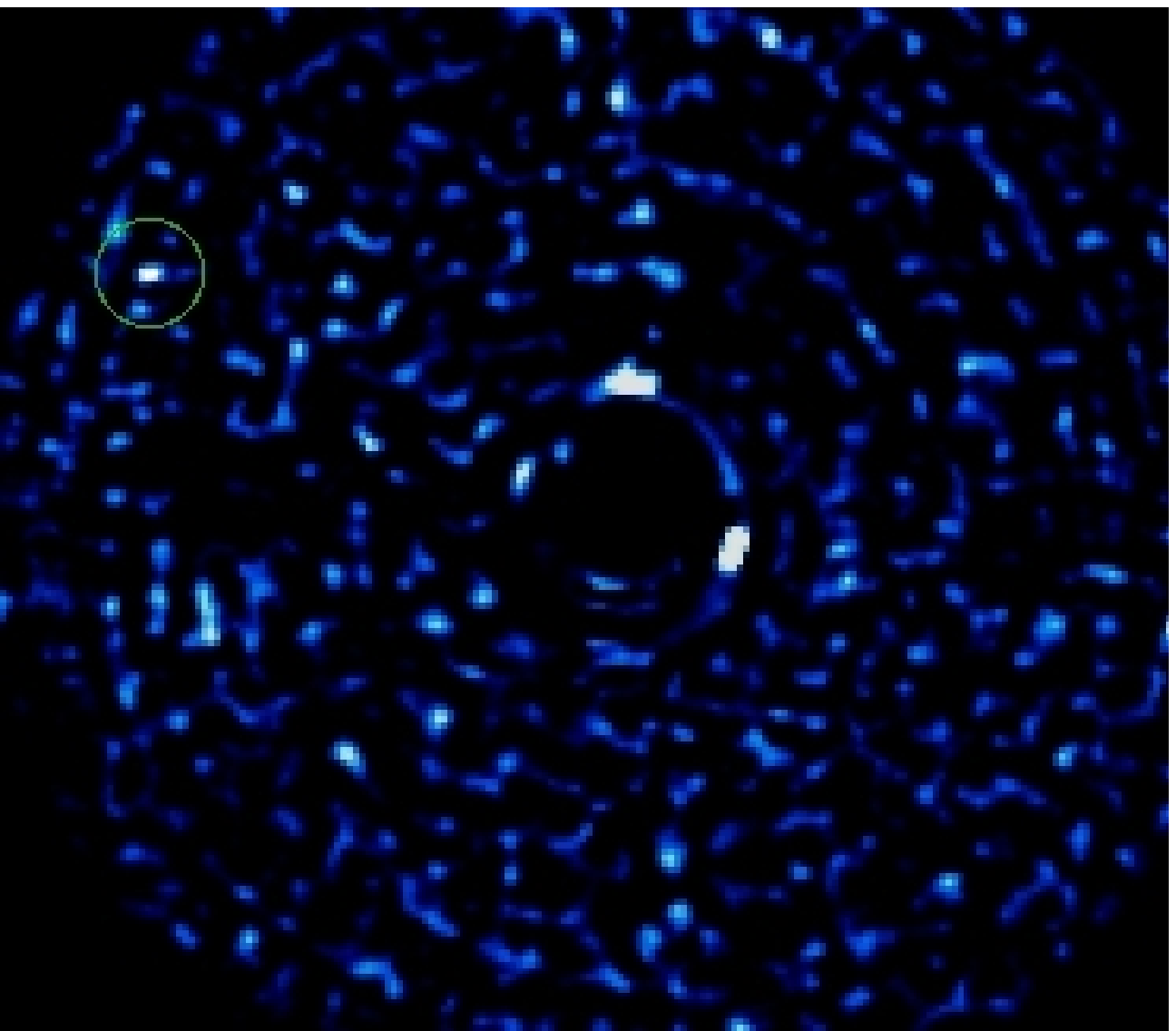}
\caption{Subaru/IRCS images (left panels) and signal-to-noise maps (right panels) 
obtained at J band (top panels) and z band (bottom panels).  The poor field rotation and short 
integration time in J limit our detection to HR 8799b.  Despite over an hour of integration time 
in z band, we marginally detect HR 8799b but fail to detect the other planets.
}
\label{ircsjz}
\end{figure}

Figure \ref{ircsjz} show reduced images at J and z obtained with IRCS.  In spite of poor field rotation 
severely limiting the performance of the ADI/LOCI processing and precluding 
detectability of objects within $\sim$ 1", we clearly detect the b planet in 
our J band data at a $\sim$ 10 $\sigma$ significance (top panels).  
In spite of good seeing conditions, good field rotation, and a 70-minute integration time, we fail 
to detect any of the planets at a $>$ 5 $\sigma$ significance in z band (bottom panels).  
Our reduced image reveals a weak detection of HR 8799b with SNR $\sim$ 3.7 (bottom-right panel) and a centroid 
within 0.25 pixels of its centroid in the J-band data obtained one hour later with the same instrument.  
However, we do not detect HR 8799c or d in our z data.
To verify that our low signal-to-noise detection of HR 8799b and nondetections for the other planets 
do not result from errors in derotation or a jump in parallactic angle near transit\footnote{For an example of 
this phenomenon, see http://www2.keck.hawaii.edu/inst/nirc2/vertAngJump.html}, 
we introduced fake planets into each image with a flux equal to $\sim$ 10 times the local noise of the final image,
reran our reduction pipeline separately for frames before and after transit, and recovered their detections.  

\subsubsection{MMT/Clio Detections}
\begin{figure}
\epsscale{0.65}
\plotone{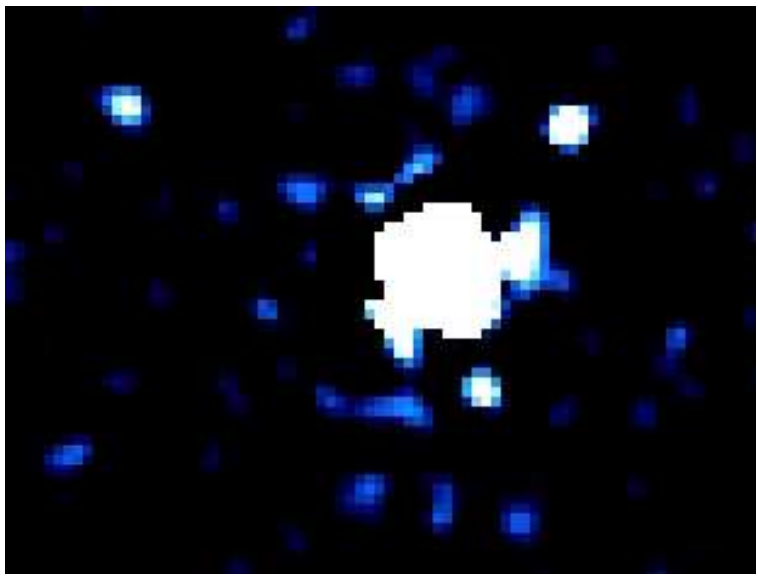}
\caption{MMT/Clio image of the HR 8799 system at L'/3.8 $\mu m$.  The three 
planets are clearly visible and all are detected at SNR $>$ 5.}
\label{cliolp}
\end{figure}

\begin{figure}
\epsscale{0.9}
\centering
\plottwo{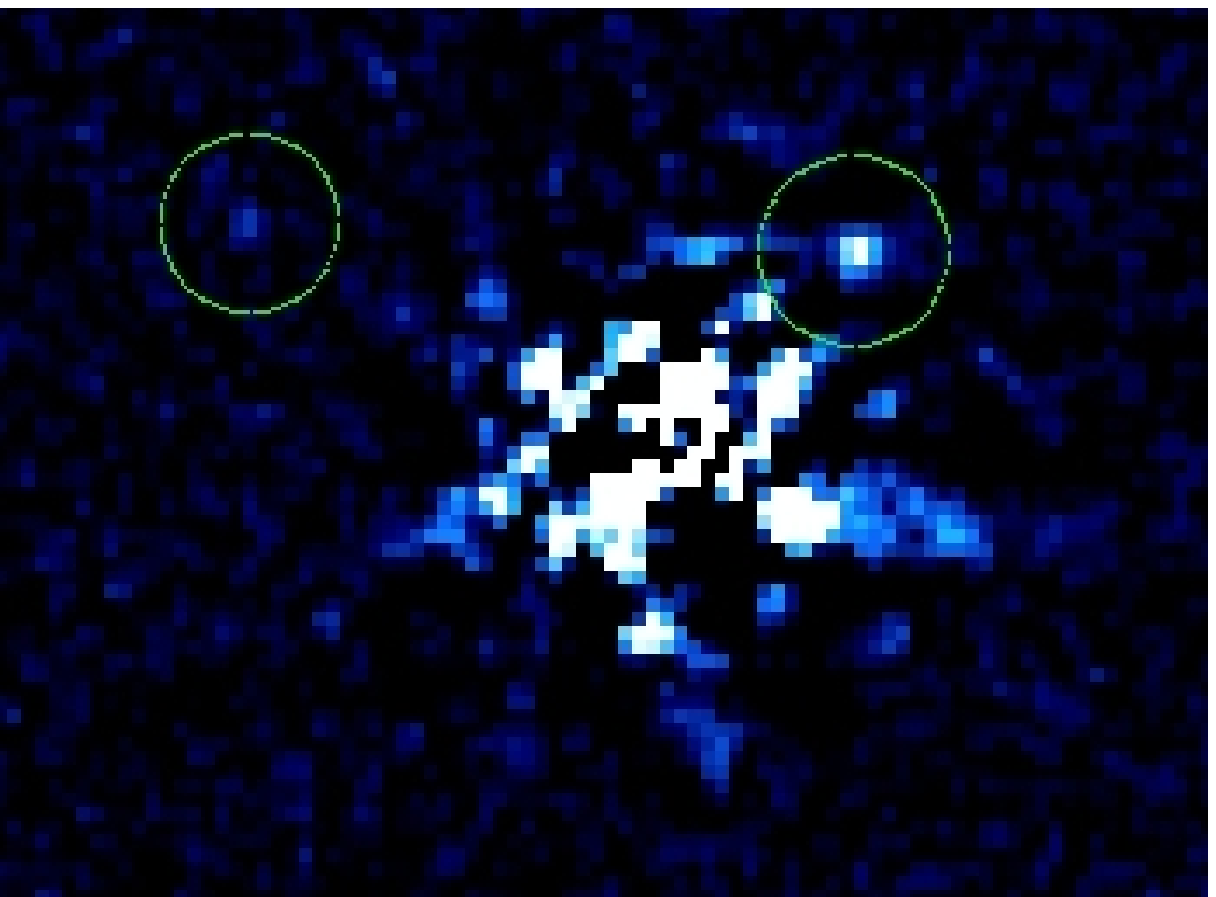}{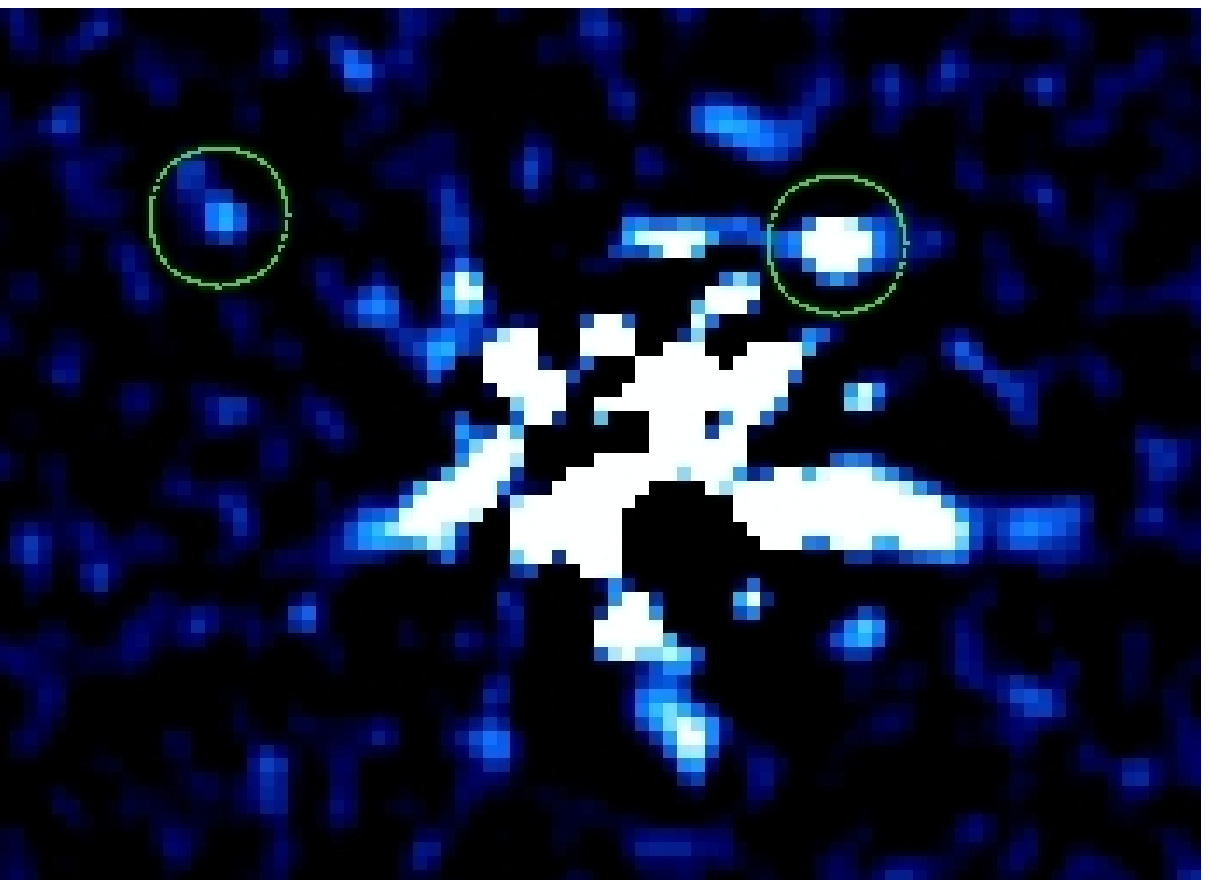}
\caption{MMT/Clio image at 3.3 $\mu m$.  (Left panel) The image shown with a high 
dynamic range to more clearly show the detection of HR 8799c.  (Right panel) The image 
with a smaller dynamic range to better illustrate the marginal detection of HR 8799b.}  
\label{cliols}
\end{figure}

\begin{figure}
\plottwo{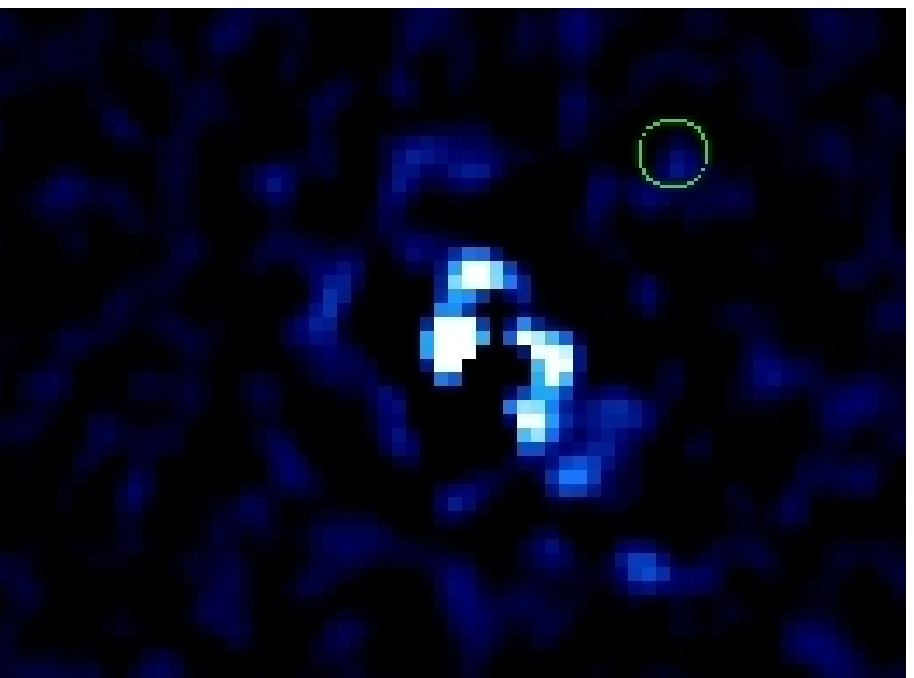}{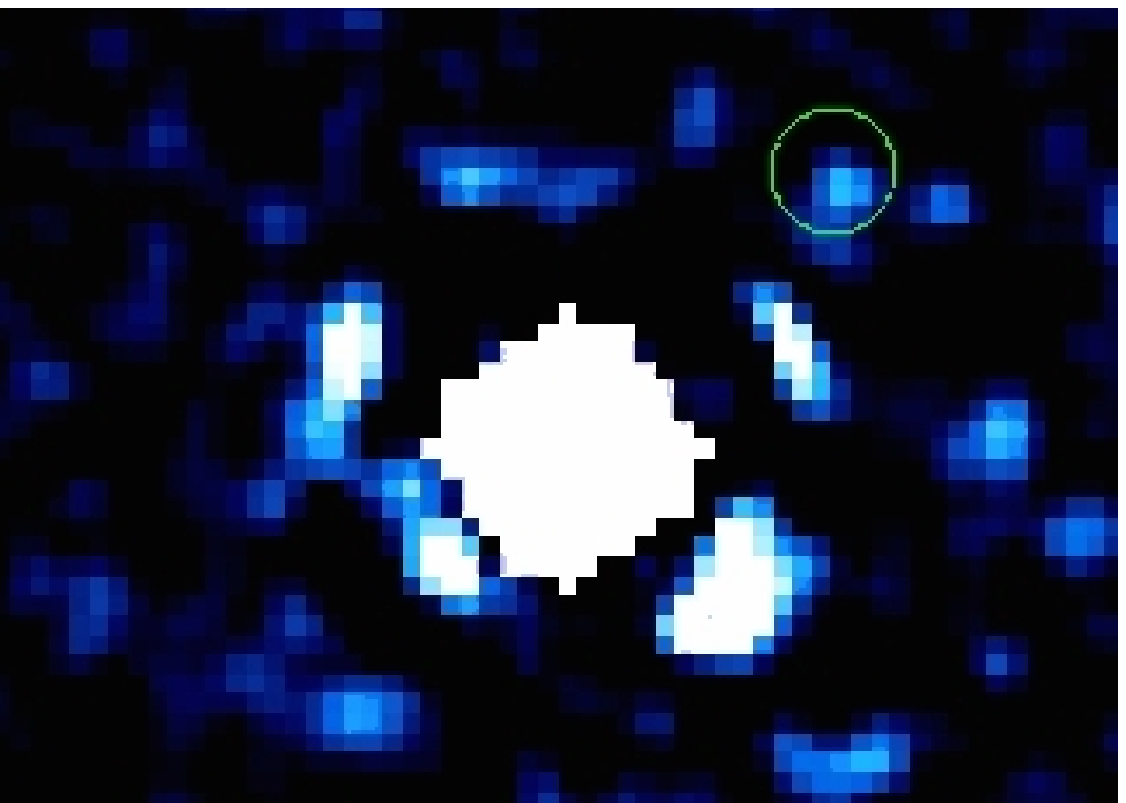}
\caption{MMT/Clio image at M'/4.8 $\mu m$ shown with slightly different procedures for PSF subtraction 
(left panel, simple ADI reduction; right panel, radial profile subtraction with LOCI reduction) and 
different dynamic ranges (left panel, high dynamic range; right panel, low dynamic range to show residual 
noise).  The circle identifies the centroid position of HR 8799c in the L' 
image obtained on the same night.  While a local peak appears near the position of HR 8799c, 
we do not identify any $>$ \textit{3-$\sigma$} peaks consistent with any of the planets in these images.}
\label{cliomp}
\end{figure}

Figures \ref{cliolp}, \ref{cliols}, and \ref{cliomp} show reduced images obtained 
with MMT/Clio in the L', [3.3], and M filters.
In the L' filter, we detect HR 8799bcd with signal-to-noise higher than that 
reported by \citet{Hinz2010} and comparable to that obtained in shorter Keck/NIRC2 exposures 
by \citet{Marois2008}.  In the [3.3] filter, we detect the c planet at SNR $>$ 5.  
We marginally detect the b planet at SNR $\sim$ 3.8.  \citet{Hinz2010} formally report a nondetection 
for b at [3.3] as they adopt a 3$\sigma$ threshold for detections, though
they identify a cluster of pixels $\sim$ 2.8 $\sigma$ above the background consistent with b 
and roughly coincident with our MMT/Clio and VLT/NaCo detections.

Conversely, we do not detect HR 8799d in [3.3], while \citet{Hinz2010} report 
a low-significance detection.  This disagreement is surprising since LOCI greatly improves the planet sensitivity 
at small separations such as that for the d planet \citep{Lafreniere2007a}.   
Furthermore, there is a 40mas offset between the reported HR 8799d centroid from \citeauthor{Hinz2010} and that 
from our 10 $\sigma$ VLT/NaCo detection obtained three weeks later.  While their detection is likely instead residual 
speckle noise, our qualitative conclusion that HR 8799d is very faint at 3.3 $\mu m$ is consistent 
with theirs.  As with \citet{Hinz2010}, we do not detect any of the planets at M band.  

\subsection{Photometry for Detections and Upper Limits for Non-Detections}
Photometry for each dataset was performed with IDLPHOT with the aperture radius set to 
the 0.5$\times$FWHM$_{image}$.  In all exposures, the stellar PSF core is either saturated 
or obscured.  For initial photometric calibration, we obtained observations of the stellar primary 
viewed through a neutral density filter (MMT/Clio, VLT/NaCo) or observed standard 
stars immediately prior to and after our science exposures (Subaru/IRCS).

\begin{figure}
\centering
\plotone{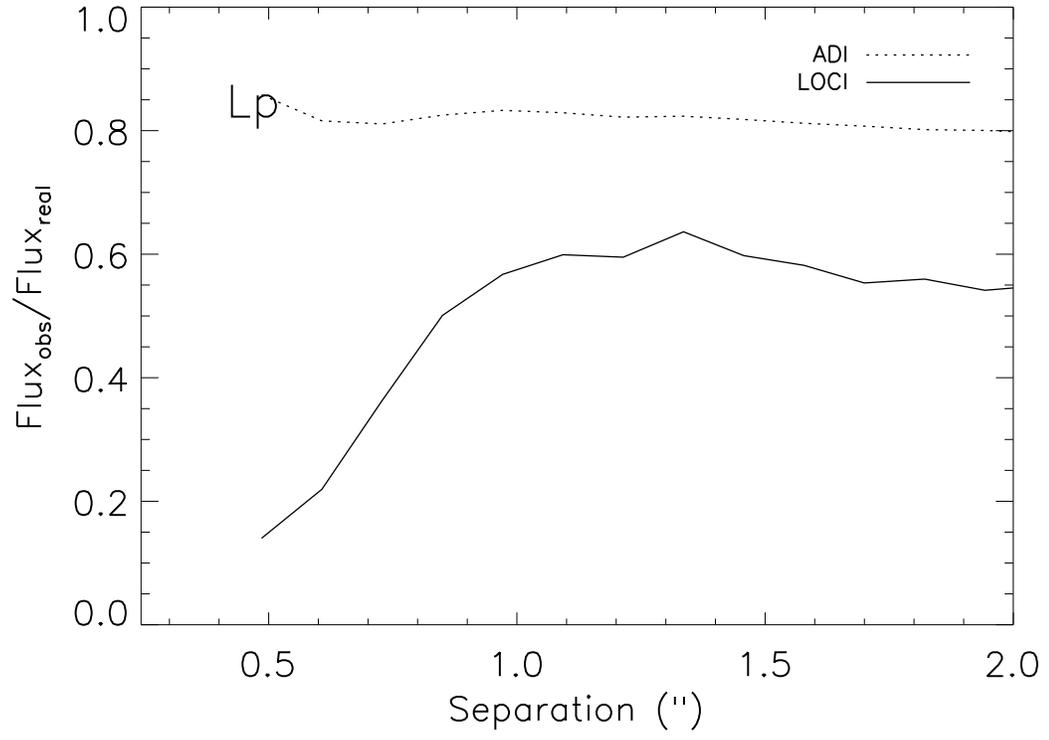}
\caption{Plot of the point source self subtraction for LOCI as a function of separation 
for our MMT/Clio data for a simple ADI reduction and our LOCI reduction.  LOCI attenuates 
more flux, especially interior to 0.75".}
\label{attenuate}
\end{figure}

Faint companions to stars observed in ADI and processed with LOCI 
lose flux due to field rotation and self-subtraction.
To further calibrate our photometry, we introduce and measure the flux for faint point 
sources at random position angles over separations encompassing to the HR 8799 planets (0.25"--2") 
in each registered frame, rerun our ADI/LOCI pipeline, compute the attenuated flux in the final, 
processed images, and correct for this attenuation.  Figure \ref{attenuate} illustrates 
this flux loss, comparing the input and output flux for fake points sources for our MMT/Clio 
L' data.  While images processed using a simple ADI reduction lose $\sim$ 20\% of their flux, 
self subtraction is stronger with LOCI, especially at separations less than 0.75".
 The attenuation curves obtained for data in other filters do not differ qualitatively: LOCI always 
attenuates more flux than a simple ADI reduction and attenuation is significantly more severe at small separations.

To place limits on our nondetections, we compute 3$\sigma$ upper limits where we correct 
our nominal sensitivity limits for point source self-subtraction inherent in ADI/LOCI.
The noise is defined in concentric annuli as before, since in most cases (for HR 8799 c and d) 
radially-dependent speckle noise dominates over radially independent photon noise.
Despite using LOCI, our detection upper limit at 3.3 $\mu m$ for HR 8799d is brighter than the 
magnitude listed by \citet{Hinz2010}.  Moreover, our upper limits for HR 8799bcd 
at M are consistently brighter than those reported by \citeauthor{Hinz2010}.  

In both cases, the disagreement is likely 
explainable by our correction for point source self subtraction in deriving upper limits 
from the standard deviation of pixel values.
  \citeauthor{Hinz2010} construct 
a reference PSF by median-combining all frames and then subtract this reference PSF 
from each image.  For the 3.3 $\mu m$ data, our reduction pipeline predicts that 
this processing should attenuate about half of the point source flux at HR 8799d's position\footnote{While 
the total field rotation is large, $\sim$ 127 degrees, the vast majority of 
the frames were taken over a time interval with only 30 degrees of field rotation.}.
For the M band data, field rotation is poorer and thus self subtraction with this 
reduction procedure is severe, reaching over 75\% at the position
of HR 8799d as nearly half the frames are obtained $\sim$ 3 hours after transit and thus 
at essentially one position angle.  Thus, the gain in sensitivity due to LOCI 
is reduced by self subtraction, resulting in brighter 3 $\sigma$ upper 
limits.

\section{Photometric Analysis: Constraining the Atmospheric Properties of the 
HR 8799 Planets}
Combining our data with that from \citet{Marois2008} yields planet flux measurements 
at nine separate wavelengths from 1 to 5 $\mu m$.  In this section, 
we use this rich multiwavelength 
sampling of HR 8799's planet SEDs to provide an empirical comparison 
with other cool, substellar-mass objects and simple atmospheric modeling 
constraints on the planets' properties.

\subsection{Near-to-Mid IR Colors of the HR 8799 Planets}
\subsubsection{Methodology}
To compare the near-to-mid IR properties of the HR 8799 planets with those 
for other cool, substellar objects, we primarily use the sample of L/T dwarfs compiled by 
\citet{Leggett2010}.  The L/T dwarf sequence defined by the \citeauthor{Leggett2010} 
sample allows us to determine how the HR 8799 planet SEDs deviate from 
those for brown dwarfs of similar temperatures.  
To explore how the HR 8799 planet SEDs 
compare to those for other planet-mass objects and very low-mass brown dwarfs with 
T$_{eff}$ = 800-1800K, we include 2M 1207b (5 M$_{J}$), 1RXJ1609.1-210524 (9 M$_{J}$), AB Pic (13.5 M$_{J}$), 
and HD 203030b ($\sim$ 25 M$_{J}$) \citep{Chauvin2004,Lafreniere2010,Chauvin2005,Metchev2006}.  
Table \ref{photcomptable} lists photometry for these objects.

We use color-magnitude diagrams constructed from the Y, J, H, K$_{s}$, and L' 
filters to determine whether the HR 8799 planets are similar to or under/overluminous 
compared to the Leggett L/T dwarf sequence.
  For simplicity and because there is no published response function for 
the IRCS z band filter, we treat the IRCS z-band magnitudes/upper limits for 
the HR 8799 planets as synonymous with its Y-band magnitude.  

To provide a physical point of reference for the L/T dwarf sequence and the HR 8799 
color-magnitude positions, we overplot loci for standard, chemical equilibrium atmosphere models 
from \citet{Burrows2006}.  We specifically choose the Model E case, which 
assumes that the clouds are confined to a thin layer, where the thickness 
of the flat part of the cloud encompasses the condensation points of different 
species with different temperature-pressure intercept points.  Above and below the 
flat portion, the cloud shape function decays to the -6 and -10 power.  
Thus, above and below the flat portion, the clouds have scale heights $\sim$ 1/7th and 
1/11th that of the gas.  See \citet{Burrows2006} for more details.

\subsubsection{Results}
\begin{figure}
\centering
\epsscale{0.999}
\includegraphics[scale=0.46,clip]{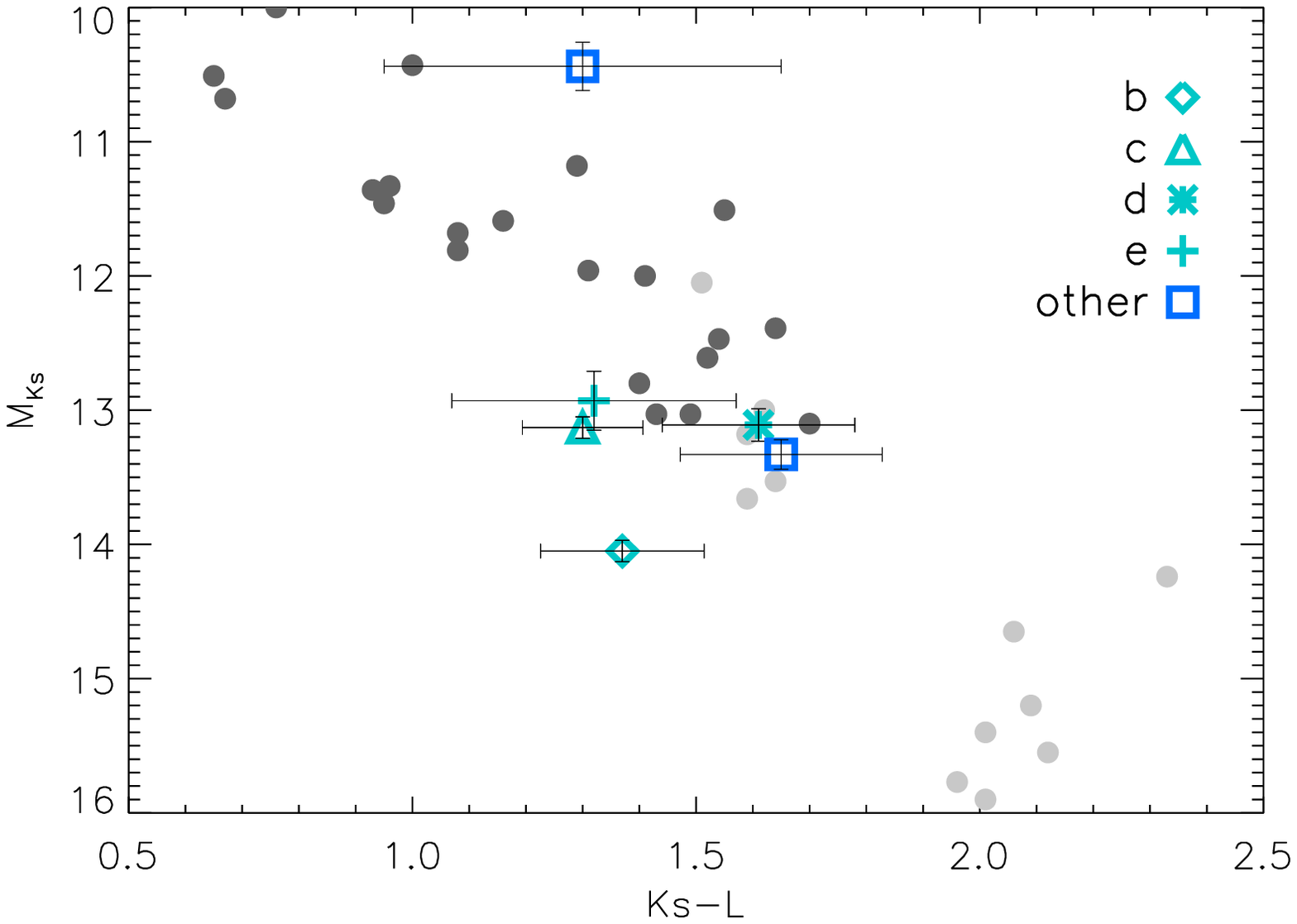}
\includegraphics[scale=0.46,clip]{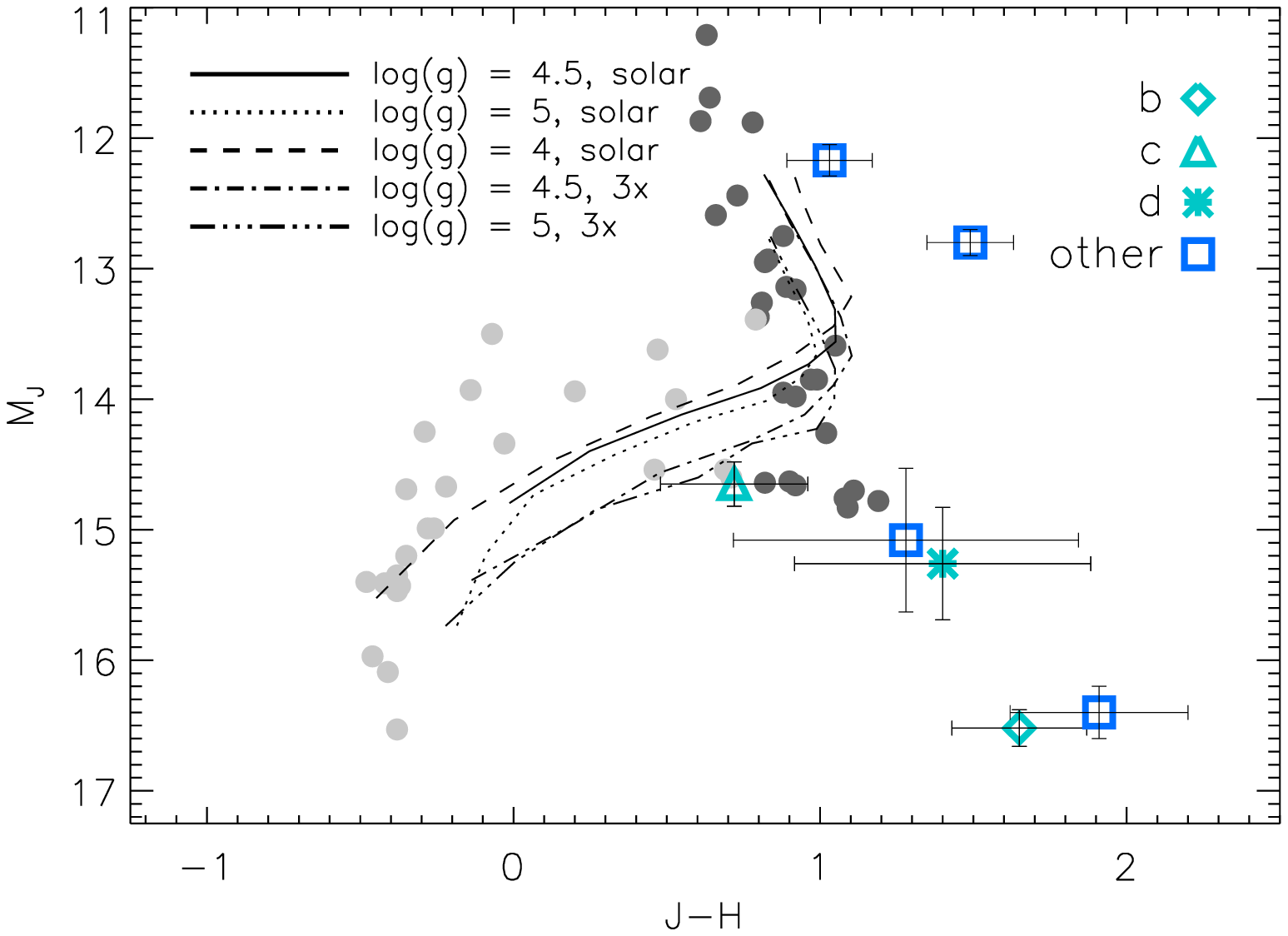}
\\
\includegraphics[scale=0.46,clip]{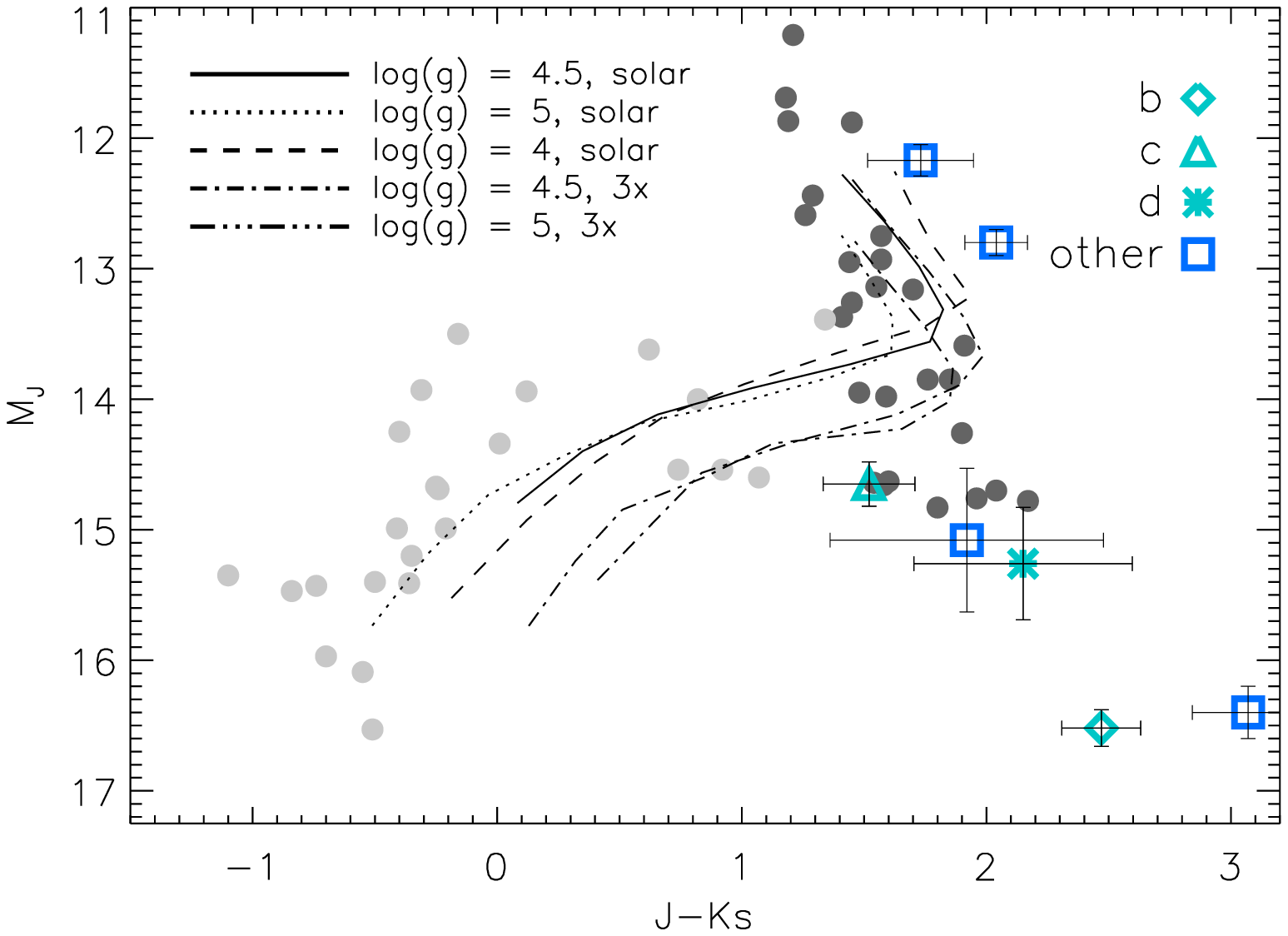}
\includegraphics[scale=0.46,clip]{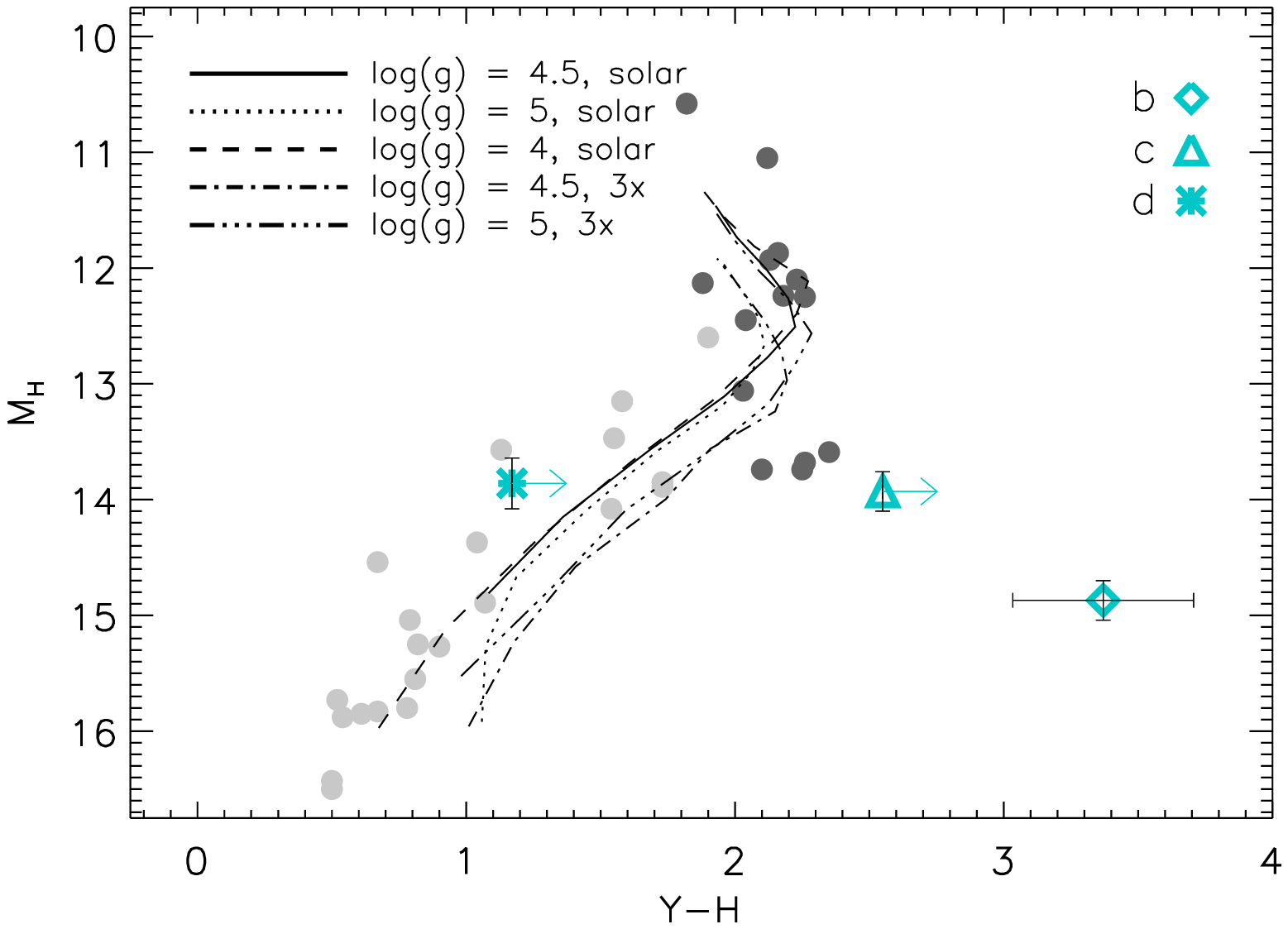}
\caption{Color-magnitude diagrams comparing the HR 8799 planets with field L dwarfs (black dots) and T dwarfs 
(grey dots) and other planetary or very low-mass brown dwarf companions (squares).  
In K$_{s}$/K$_{s}$-L' (top-left panel), the planets 
follow the L/T dwarf sequence.  In at least one of the diagrams including Y, J, and H-band data (top-right panel; 
bottom panels), the planets are red/underluminous compared to the empirical L/T dwarf sequence and the 
synthetic L/T dwarf colors from \citet{Burrows2006} for a range of metallicities and gravities. 
The positions for other planetary-mass/low-mass brown dwarf companions also depart from the L/T dwarf sequence, 
especially 2M 1207b.}
\label{ltcolseq}
\end{figure}

Figure \ref{ltcolseq} shows our color comparisons.  At least three of the HR 8799 planets have K$_{s}$/K$_{s}$-L' 
positions (upper-left panel) roughly consistent with those for the Leggett L/T dwarf 
sequence and with 2M 1207b.  HR 8799cde have positions overlapping with objects 
near the L/T dwarf boundary; HR 8799b has a similar K$_{s}$-L' color but is 
underluminous compared to the three other companions and 2M 1207b by a factor of two.  
It is unclear how its position compares to those for field L/T dwarfs because the 
sequence is poorly sampled at HR 8799b's K$_{s}$ band magnitude.

The other three panels of Figure \ref{ltcolseq} clearly show that HR 8799c, d, and especially b 
have near-IR colors that depart from the L/T dwarf sequence.  In J/J-H and J/J-K$_{s}$, 
the L/T dwarf sequence turns towards blue colors by up to 1.5 mag starting 
at the L/T transition.  
While HR 8799c's position is roughly coincident with T0 dwarfs, 
HR 8799b and d follow an extension of the slope of the L dwarf sequence 
between J/[J-H,K$_{s}$] = 11/[0.6,1.2] and 15/[1.2,2] towards fainter magnitudes 
and redder colors.  HR 8799d's position coincides with that of HD 203030b, while 
HR 8799b is located closest to 2M 1207b. 
The H/Y-H color-magnitude diagram shows that HR 8799c also is likely red/underluminous; 
HR 8799b is 2.5 magnitudes too red in Y-H for its H-band magnitude, indicating that 
it is underluminous compared to the L/T dwarf sequence at both Y and J.

Figure \ref{ltcolseq} overplots loci of standard \citeauthor{Burrows2006} models 
for parameters covering a range expected for low-mass, cool brown dwarfs -- 
T$_{eff}$ = 800-1800K, log(g) = 4--5 -- and two metallicities (solar and 3$\times$ solar)\footnote{We 
include the 3$\times$ solar models because they produce redder near-IR colors and the HR 8799 
planets are red in the near-IR compared to the L/T dwarf sequence.}.
With the exception of some L/T dwarf transition objects, the dispersion in color-magnitude positions 
for L/T dwarfs is well reproduced by model atmosphere loci.  This indicates 
that L/T dwarf atmospheres can be explained within the phase space 
encompassed by the models' assumed cloud structure and 
range in temperature, gravity, and metallicity \citep[]{Burrows2006}.

The HR 8799 planets, especially HR 8799b, are different.  They consistently 
lie below the region enclosed by the standard model atmosphere loci, indicating 
that their near-IR luminosities are weaker compared to luminosities expected 
if their cloud structure were well represented by the models.  HR 8799b 
in particular probes a completely different range of parameter space, lying 
0.75 magnitudes or more redder than \textit{any} standard atmosphere prediction 
regardless of temperature.
Thus, Figure \ref{ltcolseq} suggests a strong contrast between the atmospheric 
properties of L/T dwarfs and the HR 8799 planets.  

To summarize, all three HR 8799 planets -- especially HR 8799 b -- have near-IR 
colors that cannot be easily understood within the field L/T dwarf sequence.  They are 
consistently red and underluminous at Y and J, indicating that the 1--1.25 $\mu m$ 
portion of their SEDs are suppressed in flux.  The HR 8799 planets also lie 
well outside the loci of standard atmosphere models used to interpret the 
physical properties of L/T dwarfs.  Thus, the HR 8799 planet 
atmospheres are not simply 'scaled down' (in mass) versions of the atmospheres of 
field brown dwarfs defining the L/T dwarf sequence.

On the other hand, the planets' atmospheres show strong similarity to 
those for planetary-mass/low-mass brown dwarf companions to nearby stars.  
Specifically, HR 8799 c and d have similar near-IR colors to HD 203030b, while 
HR 8799b consistently shows near-IR colors similar to 2M 1207b. 
The planetary-mass companions 1RXJ1609.1-210524b and AB Pic b are also redder 
in near-IR colors compared to the L/T dwarf sequence but not underluminous.
Within the narrow context of our analysis, planetary-mass companions \textit{in general} 
might not follow the L/T dwarf sequence.

\subsection{Fiducial Model Atmosphere Fits to the HR 8799 Planet SEDs}
Our color comparisons motivate a further investigation of the HR 8799 
planet SEDs to better understand the source of the differences between their 
near-IR colors and those for field L/T dwarfs.
To further explore the physical properties of the HR 8799 planets we compare their photometry 
to atmospheric models.  Because the color-magnitude comparisons indicate that standard 
model atmospheres provide poor fits to the planet data, we introduce a new set of models 
to explore additional phase space not covered by the standard models, specifically a different cloud structure:

\begin{itemize}
\item \textbf{The \citet{Burrows2006} Model A Thick Cloud Layer prescription} -- Like the Model E case, this model 
defines a cloud base at the high temperature interception point with the shape function 
at higher temperatures/pressures decaying to the -10 power.  However, the cloud density 
tracks the gas density at lower temperatures/pressures (s$_{1}$ = 0 in their terminology).  
Thus, clouds in this case are far more extended high in the atmosphere 
 than in the standard Model E case.

As noted in \citet{Burrows2006}, 
these models are qualitatively similar to the AMES-DUSTY models \citep{Allard2001}.  
However, they are bluer and brighter than AMES-DUSTY in the near IR because 
\citet{Allard2001} adopts the interstellar medium particle size distribution. 
The Model A case fails to reproduce the L/T dwarf sequence as it is 
consistently too red and underluminous in IR color-magnitude diagrams \citep{Burrows2006}.  
If the HR 8799 planets have thicker clouds than L/T type 
brown dwarfs, these models -- or some hybrid between them and the "E" models -- should 
reproduce the planets' SEDs far better than the Model "E" case alone.  
\end{itemize}

Changing the cloud prescription radically alters the entire shape of the SED 
(Figure \ref{modseq}).  The K and L' band fluxes are similar.  However, the 
Model A/thick cloud prescription is underluminous over the Y and J passbands by an order of magnitude, underluminous 
at 1.65 $\mu m$ by a factor of two but overluminous in the 3.3 $\mu m$ region covering 
the trough produced by methane absorption in the Model E cloud prescription.  
Overall, the Model A SED is much flatter from 1 to 4 $\mu m$.  Additionally, 
the Model A prescription washes out the methane absorption feature at 1.65 $\mu m$ used to 
identify the L/T dwarf transition \citep[see also discussion in ][]{Burrows2006}.

Both the standard models and thick cloud layer models use the formalism
described in \citet{Burrows2006} for temperatures T$_{eff}$ = 700--1800 K, gravities 
with log(g) = 3.75--5, and solar/super-solar abundances of metals.  For both models, we assume 
modal particle sizes of 60 $\mu m$--100 $\mu m$ and a 
particle size distribution appropriate for clouds \citep{Deirmendjian1964}.  
For both models we also assume radii from \citet{Burrows1997}.  

\subsubsection{Fitting Method}

Our atmosphere model fitting follows a simplified version of the fitting procedure 
employed by \citet{Bowler2010} to model the near-IR spectrum and photometry for HR 8799b.  
Nominally, we quantify the model fits with the $\chi^{2}$
statistic, 
\begin{equation}
\chi^{2} = \sum\limits_{i=0}^{n} (f_{data,i}- 
F_{model,i})^{2}/\sigma_{data,i}^{2}.  
\end{equation}
We weight each datapoint equally.  To account for variability in emission and 
absolute calibration uncertainties, we set a 0.1 mag floor to $\sigma$ for 
each datapoint \citep[see][]{Robitaille2007}.  Because of incomplete line lists 
near the 1.6 $\mu m$ CH$_{4}$ band, we do not compare the models to data at the 
CH$_{4}$l filter \citep[see][]{Bowler2010,Saumon2007,Leggett2007}.  However, 
we confirmed that this choice has no consequential bearing on our results.

The z, [3.3], and M photometry include many nondetections.  We quantitatively 
incorporate nondetections in the following way.  For model predictions 
consistent with the 3$\sigma$ upper limits estimated for each nondetection, 
we treat the model as perfectly consistent with the data and do not 
penalize the $\chi^{2}$ value.  For model predictions inconsistent with the 
3$\sigma$ upper limits, we do not automatically discard the model.  Rather, 
we penalize the $\chi^{2}$ value by 
determining the flux ratio between the model prediction and the 3 $\sigma$ 
upper limit.  Specificially, a model prediction 2 and 4 times
brighter than the 3 $\sigma$ upper limit will be contribute 12 (3$\times$4) 
and 48 (3$\times$16) to the final 
$\chi^{2}$ value, respectively.

We fit atmosphere models in two cases.  First, to provide a straightforward 
comparison between our data and the luminosity and colors predicted from 
atmosphere models we keep the radii fixed to the \citet{Burrows1997} dwarf 
radii.  Second, we vary the radius and identify 
the scaling factor, C$_{k}$ = (R$_{scaled}/R_{nominal}$), 
that minimizes $\chi^{2}$ for a particular model: 
\begin{equation}
C_{k}^{2}=\frac{\sum\limits_{i=0}^{n} f_{data,i}F_{model,i}/\sigma_{data,i}^{2}}
{\sum\limits_{i=0}^{n} F_{model,i}^{2}/\sigma_{data,i}^{2}}. 
\end{equation}
We nominally only allow the radius to vary by $\pm$ 10\% from the 
assumed \citet{Burrows1997} values to encompass the range of radii for 
  5--20 M$_{J}$ objects at 30--300 Myr ($\sim$ 1.1--1.3 R$_{J}$).  

We determine which models are \textit{formally} consistent with the data 
by comparing the resulting $\chi^{2}$ value to that 
identifying the 3 and 5 $\sigma$ confidence limits.
For the first case, where the planet radius is fixed, the 
appropriate $\chi^{2}$ limits are 21.85 and 41.80 for 8 datapoints and 
seven degrees of freedom.  For the second case -- a variable 
planet radius -- the limits are 20.1 and 39.4 for 8 datapoints 
and 6 degrees of freedom.

To select the \textit{best-fit} models, we follow \citet{Bowler2010}
by identifying the model with the smallest $\chi^{2}$ and computing
 the $\Delta \chi^{2}$ limit for a 3 $\sigma$ confidence limit. 
'Best-fit' models satisfy $\chi^{2}_{model}$-$\chi^{2}_{best}$ $<$ 
$\chi^{2}_{99.73\%}$.  We do this separately for the Model A and E 
cloud prescriptions.

\subsubsection{Results for Standard Cloud-Deck Models}
Table \ref{standardfittable} summarizes our entire fitting results for models 
with the standard cloud deck prescription.
Figure \ref{standardfit} displays some of these fitting results 
 with the planet radii fixed to the \citet{Burrows1997} values.  
The top-left panel shows the distribution of $\chi^{2}$ values for HR 8799b; the 
top-right panel compares the HR 8799b SED to the 'best-fit' model.  

For each planet, the models with the lowest $\chi^{2}$ values 
have temperatures within 100K of those 
derived from cooling models: T$_{eff}$ = 900K, 1200K, and 1100K for 
HR 8799b, c, and d \citep[see][]{Marois2008}.  Models with a 3$\times$ solar 
abundance of metals have marginally smaller $\chi^{2}$ values.
Adopting the $\Delta \chi^{2}$ criterion, $\chi^{2}_{min}+\chi^{2}_{99.73\%}$, 
the minimum $\chi^{2}$ values for modeling b, c, and d are  300.9, 133.2, 
and 38.9.   The range of temperatures and gravities fulfilling this criterion are
T$_{eff}$ = 900--1000K, 1100--1300K, 1000--1300K and  log(g) = 4.5--5, 4.5--5, 
and 4--5 for the b, c, and d planets.  

\textit{However}, the fits are quantitatively very poor for HR 8799c and (especially) b.  
As shown by Figure \ref{standardfit} (top-left panel), the minimum $\chi^{2}$ value 
for HR 8799b is a factor of $\sim$ 5.5 times higher than the 
formal 5 $\sigma$ confidence limit.  The minimum $\chi^{2}$ value for HR 8799c 
is twice as large.
The large $\chi^{2}$ difference between that for 'best-fit' models and the 
formal 5-$\sigma$ confidence limit suggests that the models do not provide 
meaningful fits to the data.  Fits to the HR 8799d SED are not quite as poor 
but include only one model with $\chi^{2}$ $<$ $\chi^{2}_{99.73\%}$.  
Allowing the planetary radii to vary over the range plausible for 5--20 M$_{J}$ objects 
does not qualitatively improve the model fits for the b and c planets 
(Table \ref{standardfittable}).  

The top righthand panel and lower panels of Figure \ref{standardfit} illustrate how 
the models fail to reproduce the SEDs of HR 8799bcd.  For example, 
for HR 8799b the 'best fit' model provides a good estimate of its 
K$_{s}$ band and L' band fluxes and is consistent with its upper limit 
at M band.  At [3.3], however, the model predicts too deep of a trough 
due to methane absorption, underpredicting the flux by a factor of $\sim$ 3--4.   
Most strikingly, the model overpredicts the flux at Y and J band 
\textit{by over an order of magnitude}.  The model overestimates 
the H band and CH$_{4}$s flux by a factor of $\sim$ 2.  
Compared to the best-fitting models,
HR 8799c also has too strong of a 3.3 $\mu m$ flux and 
too low of a Y band upper limit.

For modeling results discussed in Figure \ref{standardfit}, 
the scaling factors for the radii are almost always C$_{k}$ =0.9
 for temperatures greater than those predicted from cooling models 
and 1.1 for lower temperatures.  
To see which radii formally yield the smallest 
$\chi^{2}$ values, we allow the radius to vary between 0.2 and 2 
times the \citet{Burrows1997} values.
The resulting trend of 
$\chi^{2}$ vs. T$_{eff}$ for all planets changes, as 
the minima are systematically pushed towards higher 
T$_{eff}$ (T$_{eff}$ = 1300--1400 K).  However, radius scale 
factors for the best-fit models imply that the planets 
are unphysically small -- R$_{b,c,d}$ $\sim$ 0.4, 0.6, and 0.7 R$_{J}$.



In summary, atmosphere models with standard, cloud deck prescriptions 
appropriate for brown dwarfs only provide statistically meaningful 
fits to HR 8799b and c for unrealistically 
small radii \citep[see also][for HR 8799b]{Bowler2010}.  
Assuming radii characteristic of planet-mass 
objects, we fail to find a single model that provides 
a statistically meaningful fit to the HR 8799b and c data 
indicating that such models provide a poor description of 
the planets' atmospheres \citep[see also][]{Marois2008,Janson2010,Hinz2010}.  
These results are independent of surface gravity for log(g) = 4--5 
and whether the planets have solar or 3$\times$ solar metallicity. 
These results then motivate 
us to see if models with different cloud prescriptions 
fare better in reproducing the SEDs of HR 8799bcd. 
 
\subsubsection{Results for Thick Cloud Layer Models}

Figure \ref{thickfit} shows and Table \ref{thickfittable} summarizes 
our fitting results for the thick cloud layer models.
Best-fit models for the HR 8799 planets cover a similar 
range in T$_{eff}$ as the standard model fits and cooling 
model predictions.  For HR 8799b, the best-fit model 
assumes T$_{eff}$ = 900K and log(g)=4.25; the range 
of best-fit models cover log(g)=4--4.5 and T$_{eff}$ 
= 900--1000K. The range in log(g) for HR 8799c and d 
are similar to that for b (log(g)=4.25--4.5 and 4--4.5), 
whereas their temperatures are slightly higher (1100--1200K 
and 1000--1200K).

As illustrated by Figure \ref{thickfit}, models with 
thick cloud layers provide \textit{far} better fits to the 
SEDs of \textit{all three} planets.  Quantitatively, the 
$\chi^{2}$ minima shrink by factors of 6, 2, and 5 for 
HR 8799b, c, and d compared to those for Model E fits.
For HR 8799b and c, the minima approach the formal 5-$\sigma$ 
confidence limit.  For HR 8799d,  multiple models have $\chi^{2}$ minima 
less than the formal 3-$\sigma$ confidence limit.

The righthand panels of Figure \ref{thickfit} illustrate 
why the thick cloud layer models are more accurate.  
For HR 8799b, the best-fit models predict a flat, rising 
SED from 1 to 1.5 $\mu m$, consistent with the planet's 
weak Y and J band emission.  The best-fit models also 
predict stronger 3.3 $\mu m$ emission than in the standard 
model case and in better agreement with HR 8799b's measured [3.3] 
flux.  While the best-fit model for HR 8799c underpredicts 
its J-band flux while overpredicting its [3.3] and L' band 
flux, the discrepancies are weaker than in the standard 
cloud model case.  With the exception of the CH$_{4}$l 
filter data, which was not incorporated into our fitting, 
the best-fit thick cloud model (log(g) = 4.25, T$_{eff}$ = 1100K) 
for HR 8799d accurately reproduces the planet's flux 
at every datapoint.

Allowing the planet radii to vary by $\pm$ 10\% slightly 
improves the model fits.  More importantly, 
results in more models with $\chi^{2}$ values 
below the formal 3$\sigma$ and 5-$\sigma$ confidence limits 
(Figure \ref{thickscalefit}).  For these models, the 
HR 8799b's range of best-fit models have log(g) = 4.25--4.5, 
and T$_{eff}$ = 900--1000K, and C$_{k}$ = 0.9--1.02; 
HR 8799c's have log(g)=4.25--4.5, T$_{eff}$ = 1100--1200K, 
and C$_{k}$ = 0.9--0.975; and HR 8799d's have 
log(g) = 3.75--4.5; T$_{eff}$ = 1000--1200K; and 
C$_{k}$ = 0.9--1.09.  As before, the scaling factor 
for each model is correlated with the model's temperature 
compared to the cooling model estimates.

\subsection{Estimates for ``Patchy"/Partly Cloudy Models}
The two models used to fit our data define limiting cases for the 
cloud structrure in planet atmospheres.  
  The Model A thick cloud layer 
prescription fits the data for each planet far better.  
However, intermediate cases -- with far thicker clouds than the Model E case 
but slightly thinner than Model A or a ``patchy" cloud coverage -- may be more 
physically realistic.  The two processes may be tied together: 
\citet{Ackerman2001} show that clouds may become patchy as they 
sediment below photospheric pressures.   Near-IR photometric variability 
detected from the T2.5 brown dwarf SIMP J013656.5+093347 is consistent 
with grain free, cloudless regions and grain-bearing cloudy regions 
rotating in and out of view \citep{Artigau2009}.  Cloud patchiness 
may also be important for defining the L/T dwarf transition \citep[e.g.][and 
references therein]{Marley2010}.

We leave a detailed construction 
of such models to a future paper (Madhusudhan et al., in prep.) but 
here we qualitatively explore how intermediate cases may affect the predicted 
planet spectrum \citep[see also][]{Marley2010}.  Similar to \citet{Burgasser2002}, we follow a highly 
simplified, crude approach by combining weighted sums of Model A and E cloud 
prescriptions to approximate an atmosphere whose cloud thickness varies 
over the seeing disk of the planet\footnote{Technically, this is not physically 
realistic as the temperature-pressure profiles for cloud layer and cloud deck 
regions would be discontinuous.  On the other hand, for a given T$_{eff}$ 
 self-consistent models with intermediate cloudiness \citep{Marley2010} 
have color-magnitude positions intermediate 
between the two extremes, broadly consistent with simple parameterizations
 \citep[e.g.][]{Burgasser2002}.}.  
For simplicity, we compare two parameterizations: a ``partly cloudy" approximation 
where we weight the thick cloud model by 60\% and a ``mostly cloudy" approximation
 where we weight 90\% of the surface by the Model A case.

Figure \ref{patchyfit} shows modeling results for these two cases compared against 
the thick cloud layer results for log(g) = 4 and 4.5.  Our approximations 
yield smaller $\chi^{2}$ minima for HR 8799b and c; models with 
partly/mostly cloudy approximations have the smallest $\chi^{2}$.  
The best-fit model for HR 8799b has T$_{eff}$ = 900K, consistent with the thick 
cloud layer model, while temperatures for HR 8799c and d are lower by 100K.

   While our approach is entirely ad hoc, it indicates that slightly weakening 
clouds compared to the limiting Model A case may provide better fits, 
at least for low surface gravity models (log(g) = 4).  
Madhusudhan et al. (2011) present a set of new atmosphere models with 
a range of cloud coverages intermediate between the Model A and E cases 
to explore how varying the cloud strength between these two extremes 
affects fits to the data.

\section{Dynamical Stability Analysis}
As shown by \citet{Fabrycky2009} and \citet{MoroMartin2010}, stability analysis of the HR 8799 system 
constrains the planet masses independently of planet cooling and atmospheric modeling.  Here, 
we investigate the plausible mass range of companions imposed by dynamical stability.  
Later, we will combine the results of these simulations with the implied mass range from atmospheric 
modeling to identify planet masses consistent with both atmospheric modeling and dynamical 
stability analysis.
\subsection{Procedure}
Using the Swifter N-body code, an updated version 
of the Swift package \citep{Duncan1998}, we integrate the equations of motion for the HR 8799 planets. 
We adopt the Burlirsh-Stoer method to treat close encounters.  For all simulations we  
use an accuracy parameter of 10$^{-12}$ and dynamically evolve the system until one or more planets 
are ejected or until 100 Myr is reached.

We expand the analysis of \citet{Marois2011} who 
searched for stable orbital configurations for two sets of planet masses  incorporating HR 8799e 
-- 5, 7, 7, and 7 M$_{J}$; 7, 10, 10, 10 M$_{J}$ for b, c, d, and e.  
We assume a single-2:1 resonance between 
c and d for the former and a double-2:1 resonance for d-c and c-b pairs for the latter.
We hereafter refer to these sets of initial conditions as Cases A and B.
To expand upon the \citet{Marois2011} investigation, we considered a wider range of masses for HR 8799bcde -- 
10, 13, 13, and 13 M$_{J}$ -- with the same double resonance configuration as Case B.  We 
refer to this set of initial conditions as Case C.  In 
all cases, we simply require the system to be stable for 30 Myr -- the minimum age of HR 8799 -- 
to be consistent with the data.  

We do two sets of 8000 simulations for each case.  In the first set, we allow HR 8799e to vary 
in separation from 13.1 AU to 15.7 AU.  
This allows us to identify general trends in the time to instability vs. separation for HR 8799e.
In the second set, we more finely sample initial orbital properties for the planets assuming a range 
of 14--15 AU for HR 8799e to better identify stable solutions.

\subsection{Results}
Figure \ref{dynamics} illustrates our simulation results.  
The top panel displays the time to instability for Case A.  
The bottom-left and 
bottom-right panels show the same plot for Cases B and C, respectively.
The first set of simulations allowing HR 8799e to range from 13.1 AU to 15.7 AU are shown 
as orange lines; the second set are shown as black lines.

Our results show that the HR 8799 companions must have masses below the deuterium-burning 
limit \textit{based on dynamics alone}.
Case C configurations are typically only stable for 0.01 Myr and 
\textit{never} stable for more than 10--20 Myr.  Because HR 8799 is 
a main sequence star, it cannot be as young as 10--20 Myr.  Therefore, companion masses 
for HR 8799cde $\ge$ 13 M$_{J}$ and a mass for HR 8799b $\ge$ 10 M$_{J}$ can be ruled out.

Lower planet masses are strongly preferred on dynamical 
grounds, consistent with the results of \citet{Marois2011}.  
Only seven Case B configurations are stable for $\sim$ 30 Myr, nearly all of which require separations for 
HR 8799e more than 1-$\sigma$ different from the position implied by our astrometry.
Only two are stable for 100 Myr, and these likewise require anomalously small separations.
On the other hand, sixty Case A configurations are stable for 30 Myr.
Three are stable for 100 Myr, one of which places HR 8799e at a separation consistent 
with our astrometry. Our dynamical stability results are 
 in rough agreement with \citet{Marois2011}'s results.  They find 12 solutions 
out of 10$^{5}$ possible solutions stable for more than 100 Myr, where HR 8799e varies between 
14.35 AU and 14.56 AU.  We find 3 out of 1.6$\times$10$^{4}$ solutions are stable for 100 Myr over this semimajor 
axis range.

In summary, we can rule out companion masses greater than 10 M$_{J}$ for HR 8799b and 
13 M$_{J}$ for the others.  The companions cannot be brown dwarfs.  Systems with 
masses of 5 M$_{J}$ for HR 8799b and 7 M$_{J}$ for the others are characteristically 
far more stable than systems with larger masses.  We fail to find any stable configuration 
with 7, 10, 10, and 10 M$_{J}$ for HR 8799bcde's masses that place HR 8799e at a position 
consistent with our astometry.  While our investigation is not exhaustive, it 
implies that masses of less than 7 M$_{J}$ for HR 8799b and less than 10 M$_{J}$ for 
the others are most plausible.

\section{Discussion}
Our primary result in this paper is that the atmospheres of 
at least two and potentially all of the HR 8799 planets do not 
easily fit within the empirical IR color sequence for L/T type brown dwarfs of 
similar temperatures, nor can they be well fit by standard atmosphere 
models used to infer the properties of brown dwarfs.  Adopting realistic 
assumptions about planet radii, all atmosphere model fits to data 
for HR 8799b and c are far poorer than any meaningful threshold 
identifying models consistent with the data.  The models primarily 
fail by underpredicting the 3.3 $\mu m$ flux and badly overpredicting 
flux at 1--1.3 $\mu m$.

Our analysis suggests that having ``thicker" clouds --ones with larger 
vertical extents -- is key to reproducing the planets' SEDs.  Compared 
to cloud structures assumed in standard L/T dwarf atmosphere models, 
these clouds are more optically thick at a given T$_{eff}$, so they are 
visible (in the photosphere) at a lower T$_{eff}$ even though the 
cloud base is located far below at much higher pressures.  
Adopting a thick cloud layer prescription, we 
succeed in identifying models for each planet that quantitatively are good-fitting 
models.  Moreover, the temperatures of these models are consistent with simpler, presumably 
more accurate cooling model estimates.

\subsection{Comparisons with Previous Studies of HR 8799}
The most direct comparison to this work is the recent analysis of the HR 8799b K-band spectrum 
and 1.1--4.1 $\mu m$ photometry from \citet{Bowler2010} whose modeling formalism we largely follow.  
\citet{Bowler2010} also finds difficulties in using 
standard atmosphere models to fit HR 8799b's SED and interpret its properties 
 \citep[see also][]{Marois2008}.  
Likewise, they find that temperatures inferred from standard atmosphere models 
disagree with cooling model predictions and that the former require unphysically 
small radii.

Our results indicate that including Y/z band data only exacerbates the already serious 
disagreement between standard cloud deck model predictions and the planet's SED.  
Our analysis confirms \citet{Bowler2010}'s inference that HR 8799b's atmosphere is exceptionally 
dusty compared to field brown dwarfs.  Our results extend this inference, indicating
 that HR 8799c and, plausibly, HR 8799d are also dusty compared to field brown dwarfs.

\citet{Janson2010} noted that while standard atmosphere models -- the COND models in 
their case -- can reproduce the mean brightness of HR 8799c's L'-band spectrum they 
incorrectly predict the spectral slope from 3.9 $\mu m$ to 4.5 $\mu m$.  They 
cite greater atmospheric dust absorption and, especially, non-equilibrium 
carbon chemistry as features that may bring the models into better agreement.  \citet{Hinz2010} 
argue that incorporating non-equilibrium chemistry is necessary to reproduce the 
mid-IR photometry of HR 8799bcd since the chemical equilibrium models they use 
\citep{Saumon2006} predict M-band fluxes larger than the upper limits they report. 

Non-equilibrium carbon chemistry has little effect on the near-IR 
portion of the SED \citep[e.g.][]{HubenyBurrows2007}.
Thus, our analysis indicates that thicker clouds -- and, by implication, stronger atmospheric 
dust absorption -- are far more important than non-equilibrium chemistry in 
reproducing the HR 8799 planet 1--5 $\mu m$ SEDs.  
Nevertheless, the HR 8799 planet atmospheres are plausibly not in local chemical equilibrium.
Since departures from chemical equilibrium alter the spectral structure at 
4--5 $\mu m$, non-equilibrium chemistry incorporated into thick or ``patchy" cloud models 
may yield better fits to 1--5 $\mu m$ photometry and mid-IR spectroscopy of the planets.  
Higher signal-to-noise L' band spectra and detections/more stringent upper limits 
at M will better identify evidence of non-equilibrium chemistry in the planets' atmospheres.

\subsection{Comparisons with Planet Evolution Models and Implied Masses}
Within the context of the \citet{Burrows1997} planet cooling models, a particular combination 
of log(g) and T$_{eff}$ defines an object with a mass M and age t.  Taking the 
gravity and temperature range implied by our modeling at face value, we can then identify the 
mass and age range implied.  Our modeling efforts succeed in yielding planets with 
physically realistic radii.  However, if our range of log(g) and T$_{eff}$ were to imply 
wildly discrepant masses compared to cooling model estimates and dynamical stability requirements 
or widely varying ages our analysis would have solved one problem only to create comparably serious ones.  

Here, we combine all modeling results to identify the range of 
best-fit parameters and implied parameters -- mass and age -- 
from atmosphere models that we consider. 
We then determine whether the atmospheric and dynamical modeling 
constraints are consistent and, if so, what mass and age range they imply. 

\begin{itemize}
\item {HR 8799b} -- The minimum $\chi^{2}$ value for HR 8799b 
for thick cloud models is 27.6 if we allow the radius to vary by 
up to 10\% from the \citet{Burrows1997} values 
and 48.9 if we don't.  For the ``patchy" cloud approximation, the 
corresponding $\chi^{2}$ minima are 20.6 and 51.4.  Considering the best-fit models 
passing the $\Delta \chi^{2}$ threshold in each case, this range 
covers log(g) = 4--4.5 and T$_{eff}$ = 800--1000K.  Thus, our modeling yields 
log(g) = 4--4.5, T$_{eff}$ = 800--1000K.  Using the \citet{Burrows1997} 
evolutionary models, this implies a mass and age range of 
M, t = 5 M$_{J}$, 30 Myr to 15 M$_{J}$, 300 Myr.

\item {HR 8799c, d, and e} -- The minimum $\chi^{2}$ values here for 
thick cloud models are 
 43.5 and 60.7 for c and 5.7 and 5.3 with and without radius rescaling.  
For the ``patchy" cloud approximation, the corresponding $\chi^{2}$ 
minima are 14--14.1 for c and 2.8--7.4 for d.
For HR 8799c, the range of models passing the $\Delta \chi^{2}$ threshold 
for the thick and patchy cloud prescriptions
cover log(g) = 4--4.5 and T$_{eff}$ = 1000K--1200K.  This yields 
a mass/age range of 7 M$_{J}$, 30 Myr to 15--17.5 M$_{J}$ at 150--300 Myr.
  For HR 8799d, the range is log(g) = 3.75--4.5, T$_{eff}$ = 1000-1200K, 
yielding 5 M$_{J}$ at 10 Myr to 15--17.5 M$_{J}$ at 150--300 Myr. 
Since HR 8799e likely has a bolometric luminosity and K-L colors 
comparable to HR 8799c and d, its range of masses is plausibly 
consistent with those derived for HR 8799c and d.
\end{itemize}

Dynamical constraints require that HR 8799b is less than 7 M$_{J}$ and 
HR 8799cde are less than 10 M$_{J}$ \citep[Section 4 of this work;][]{Marois2011}.
The 5 M$_{J}$ mass estimate for HR 8799d can be ruled out because 
the primary star is on the main sequence and thus cannot be 10 Myr old.
Coupled with the range in surface gravities and temperatures, 
the implied range in masses are then 6--7 M$_{J}$ for HR 8799b, 7--10 M$_{J}$ for 
HR 8799c, and 7--10 M$_{J}$ for HR 8799 d.  If HR 8799e's atmospheric properties mirror those 
of c and d, its plausible range of masses is also 7--10 M$_{J}$.  
Conversely, for these ranges of masses, the surface gravities of HR 8799bcde should 
be no greater than log(g) $\approx$ 4.25.

These estimates are consistent with cooling model estimates 
from \citet{Marois2008,Marois2011}.  For the lower end of the mass ranges, 
the system age corresponding to these models is $\approx$ 30 Myr and 
puts HR 8799's age on the low end of the 30--160 Myr range quoted by \citet{Marois2008}.
The (disfavored) high end of the mass range corresponds to $\sim$ 100 Myr-old objects.

Despite our success in arriving at self-consistent answers for the planets' masses 
and ages, we strongly caution against overinterpreting these results.  
Our results do not \textit{prove} that, above the cloud base, the vertical density/pressure 
profile of clouds follows that of the gas as a whole (e.g. s$_{1}$ = 0), 
as opposed to being truncated at higher pressures.  Neither do our 
results prove that other models with slightly different assumptions about 
the clouds, grain particles, atmospheric chemistry, etc. provide better 
fits to the data.  In particular, slight modifications to our models may improve 
the fit at L' band, the datapoint responsible for much of the $\chi^{2}$ contribution 
for HR 8799b.  Even within the context of our adopted physical models, 
our sampling in temperature and gravity is also too coarse to precisely 
estimate best-fit atmosphere parameters.  

On the other hand, our analysis provides compelling evidence for thick clouds, 
 motivates future modeling work to 
test how different assumptions about thick clouds affect model fits 
to planetary atmospheres, and encourages further observations of 
substellar objects to test these models.
Madhusudhan et al. (2011) will develop and better assess model fits for 
varying cloud strengths and more precisely and accurately determine 
temperatures and gravities for the HR 8799 planets and other planetary-mass 
objects.

\subsection{Constraints On The Formation of the HR 8799 Planetary System}
The planets' large masses and wide orbits make them a particularly 
interesting probe of planet formation.
The favored theory invoked to explain the formation 
of gas giant planets is \textit{core accretion} \citep[e.g.][]{Mizuno1980,Pollack1996,KenyonBromley2009, 
Chambers2010}, where cores that have 
grown to $\approx$ 5--10 M$_{\oplus}$ rapidly accrete much more massive gaseous envelopes.
Alternatively, planets could form by disk instability \citep[][and later papers]{Boss1997}, where 
the protoplanetary disk is massive and gravitationally unstable, forming multiple self-gravitating 
clumps of gas that coalesce into bound, planet-mass objects.  

HR 8799's planets are often described as confounding either planet formation model 
\citep[e.g.][]{Marois2011} or being clear examples of disk instability-formed planets, 
as claimed by \citet{DodsonRobinson2009}.  
They find that cores at distances characterizing the HR 8799 planets 
cannot reach $\sim$ 10 M$_{\oplus}$ in mass to undergo runaway gas accretion 
\textit{even under the most favorable conditions}.  They 
claim that planet-planet scattering cannot create stable, wide-orbit systems like 
HR 8799's.  They conclude that massive, wide-separation gas giants like HR 8799bcd 
form by disk instability and ''can certainly rule out core accretion".

Critical to \citeauthor{DodsonRobinson2009}'s conclusion is their treatment of the core growth rate.  
The growth rate strongly depends upon the planetsimal approach velocity, which they fix at v$_{a}$ = $\Omega$R$_{hill}$.  
They claim this velocity yields an ``optimistically high" growth rate.  Their formalism implicitly 
assumes that planetesimals have an isotropic velocity dispersion (v$_{a}$ $\sim$ v$_{z}$), which 
is valid as long as the scale height of planetesimals accreted by cores (v$_{z}$/$\Omega$) 
is larger than the core's impact parameter, R$_{core}$$\sqrt{(1+\theta)}$ \citep[][]{Rafikov2004}, where 
$\theta$ is the Safranov number.  However, 
if the planetesimals are dynamically cold such that v$_{z}$ $\le$ $\sqrt{p}$$\Omega$R$_{Hill}$ 
(where p = R$_{core}$/R$_{Hill}$), this condition is violated 
\citep{Dones1993,Rafikov2004}.  The core can then 
accrete the entire vertical column of planetesimals at a vastly 
higher rate since accretion is now essentially two-dimensional \citep{Rafikov2004}.   

As a result, \citet{DodsonRobinson2009} 
catastrophically underestimate the maximum growth rate by a factor of p$^{-1/2}$, or up to 114, 
85, and 68 at the positions of HR 8799b, c, and d (cf. Equations 78, 80, and 82 in Rafikov 2004; see 
also Rafikov 2010)\footnote{At first glance, Equation (17) in \citet{Rafikov2010} appears to imply that 
the limiting distance for core accretion in shear-dominated growth is comparable to \citeauthor{DodsonRobinson2009}'s 
estimate (44 AU vs. their 20--35 AU).  However, \citeauthor{Rafikov2010}'s result of 44 AU is valid for a 
Minimum Mass Solar Nebula case \citep{Hayashi1981}.  Adopting initial assumptions more comparable to those that 
\citeauthor{DodsonRobinson2009} assumes -- e.g. a disk more massive than the Minimum Mass Solar Nebula or 
a longer-lived one with $\tau_{disk}$ = 5 Myr instead of 3 Myr-- implies that gas giants can in some cases form by core accretion 
at separations comparable to HR 8799c and b.}.  Detailed numerical simulations confirm that this rapid growth 
phase can be reached if collisional fragmentation and gas drag 
are properly treated \citep{KenyonBromley2009}.  The \citeauthor{DodsonRobinson2009} 
planet-planet scattering simulations also were conducted assuming gas free, planetesimal-free 
conditions and assumed that planets could not further grow after scattering.  
However, gas drag and dynamical friction from planetesimals are 
critically important as they promote orbit circularization and stability 
\citep[e.g.][]{Goldreich2004,FordChiang2007}\footnote{In fairness, they clearly acknowledge 
that their study does not consider planet-planet scattering \textit{in a gaseous disk}, which may 
result in a more favorable outcome for core accretion.}.
Cores with masses sufficient for rapid gas accretion can circularize after being scattered to the 
outer disk \citep[][S. Kenyon 2010, pvt. comm.]{BromleyKenyon2011}.  Simulations 
by Thommes et al. (in prep.) show that the HR 8799 planet cores could acquire most of their gas 
\textit{after} scattering.   

The mass ratio and semimajor axis distribution of wide planets and low-mass brown dwarfs 
may help constrain the formation mechanism for HR 8799's planets \citep{Kratter2010}.  Core accretion 
preferentially forms planets with smaller masses and orbital separations, while 
disk instability has difficulty producing lower-mass gas giants and forming them 
close to the star \citep[e.g.][]{Rafikov2005,Kratter2010}.  
Therefore, if HR 8799bcde formed by core accretion (disk instability), they should 
comprise the high-mass extrema (low-mass tail) of a population continuous 
with radial-velocity detected planets (brown dwarf companions).
Using our new results for the masses of the HR 8799 planets, we update \citeauthor{Kratter2010}'s 
plot comparing planet and brown dwarf distributions.  
We also add the planet-mass companions to 
1RXJS1609.1-210524, and 2M J044144b \citep[5--10 M$_{J}$, 15 AU]{Todorov2010}; 
the planet/brown dwarf companion to GSC 06214-00210B \citep[14 M$_{J}$, $\sim$ 300 AU][]{Ireland2010}; 
and the low-mass brown dwarf companion GJ 758B \citep[25--40 M$_{J}$, 44 AU][]{Currie2010a}.

As shown by Figure \ref{kratterplot}, the revised masses for the HR 8799 planets 
and the addition of HR 8799e expand the space between them and brown dwarf companions 
(asterisks).  Visually, they join with the distribution of closer-separation planets 
plausibly formed by core accretion.  The other new companions 
 (red triangles) are continuous with brown dwarfs that may form by disk fragmentation.

While core accretion -- especially when coupled with planet-planet scattering -- 
may form the HR 8799 planetary system, HR 8799-like systems are still 
plausibly uncommon.  The Gemini Deep Planetary Survey of 85 nearby, young (mostly solar-mass) stars 
was typically sensitive to 2 M$_{J}$ planets at 40--200 AU yet failed to detect any 
\citep{Lafreniere2007b}.  
Similarly, non-detections from the deep (M $<$ 1 M$_{J}$) survey from \citet{Kasper2007} showed 
that the giant planet populations detected at small separations (a $\lesssim$ 4 AU) by RV surveys
cannot extend to separations larger than $\sim$ 30 AU.
More massive stars like HR 8799 likely have more massive disks, which aid gas 
giant planet formation.  However, their disks also dissipate much more rapidly \citep{CurrieLada2009}: 
even if critical core masses are reached, the leftover mass of gas may be small.    
Moreover, rapid core growth results from being able to fragment and then dynamical cool 
the surrounding planetesimal population.   The current state-of-the-art 
simulations show that this requires Pluto-mass cores to start with 
\citep[e.g.][]{KenyonBromley2009}, yet the formation time 
for Pluto-mass objects at wide separations may be long \citep[e.g.][]{Rafikov2010}.
Thus, forming HR 8799-like systems by core accretion is difficult, though \textit{not} impossible, 
and probably happens infrequently.

\subsection{Implications for the Atmospheres of Other Substellar Companions: A Possible
Fundamental Difference Between Planetary-Mass Objects and M $>$ 15--20 M$_{J}$ Brown Dwarfs}
In some ways, the difficulty in reproducing the IR SEDs of the HR 8799 planets mirror
difficulties in modeling other planetary-mass objects detected prior to HR 8799bcde.
In particular, 2M 1207b also appears discrepant compared to 
 standard atmosphere models as noted in \citet{Mohanty2007} and discussed in this work.
Like HR 8799b, 2M 1207b is noticeably underluminous ($\sim$ 2.5 mags) in the near-IR
\citep[][ this work]{Mohanty2007}.

To explain 2M 1207b's properties, \citet{Mohanty2007}
propose that the object is occulted by an edge-on disk with large, gray dust grains.
Alternatively, \citet{Mamajek2007} propose that 2M 1207b's properties can
be explained as resulting from a recent protoplanet-protoplanet collision.
Comparing high-resolution spectra of 2M 1207b to the DUSTY atmosphere models from
\citet{Allard2001}, \citet{Patience2010} identify a problem similar to that
noted for modeling HR 8799b from \citet{Bowler2010} and this work.  Namely,
allowing the object radius to freely vary yields best-fit radii far smaller ($\approx$ 0.5 R$_{J}$)
than is physically plausible \citep[cf.][]{Burrows1997}.  \citet{Patience2010} also conclude
that extinction from an edge-on disk comprised of gray dust grains is also a viable
scenario.

For the same reasons -- underluminosity/red colors -- a disk origin also has been
proposed to explain the IR SED of HR 8799b and (to a lesser extent) c and d \citep{Marois2008}.
However, \citet{Marois2008} consider the chance alignment of an edge-on circumplanetary disk to be unlikely,
especially given that the system is viewed nearly face on.  Even more unlikely is the chance that
 circumplanetary disks or recent protoplanet collisions explain the near-IR properties of
two to four separate planets in two systems with very different ages and primary star properties.

Given the success in better reproducing HR 8799bcd's SEDs with thick cloud models
and the similarity between HR 8799b and 2M 1207b, it is more plausible that
the latter's near-IR spectrum is likewise explained by thick clouds.
If this is generally true of planetary-mass objects, thicker clouds may constitute
the primary difference between the atmospheres of massive planets and brown dwarfs,
at least over the gravity and temperature range enclosed by the
HR 8799 planets and 2M 1207b (e.g. log(g) = 3.75--4.5, T$_{eff}$ =900-1600K).
Since thicker clouds affect the color-magnitude positions of substellar objects
it is quite possible the Model A 'thick cloud' sequence extending to HR 8799b and 2M 1207b from
 the nominal L/T dwarf boundary continues on to even cooler temperatures (e.g. T$_{eff}$ $\sim$ 700--900K). 
Since thick clouds present reshape the spectral structure at $\sim$ 1.6 $\mu m$ (e.g. in the methane band), 
they may also affect the L/T dwarf transition,  which is already known to be dependent upon
surface gravity \citep[e.g.][]{Metchev2006,Luhman2007}.

\subsection{Future Work}
Our study motivates the development of a suite of new atmosphere models with clouds 
intermediate in thickness between the Model E cloud deck and Model A thick cloud layer 
prescriptions.  Adopting these models as fiducial models, we can revisit the 
(secondary) effects of surface gravity, metallicity, and non-equilibrium chemistry on the atmospheres of 
planetary mass objects, complementing similar investigations for brown dwarfs 
\citep[][]{Allard2001,Marley2002,Burrows2006,HubenyBurrows2007}.  
These models will be developed and applied to HR 8799bcde and other planetary-mass objects 
in upcoming papers (Madhusudhan et al. 2011, in prep.) and may provide 
a useful comparison to planet parameters derived from cooling models \citep[e.g.][]{Burrows1997, 
Baraffe2003,Fortney2007,Fortney2008}.

\begin{figure}
\centering
\epsscale{0.99}
\plotone{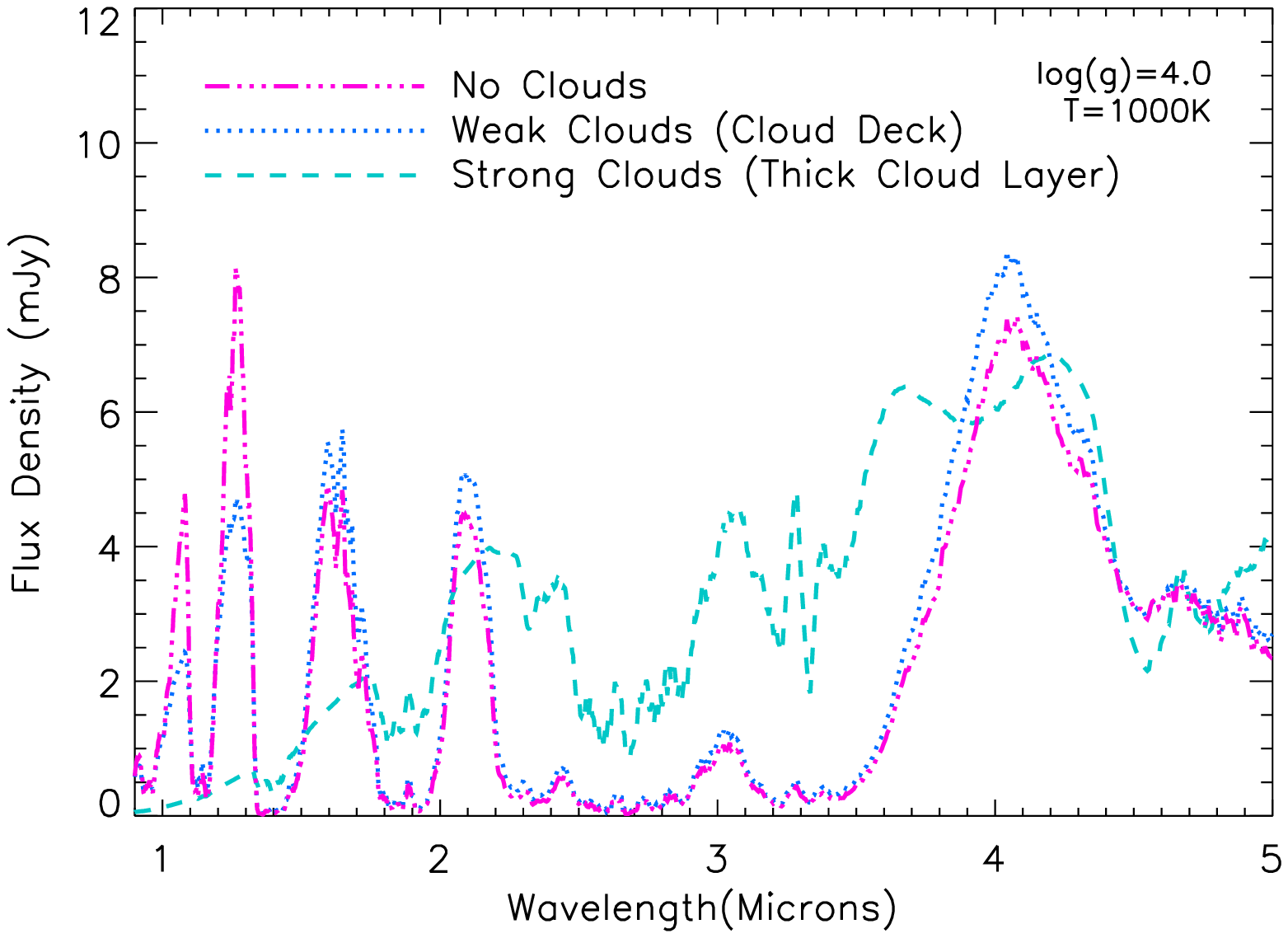}
\caption{
Comparing SEDs for different cloud prescriptions (no clouds, Model E, and Model A) at a given temperature
 and gravity.}
\label{modseq}
\end{figure}

\begin{figure}
\centering
\includegraphics[scale=0.46,clip]{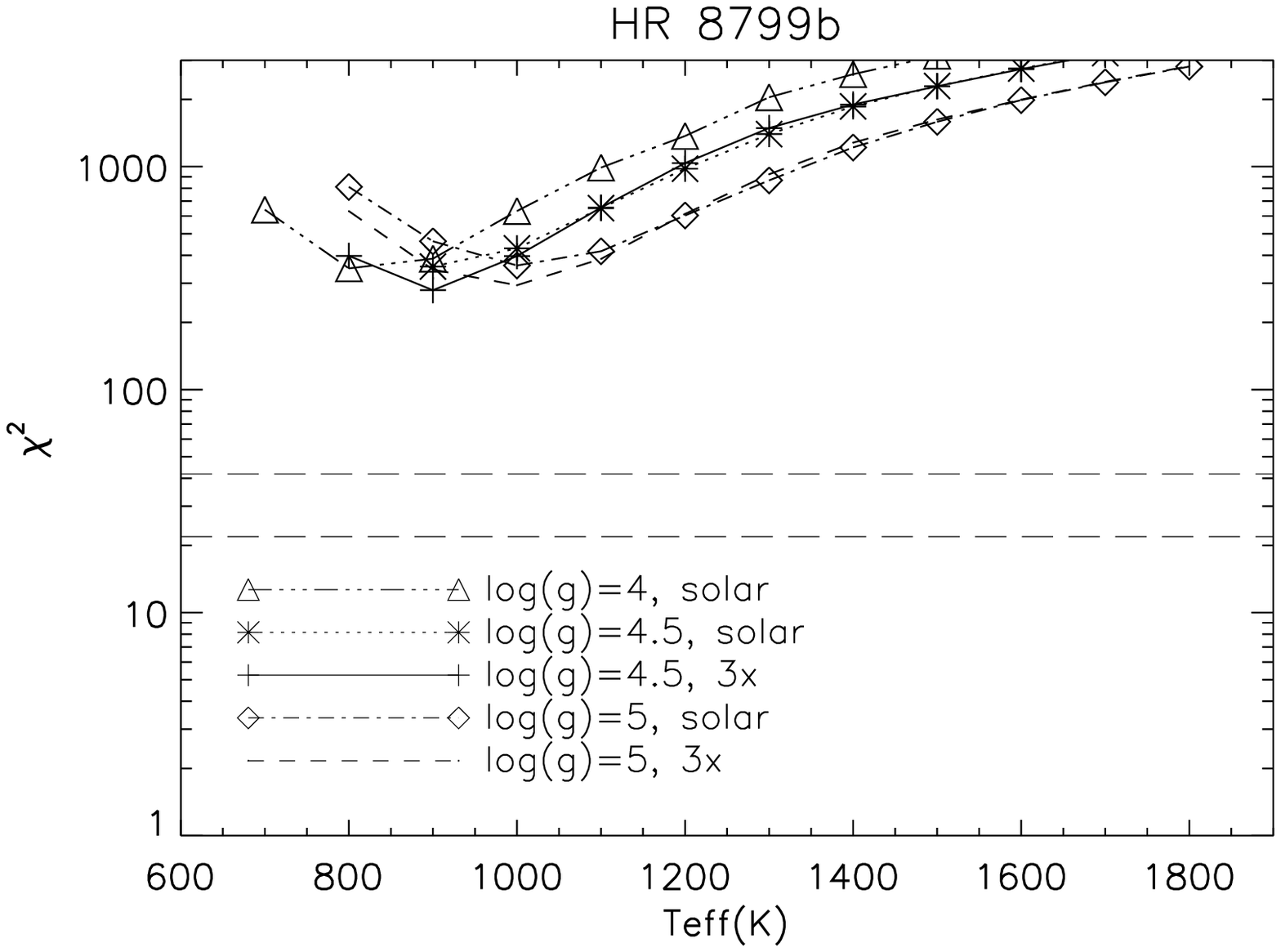}
\includegraphics[scale=0.45,clip]{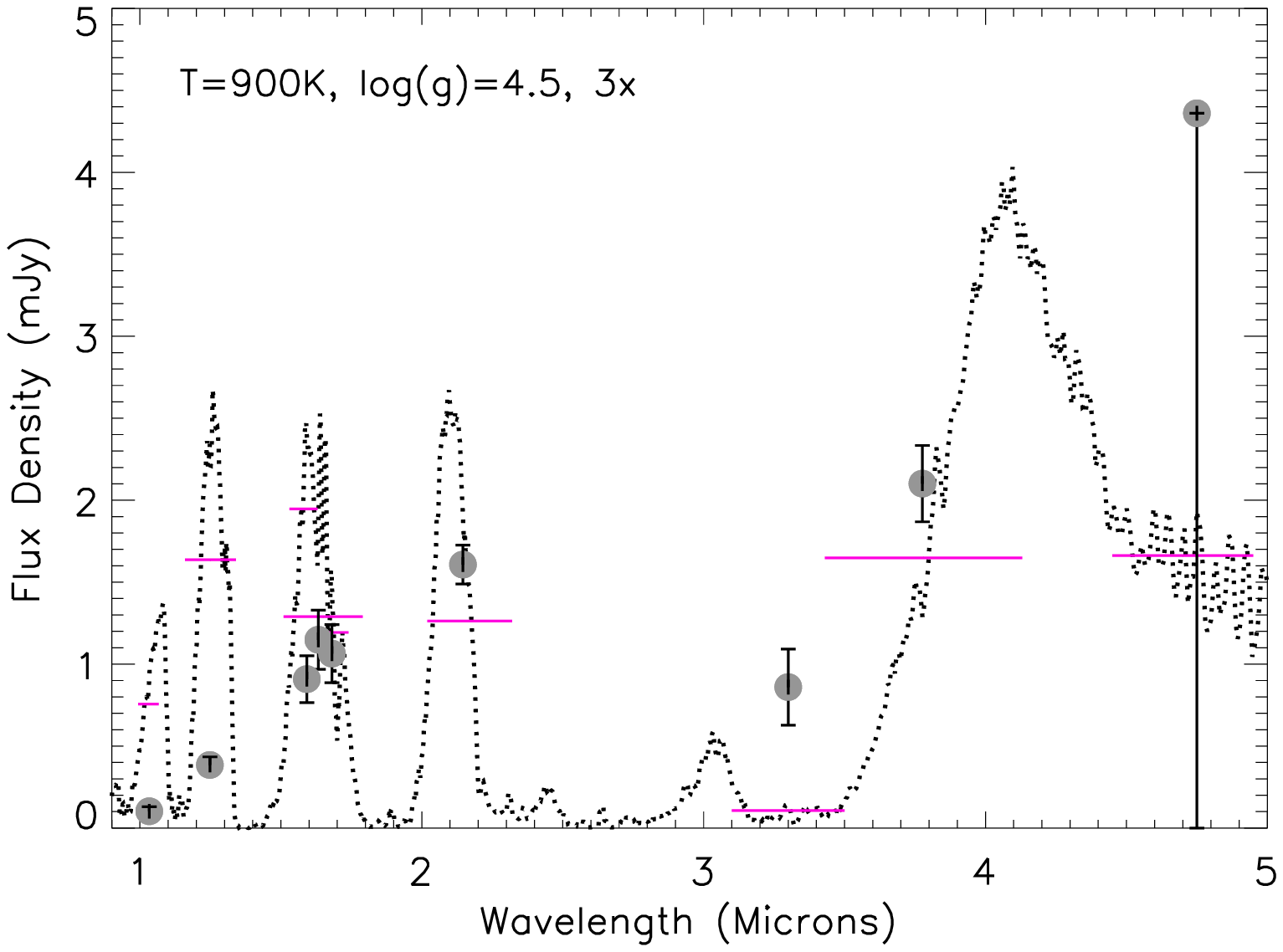}
\\
\includegraphics[scale=0.45,clip]{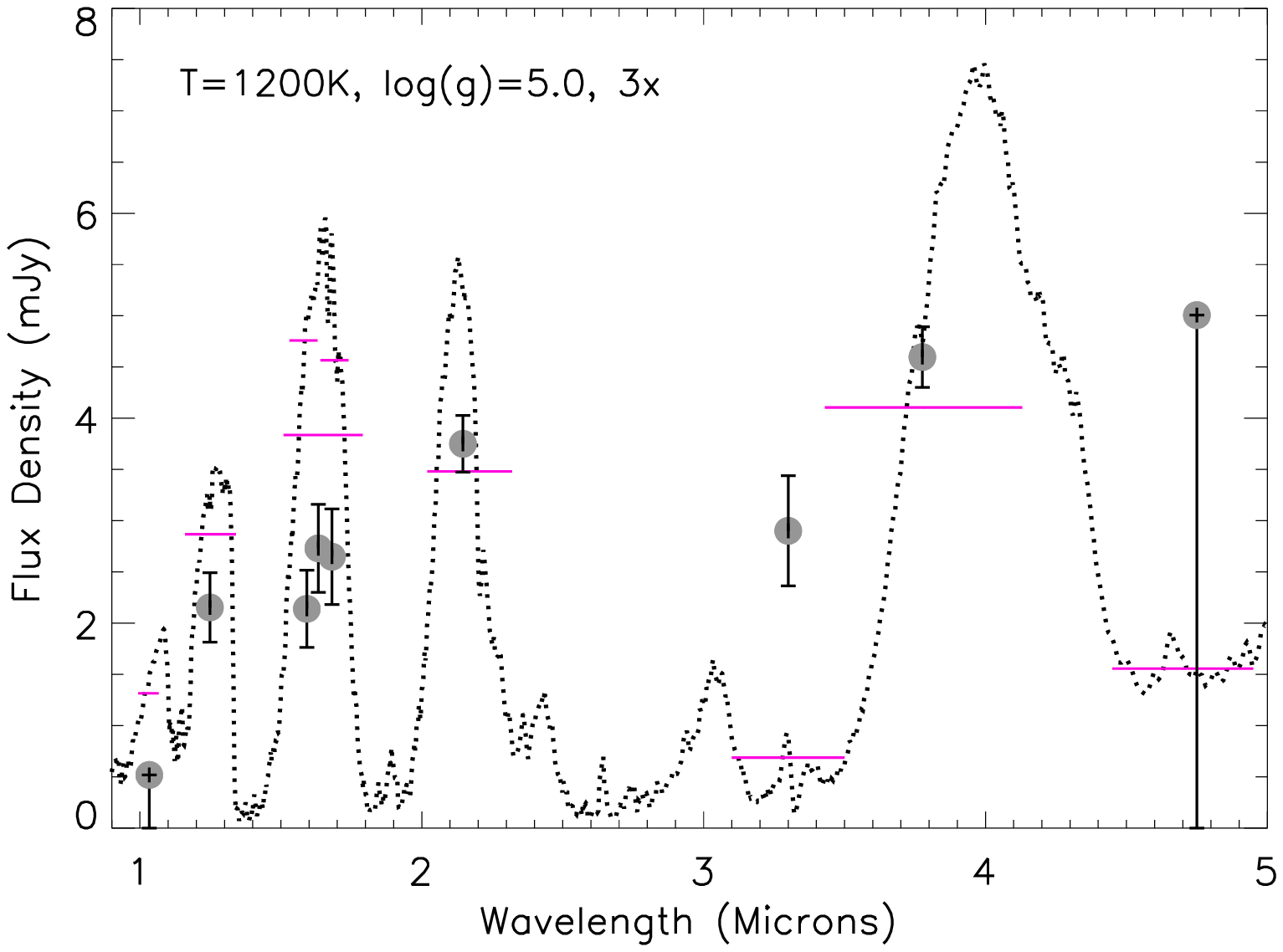}
\includegraphics[scale=0.45,clip]{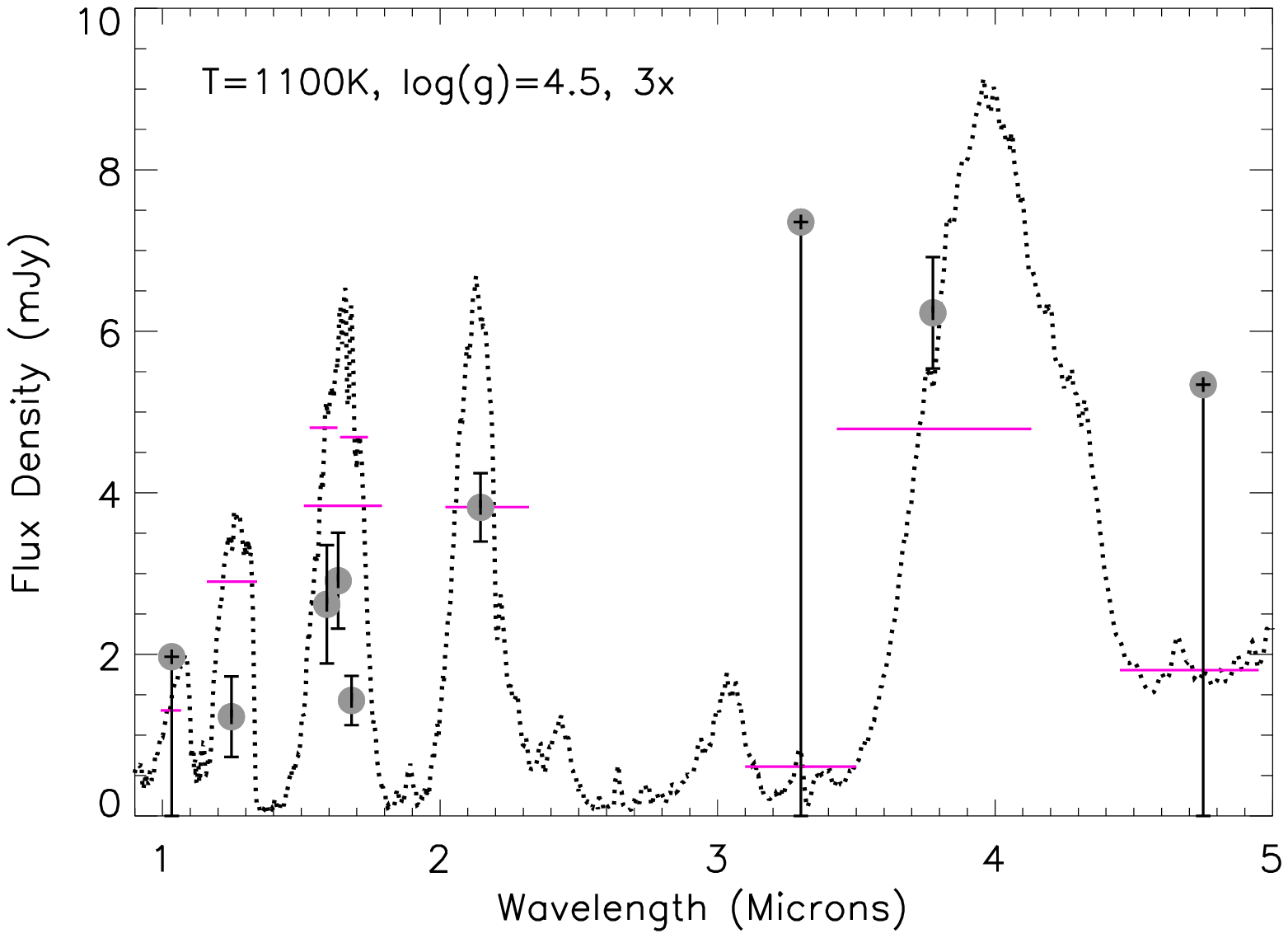}
\caption{Fitting results for the standard cloud deck models assuming the \citet{Burrows1997} 
radii.  The top-left panel show the distribution of $\chi^{2}$ vs. T$_{eff}$ for model 
fits to HR 8799b with a range of surface gravity and metallicity.  The top-right panel compares
 the HR 8799b planet SED to the model with the smallest $\chi^{2}$ value.  
The bottom panels compare the HR 8799c (left) and HR 8799d (right) SEDs to the best-fit models 
for these data.  In the SED comparisons, the horizontal magenta lines identify the flux of 
the model in the photometric filters convolved over the filter function.  The width of the magenta 
line corresponds to the width of the filter.
}
\label{standardfit}
\end{figure}

\begin{figure}
\centering
\includegraphics[scale=0.46,clip]{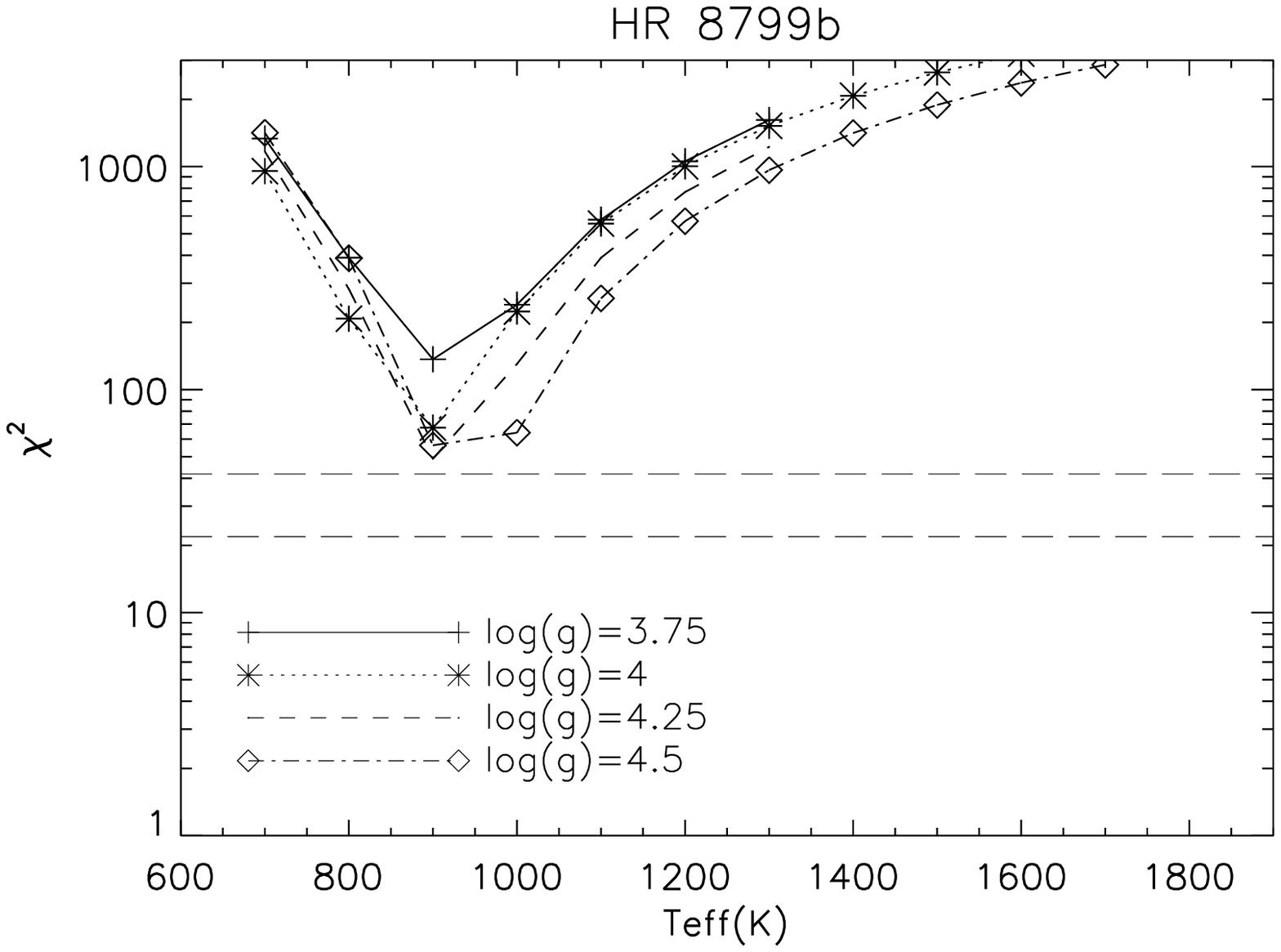}
\includegraphics[scale=0.46,clip]{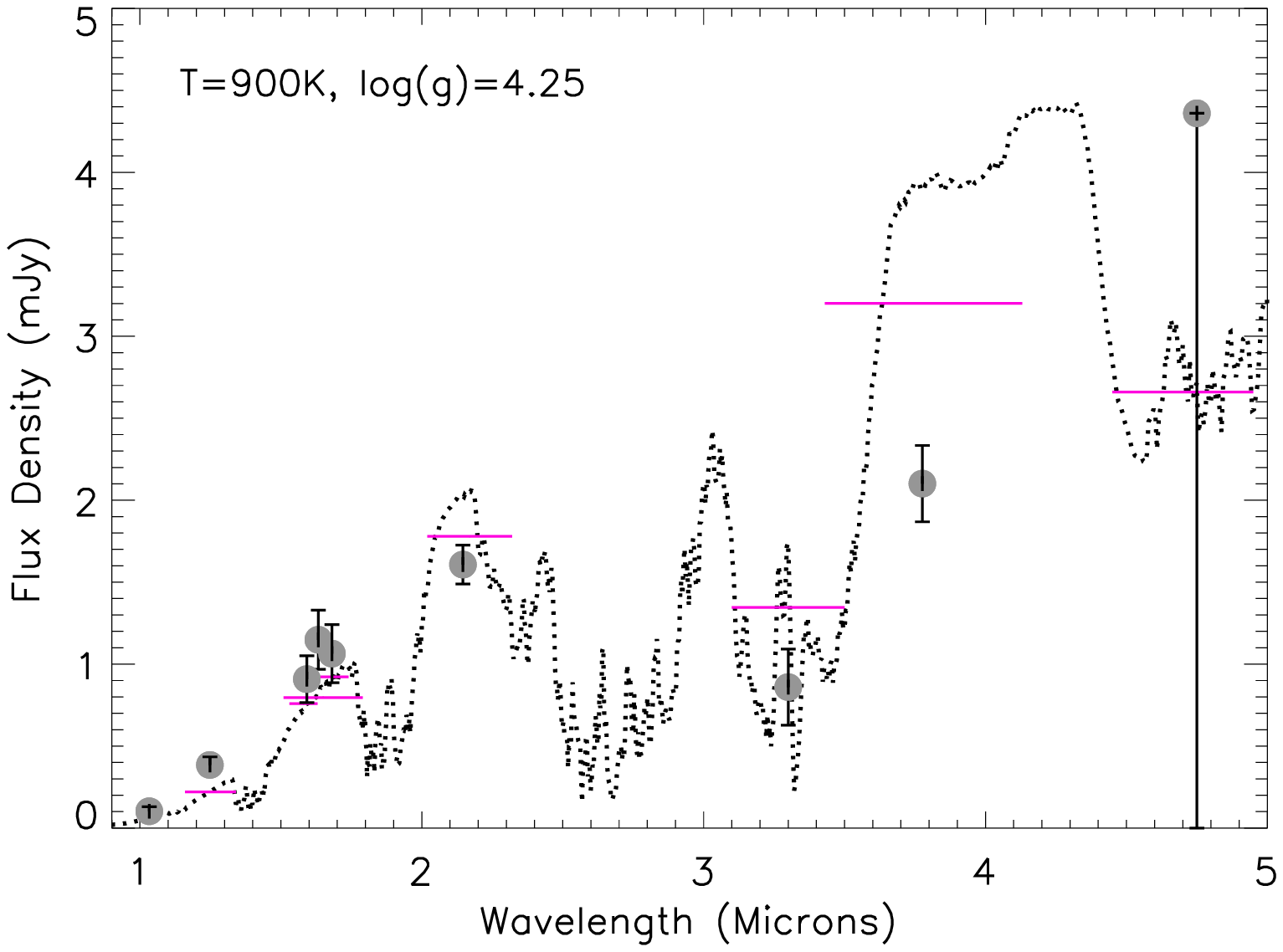}
\\
\includegraphics[scale=0.46,clip]{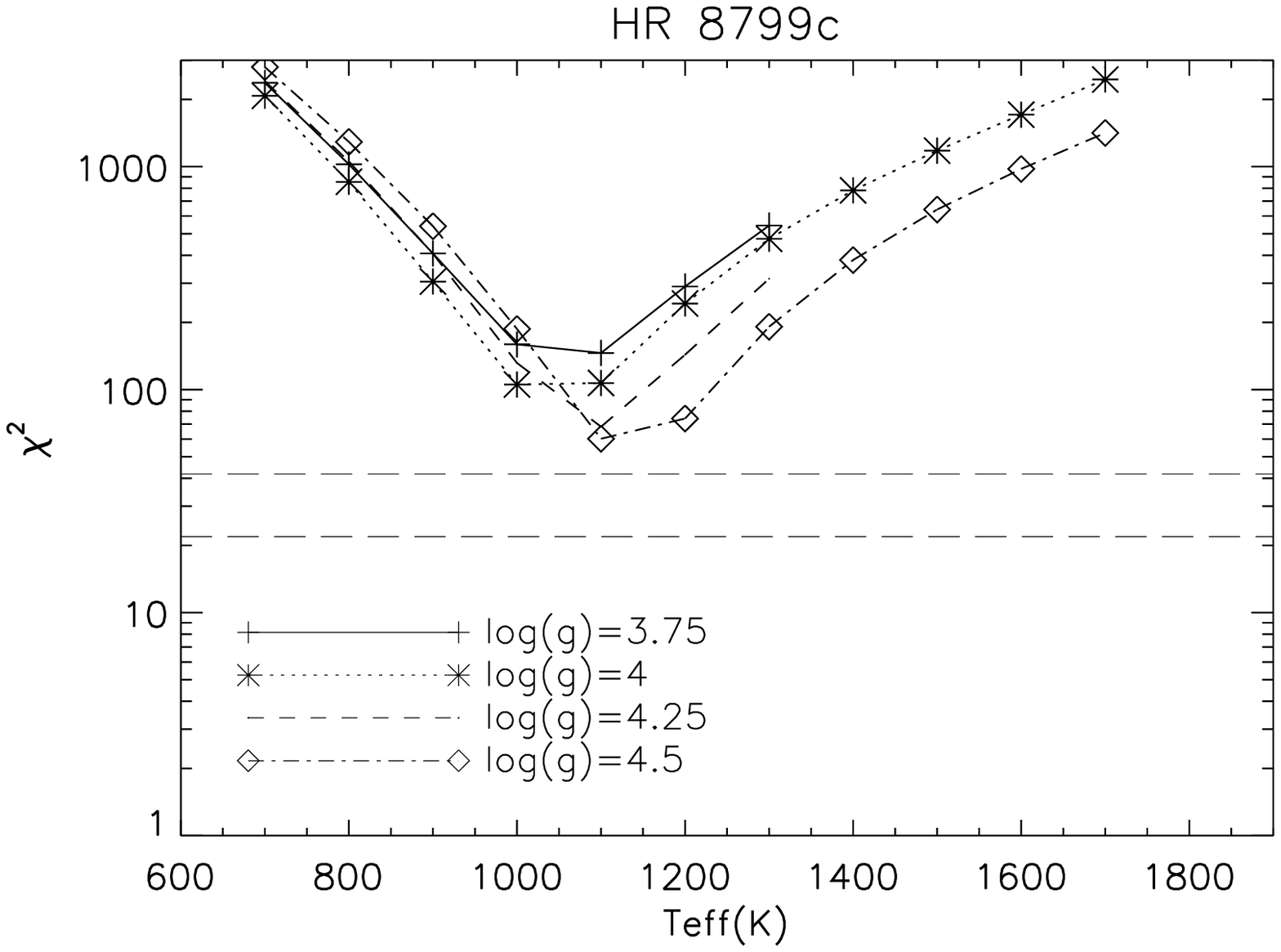}
\includegraphics[scale=0.46,clip]{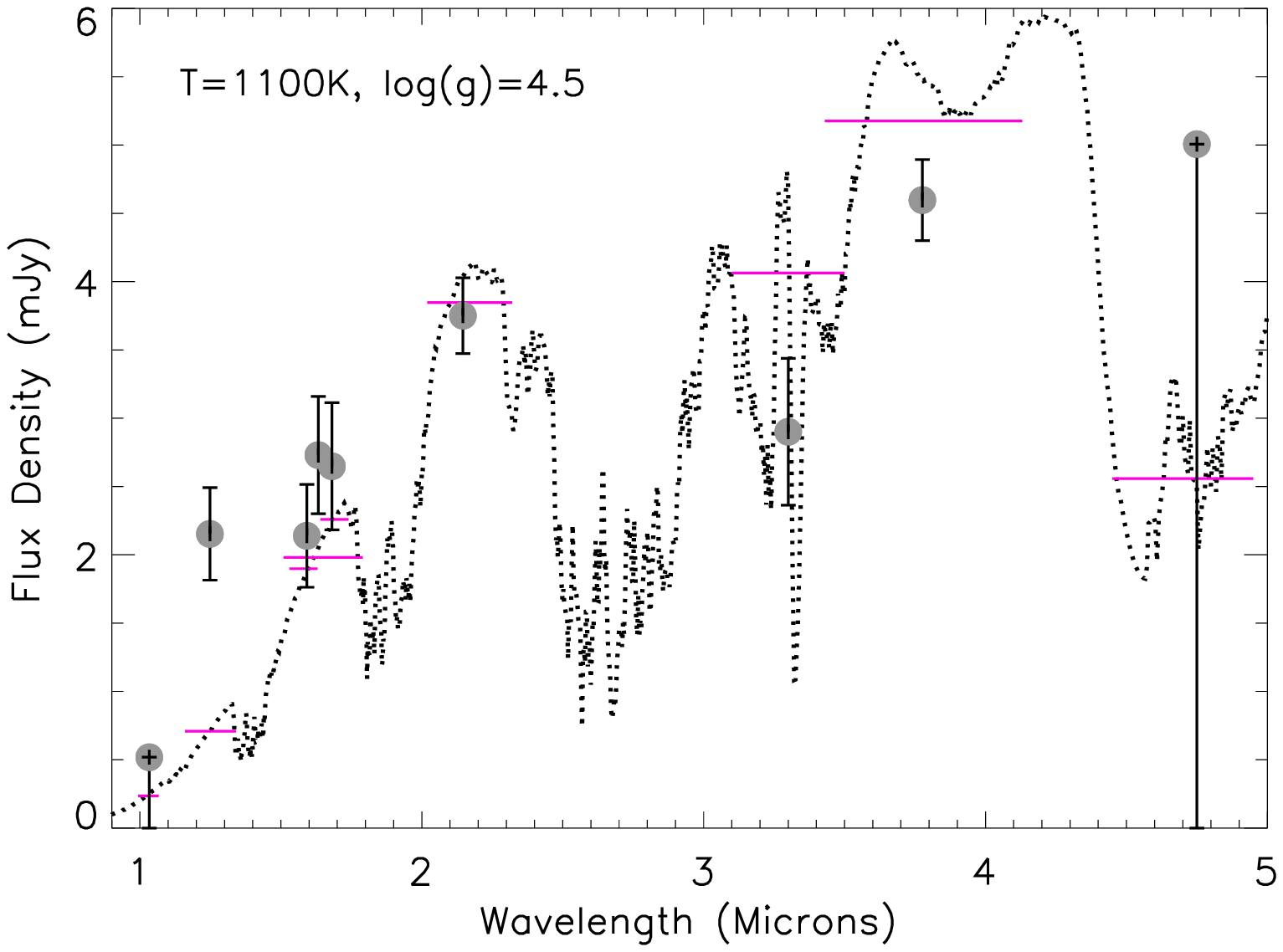}
\\
\includegraphics[scale=0.46,clip]{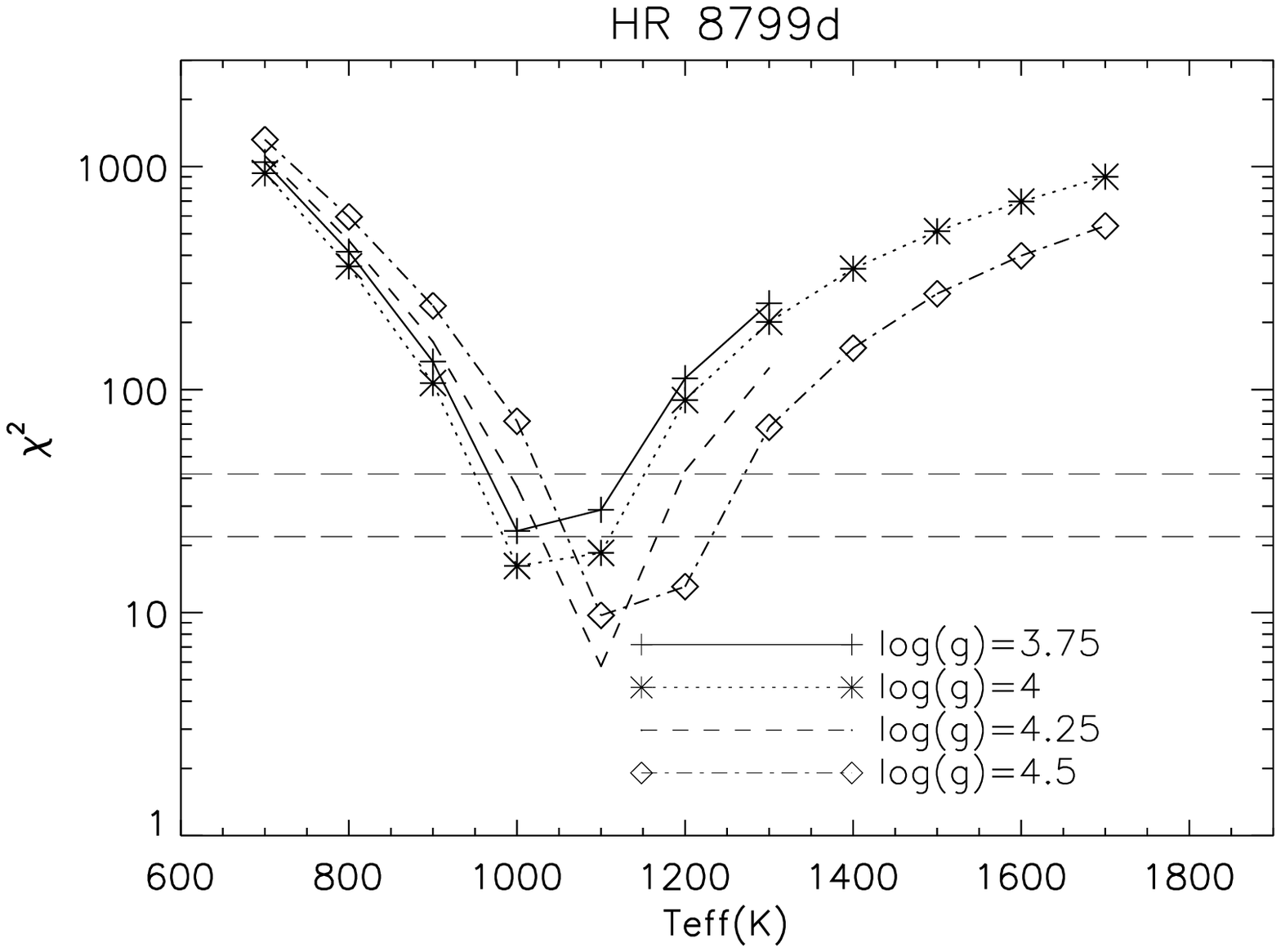}
\includegraphics[scale=0.46,clip]{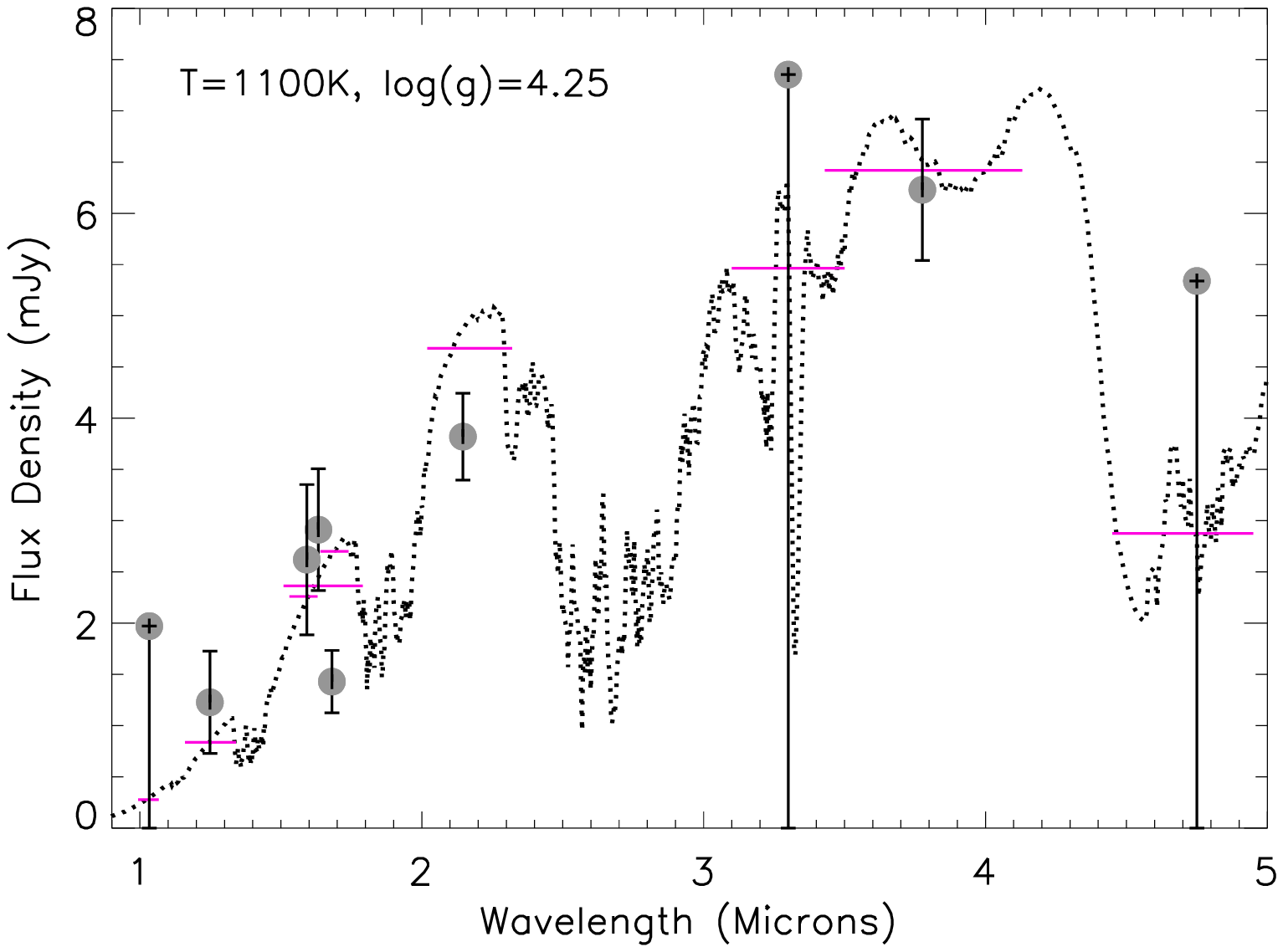}
\caption{Fitting results for the Model A, thick cloud layer prescription with 
a modal particle size of 60 $\mu m$.  The 
lefthand panels show the $\chi^{2}$ distributions for each planet while the 
righthand panels compare the planet SEDs to the best-fit models in each case.  Compared 
to the Model E, standard cloud deck fits, these models yielded smaller $\chi^{2}$ 
minima and better fits to the data.}
\label{thickfit}
\end{figure}

\begin{figure}
\centering
\includegraphics[scale=0.46,clip]{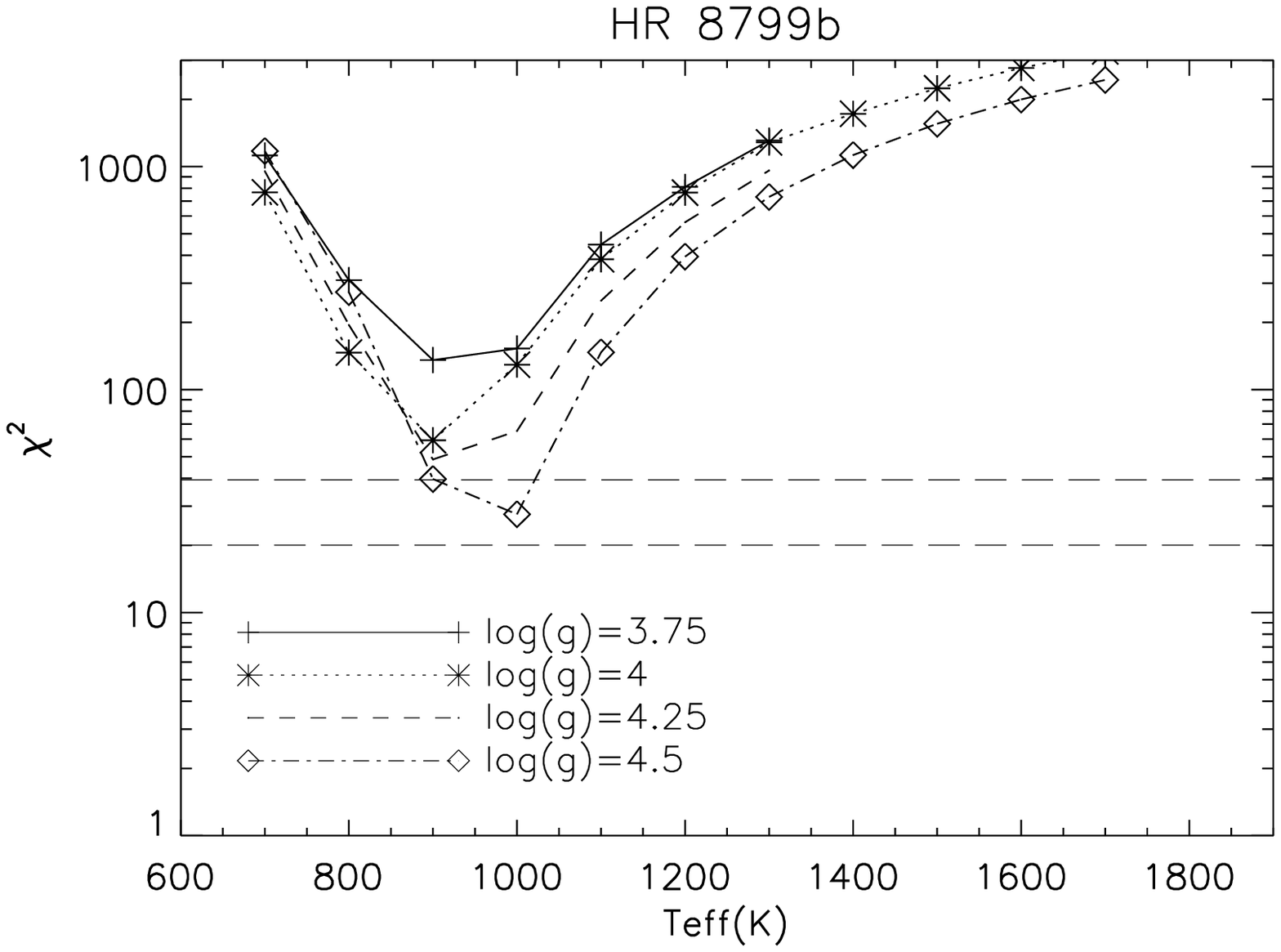}
\includegraphics[scale=0.46,clip]{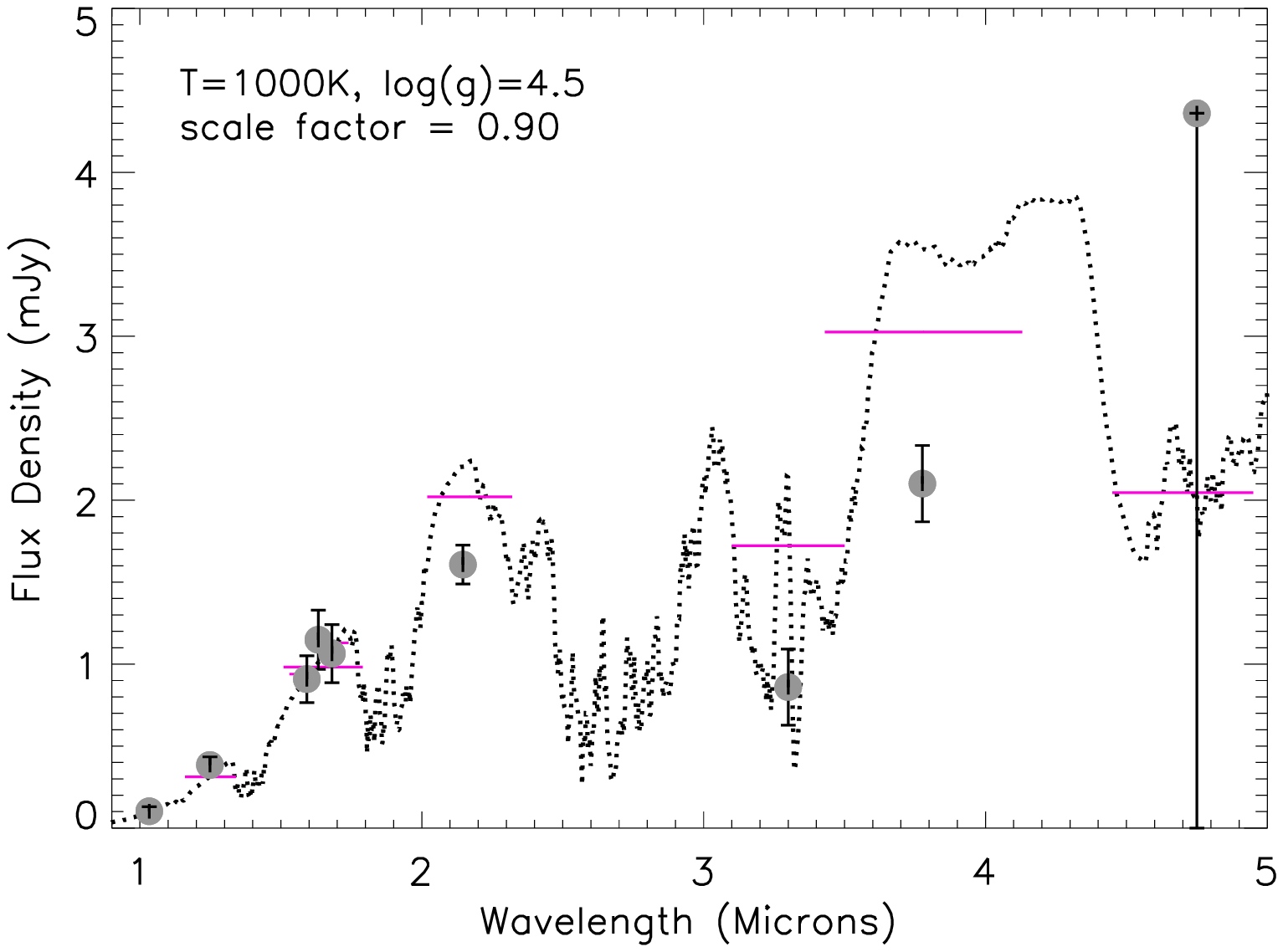}
\\
\includegraphics[scale=0.46,clip]{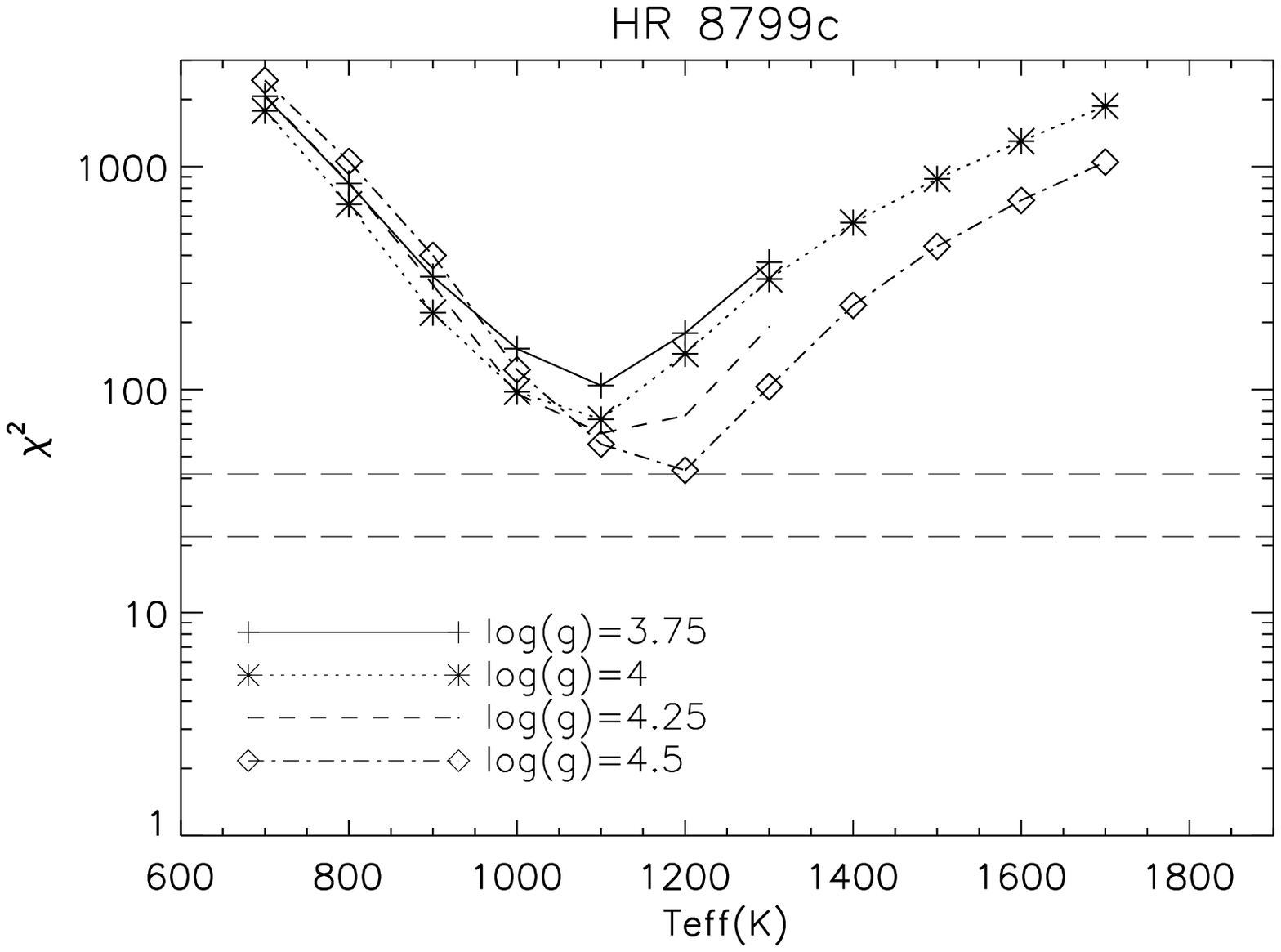}
\includegraphics[scale=0.46,clip]{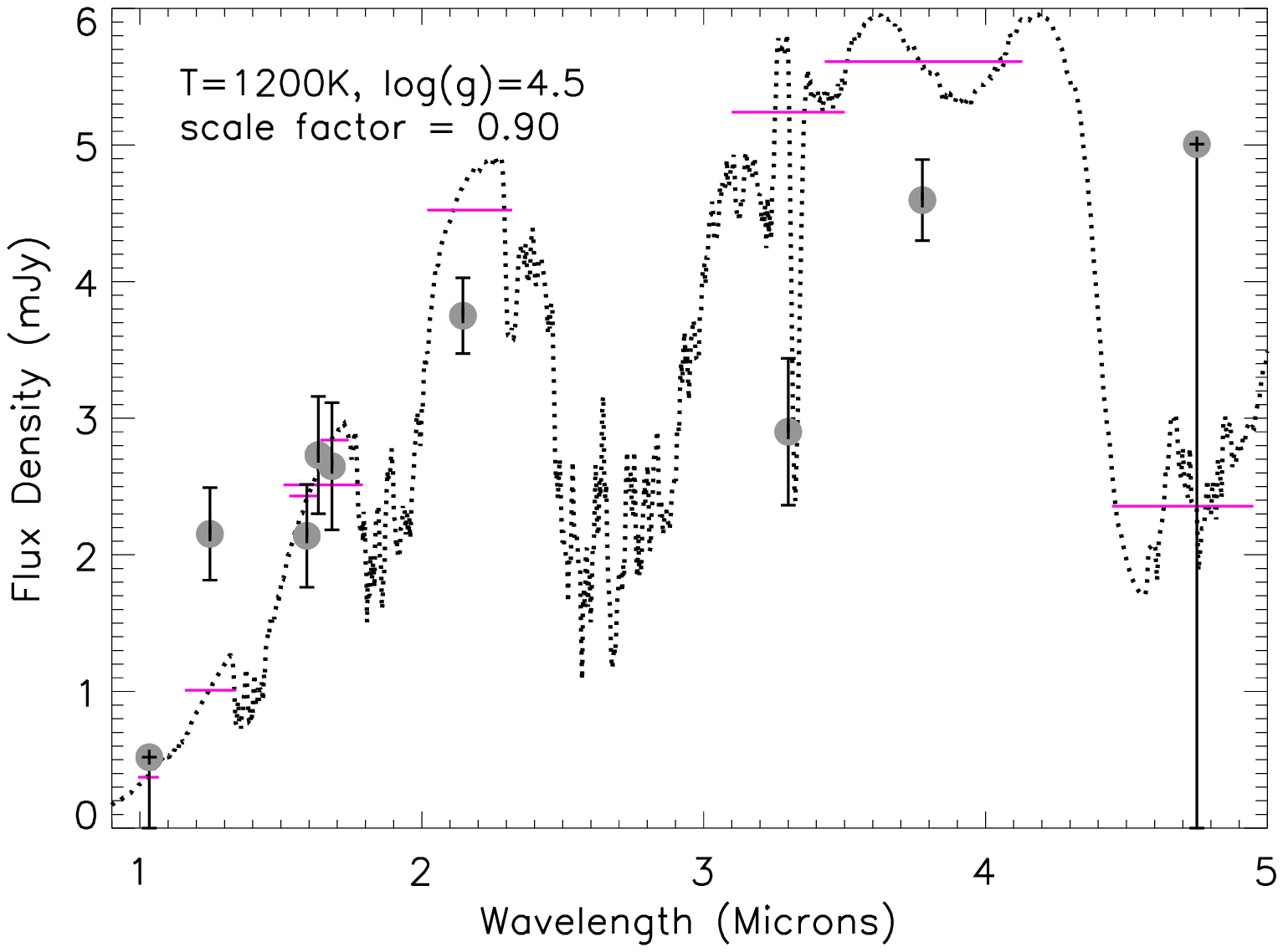}
\\
\includegraphics[scale=0.46,clip]{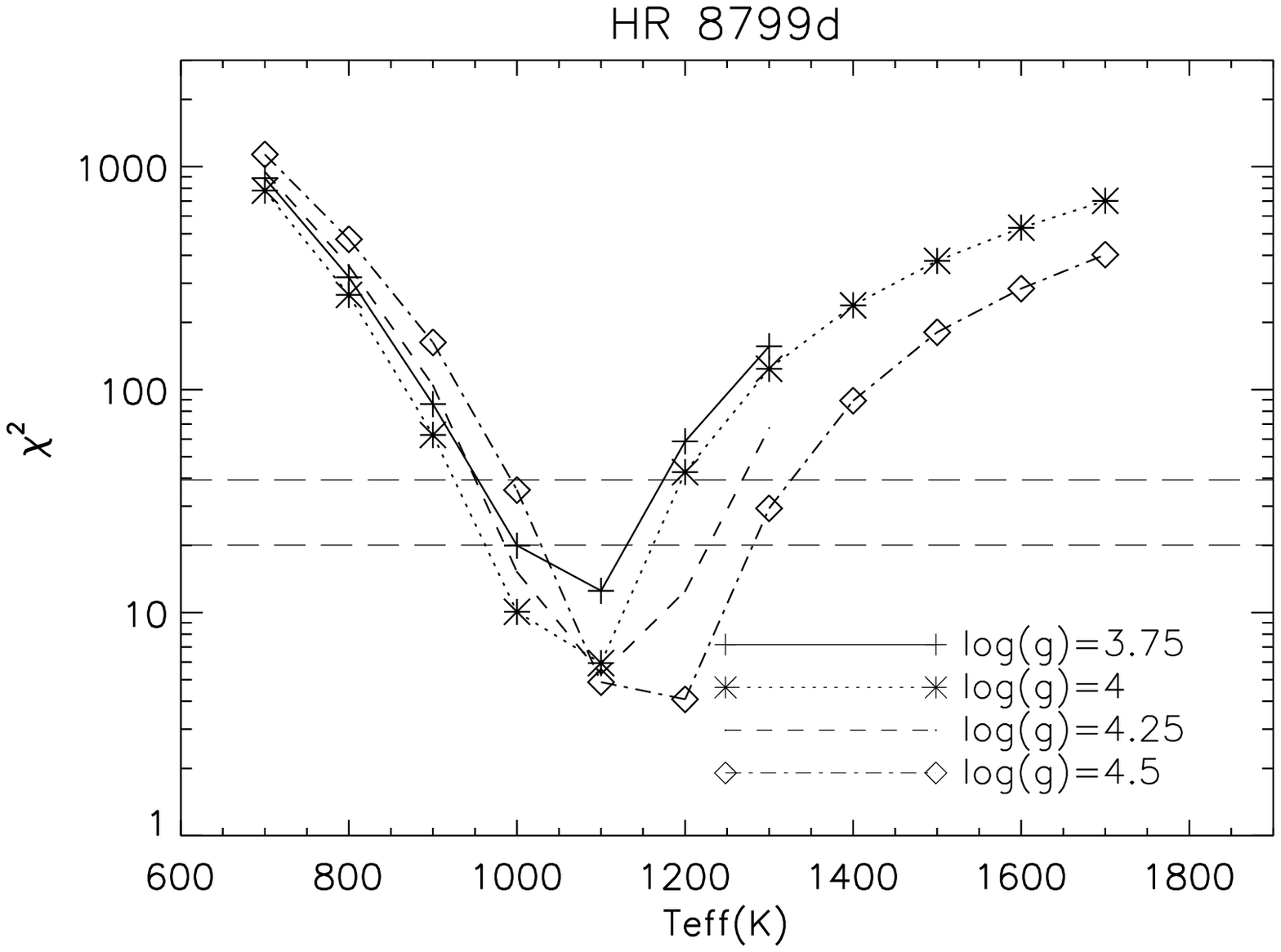}
\includegraphics[scale=0.46,clip]{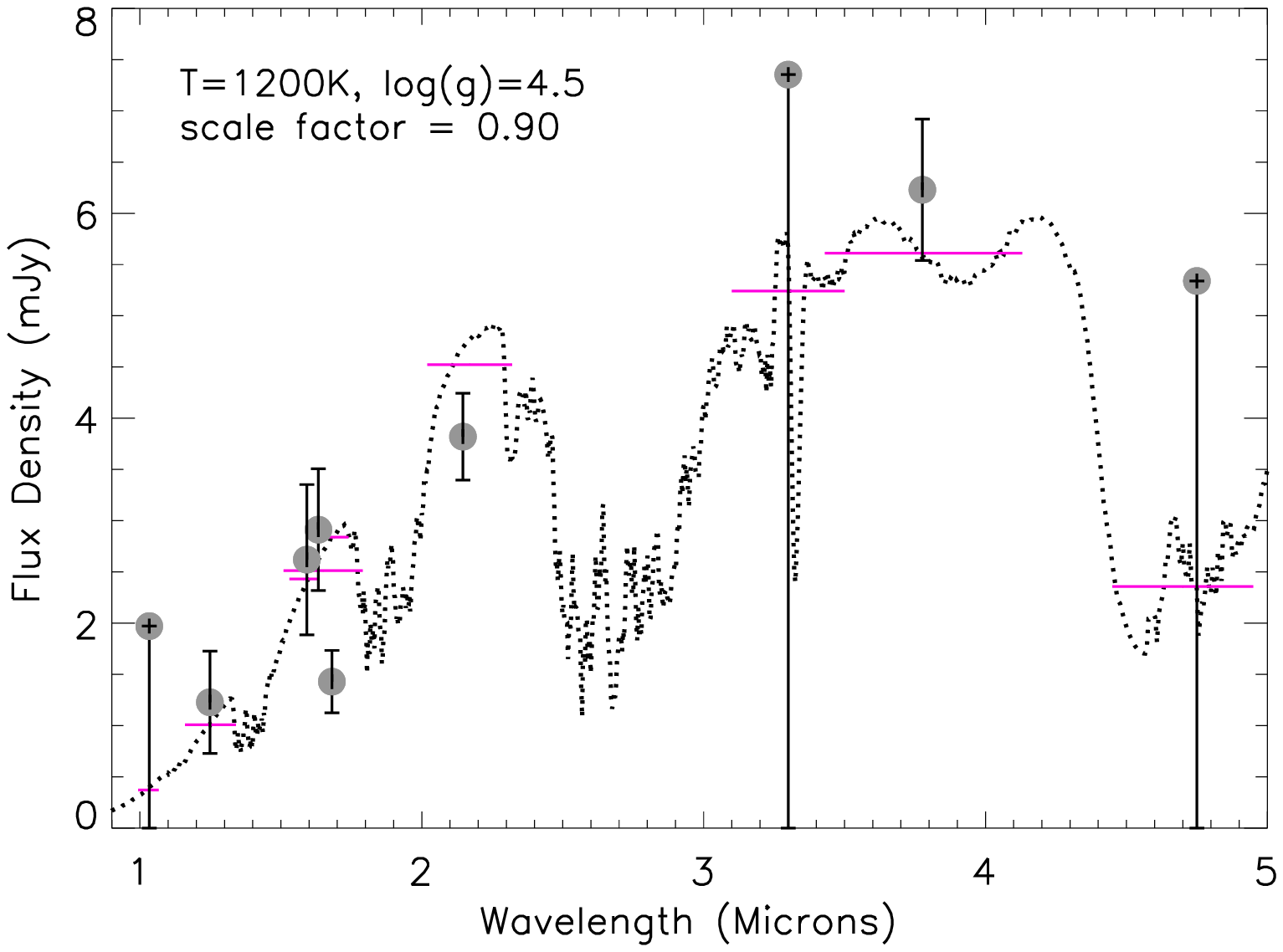}
\caption{Same as Figure \ref{thickfit} except allowing the planet radius to depart by 
$\pm$ 10\% from the \citet{Burrows1997} values.}
\label{thickscalefit}
\end{figure}

\begin{figure}
\centering
\includegraphics[scale=0.46,clip]{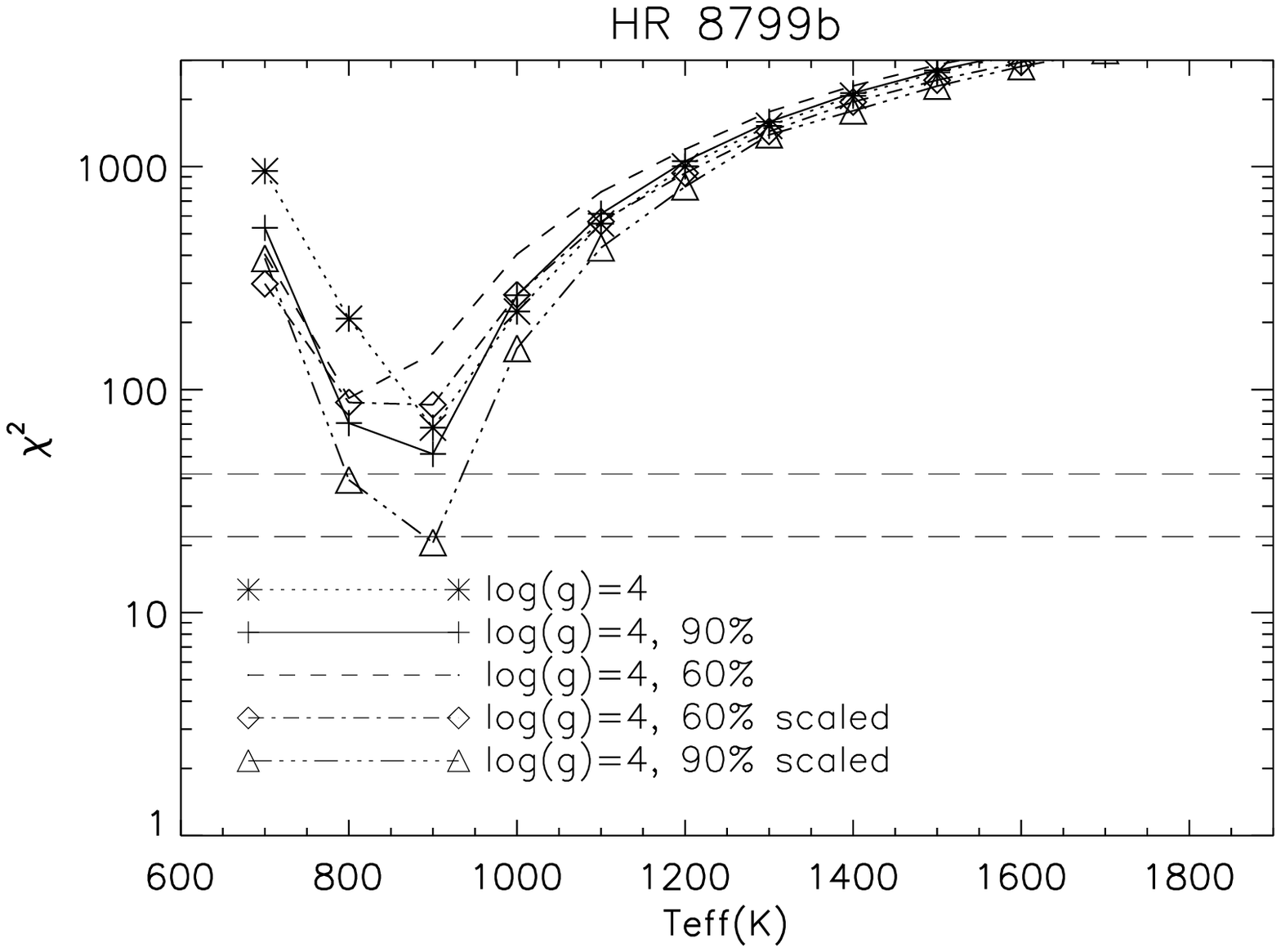}
\includegraphics[scale=0.46,clip]{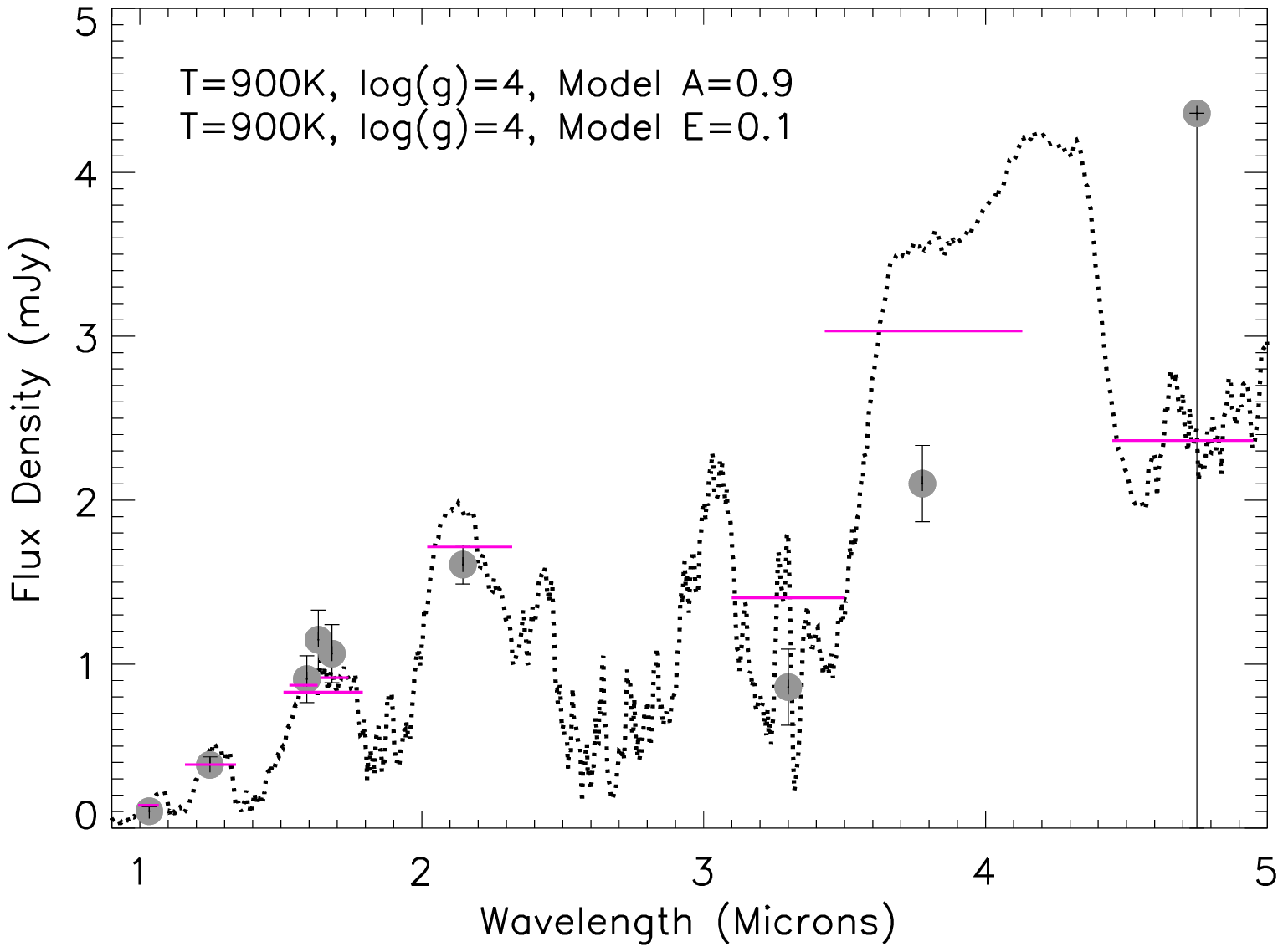}
\\
\includegraphics[scale=0.46,clip]{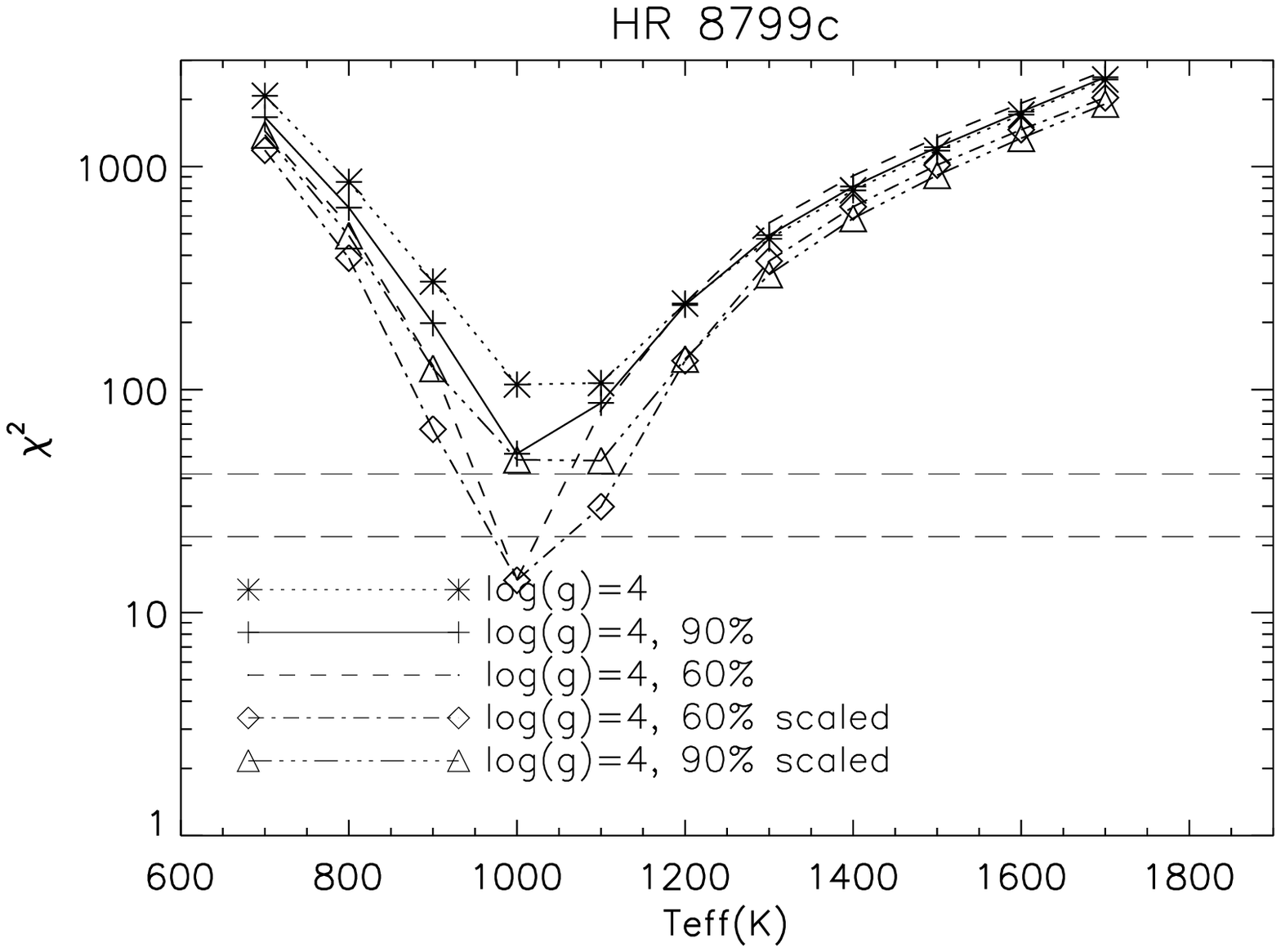}
\includegraphics[scale=0.46,clip]{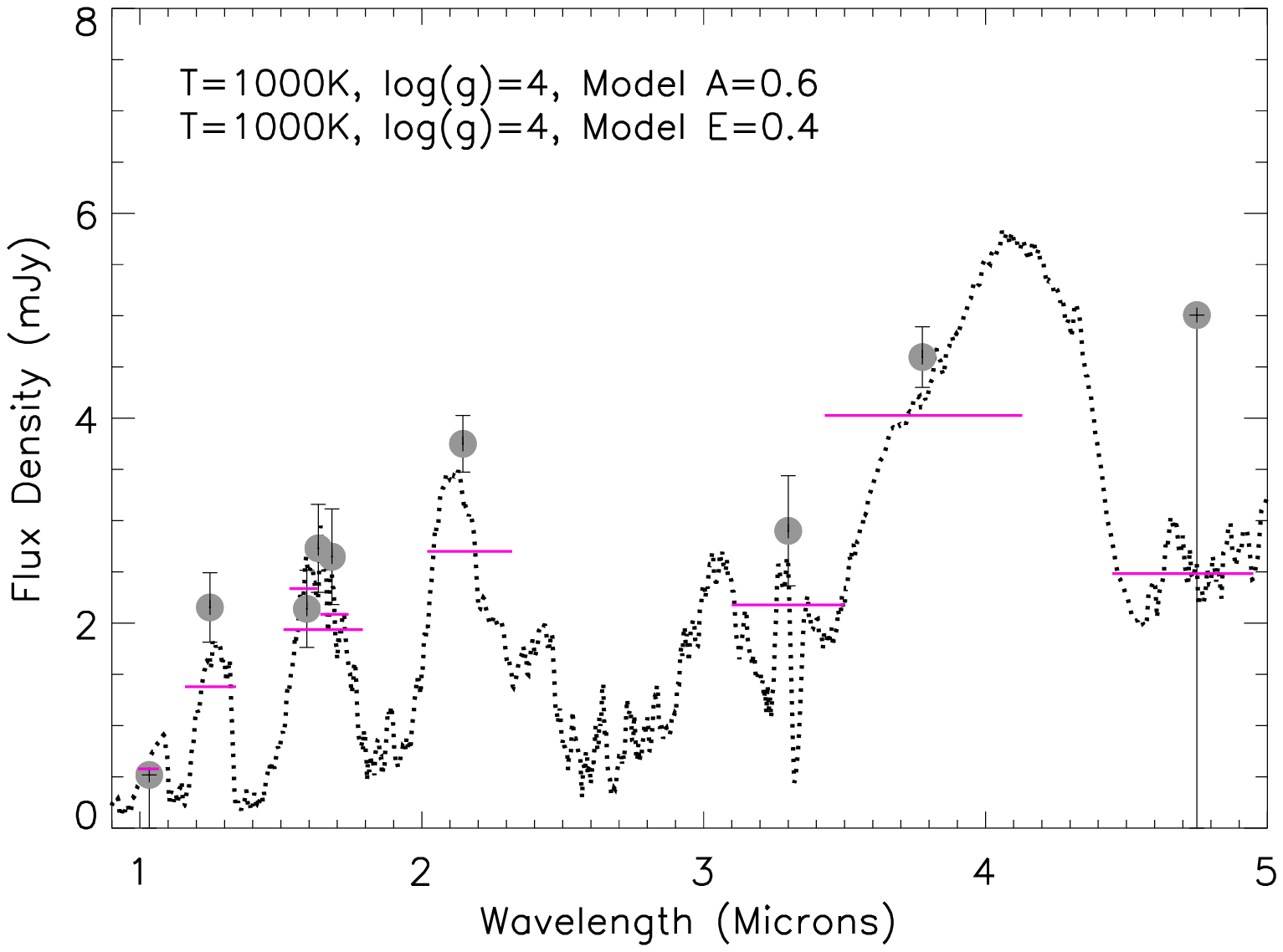}
\\
\includegraphics[scale=0.46,clip]{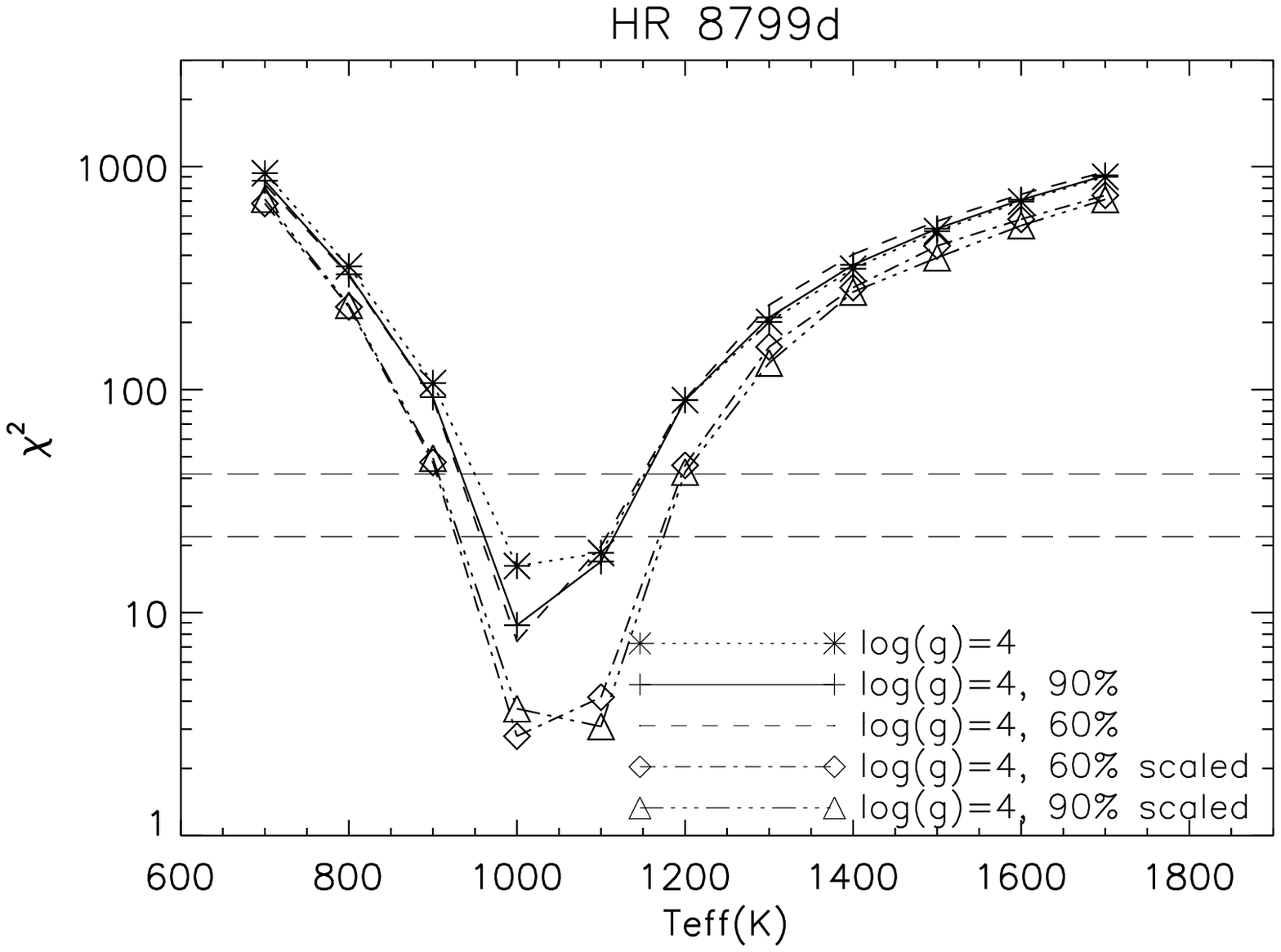}
\includegraphics[scale=0.46,clip]{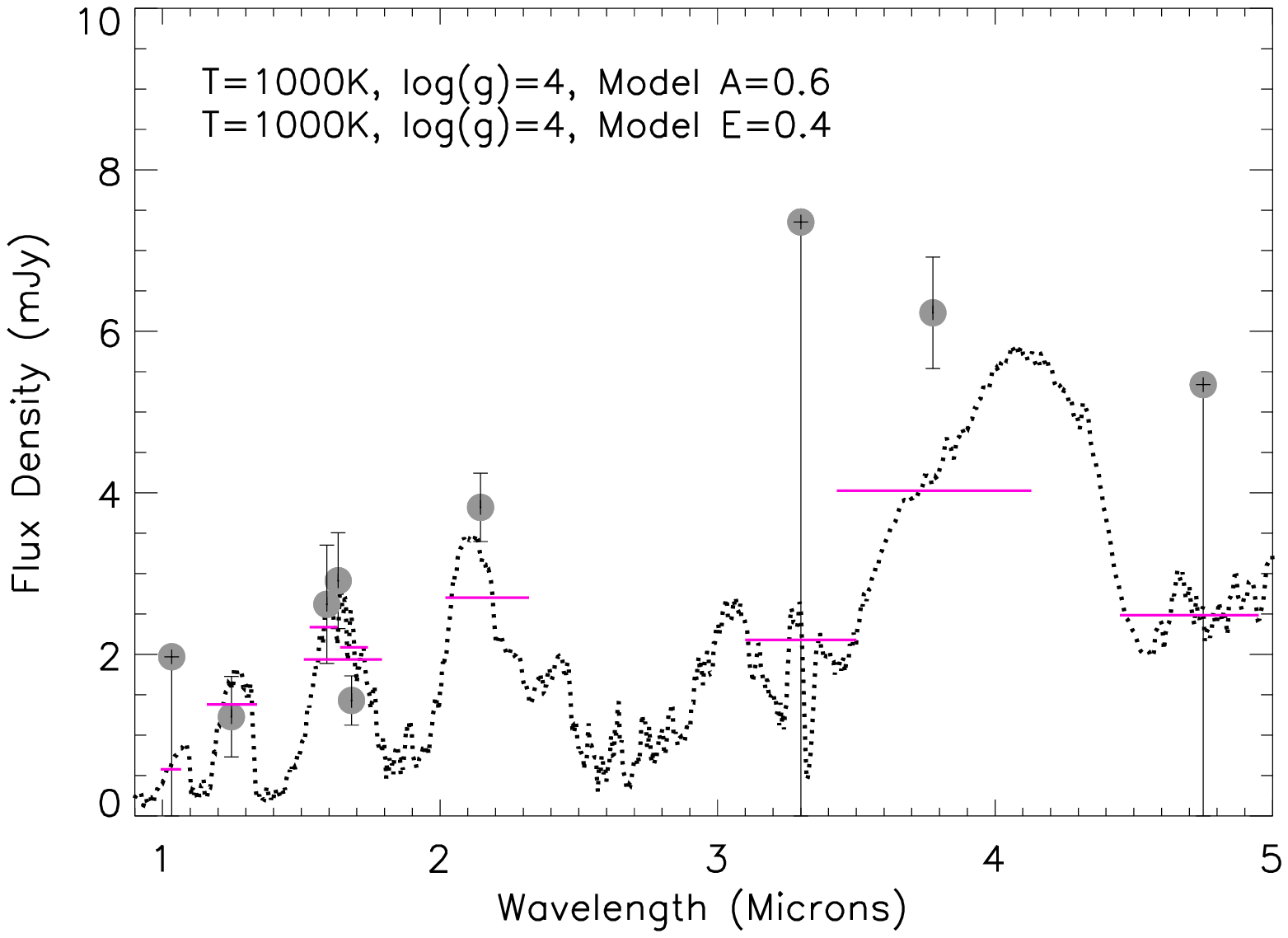}
\caption{Fitting results for our simple approximation of 
a ''patchy" cloud atmosphere.  In all right-hand panels, the displayed best-fit SEDs 
have C$_{k}$ = 0.9.}
\label{patchyfit}
\end{figure}

\begin{figure}
\centering
\epsscale{0.55}
\plotone{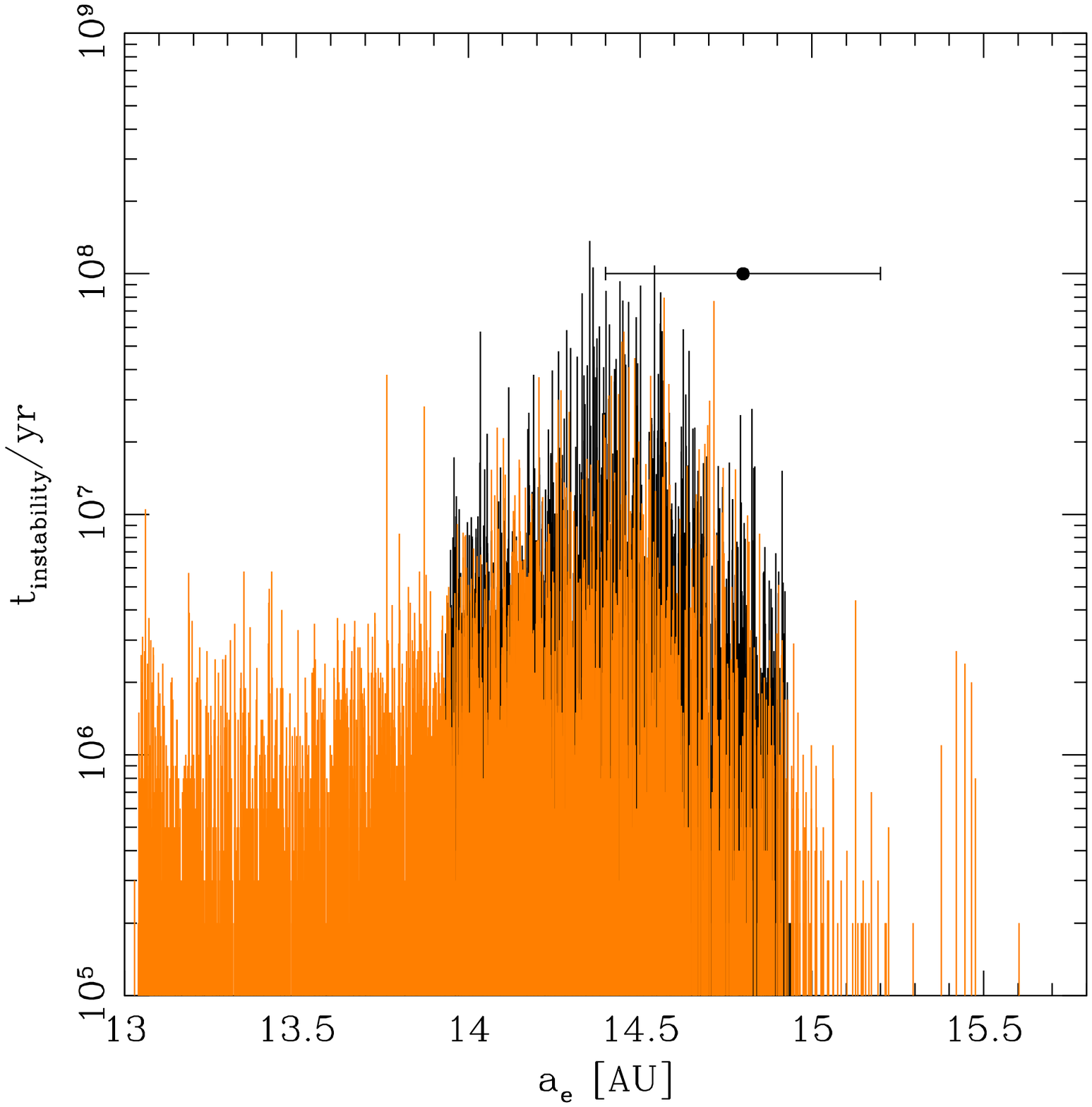}
\\
\epsscale{0.99}
\plottwo{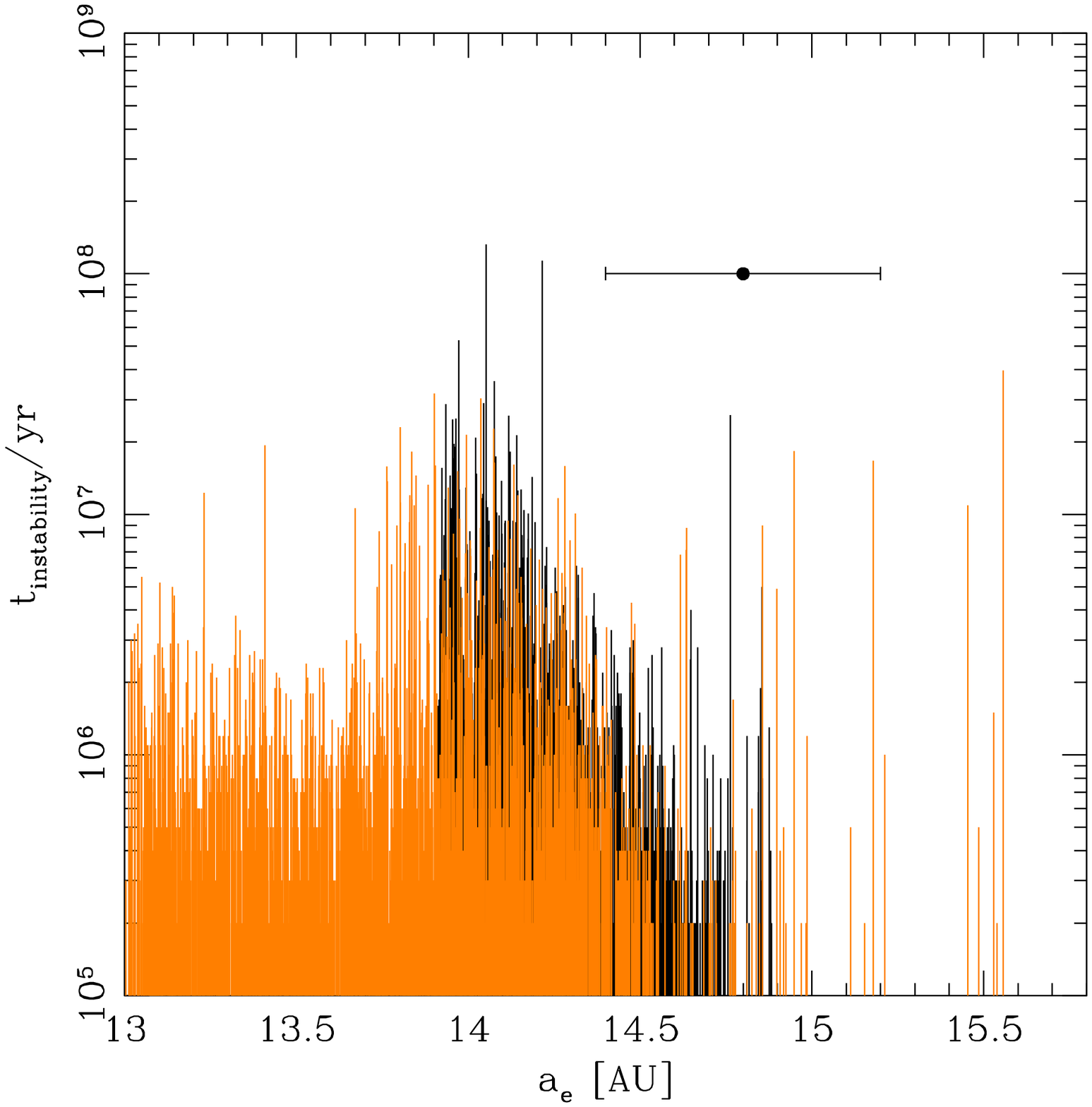}{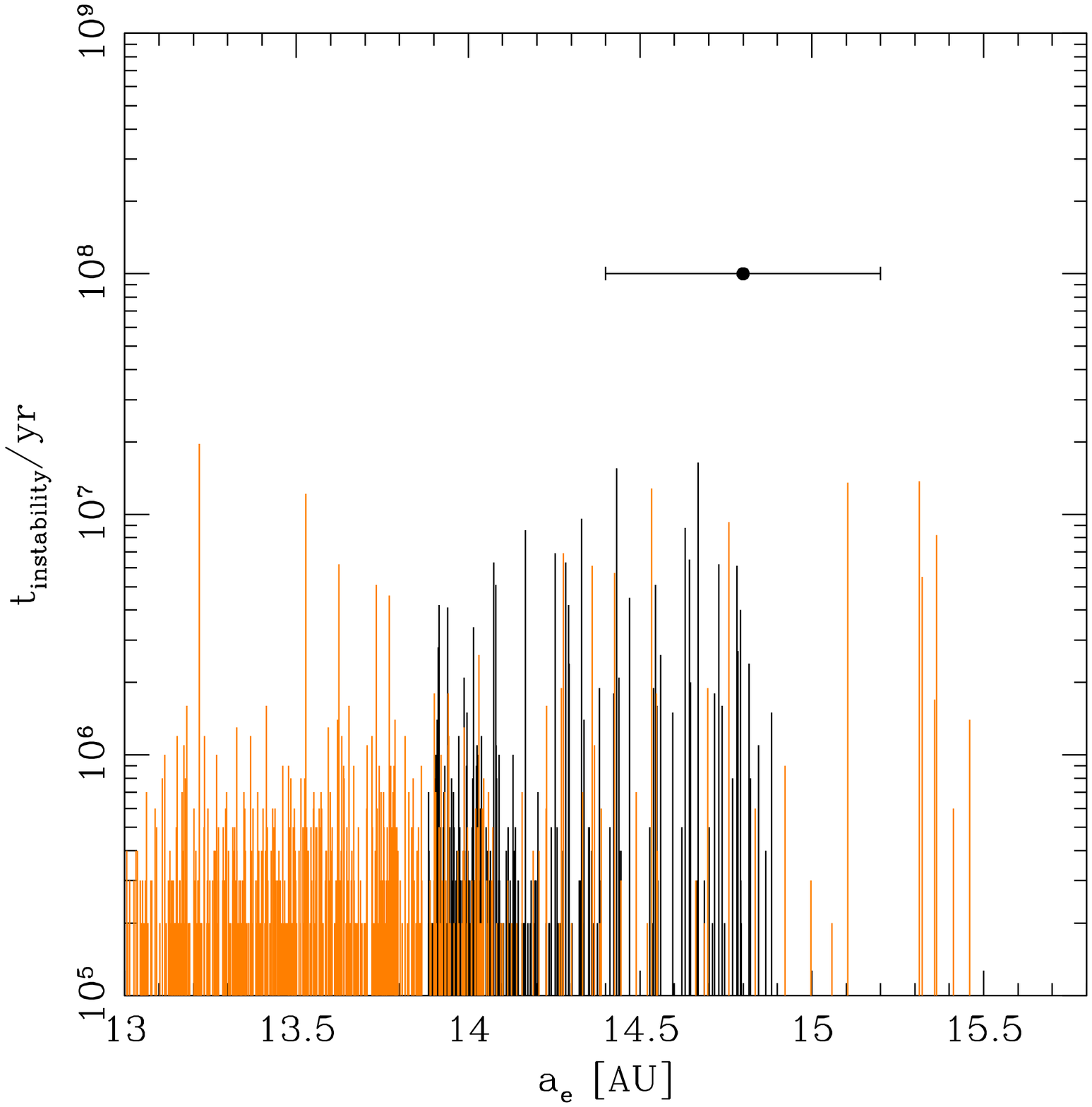}
\caption{The time to dynamical instability vs. semimajor axis of HR 8799e for three separate sets of 
masses and orbital resonances.  In all plots, the orange lines denote our set of simulations allowing 
HR 8799e to vary between 13.1 AU and 15.7 AU, while the black lines denote our simulations that restrict 
HR 8799e to be between 14 and 15 AU while more finely sampling the range of other orbital parameters (e.g. mean anomaly, 
longitude of periastron).  The horizontal dot with error bars identifies the 1-$\sigma$ range of projected separations 
for HR 8799e from our work.  In Case A (top panel), HR 8799bcde have masses of 5, 7, 7, and 7 M$_{J}$.  Case B (bottom-left) 
corresponds to planet masses of 7, 10, 10, and 10 M$_{J}$ and Case C (bottom-right) correspond to 10, 13, 13, and 13 M$_{J}$.  
The density of the bars appears anomalously low for Case C because many simulations have instability times less than 
10$^{5}$ years.}
\label{dynamics}
\end{figure}

\begin{figure}
\centering
\plotone{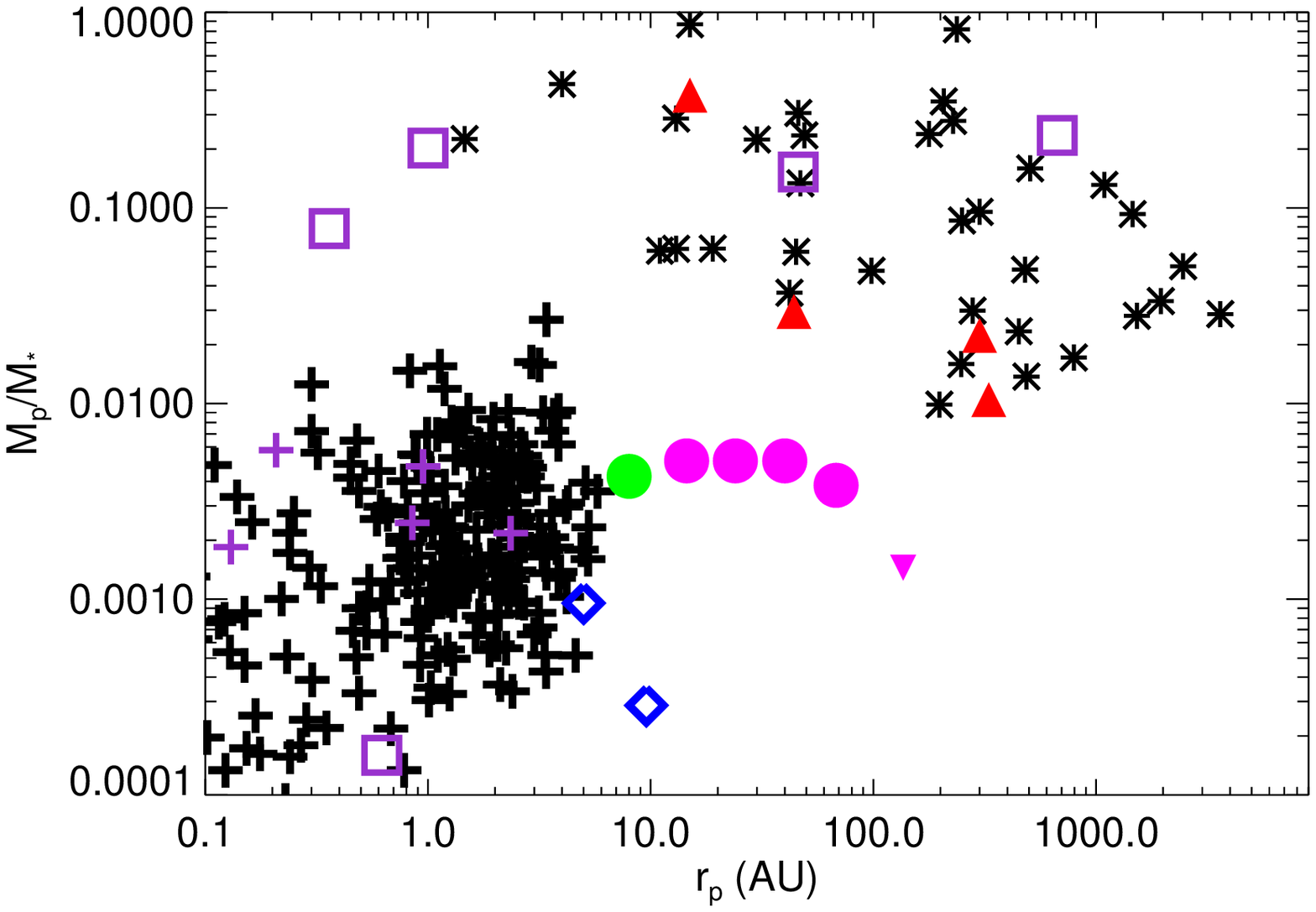}
\caption{Updated version of the mass ratio vs. orbital separation plot from \citet{Kratter2010} incorporating 
our revised masses for HR 8799bcd (magenta dots).  We also include HR 8799e, assigned a mass of 7 M$_{J}$ 
from \citet{Marois2011}, and displayed as the left-most magenta dot. The $\beta$ Pic planet is specifically 
identified as a green dot \citep{Lagrange2010}.  Fomalhaut b is a downward-pointing magenta triangle 
\citep{Kalas2008,Chiang2009}.  Substellar companions discovered after or not included in the 
\citet{Kratter2010} publication -- 1RXJS1609.1-210524, GJ 758B, 2M J044144b, and GSC 06214-00210B -- 
are plotted as red triangles.  Black crosses, purple crosses and purple squares
 denote radial velocity, transit, and microlensing-detected 
planets around stars with three mass bins: M$_{\star}$ $\ge$ 0.4 M$_{\odot}$, M$_{\odot}$ = 0.1--0.4 M$_{\odot}$, 
and M$_{\star}$ $<$ 0.1 M$_{\odot}$.   For direct comparisons and simplicity, we plot same the population of 
exoplanets \textit{not} detected by direct imaging as that used \citet{Kratter2010} (e.g. we do not include 
planets discovered by RV or transits since the publication of this paper).
 Black asterisks denote the sample of substellar companions listed in \citet{ZuckermanSong2009}.  Jupiter and 
Saturn are plotted as blue diamonds.}
\label{kratterplot}
\end{figure}

New observations at 1--5 $\mu m$ will provide better constraints on the HR 8799 
planet atmospheres.  In addition to more sensitive data at Y band and [3.3], 
Figure \ref{modseq} (lower-left panel) implies that thick-cloud 
atmospheres may have \textit{far} stronger emission at $\sim$ 2.3 $\mu m$ 
and 3.0 $\mu m$ than standard models predict.  This wavelength range can 
be probed for at least HR 8799bcd by current ground-based facilities 
such as VLT/NaCo, Keck/NIRC2, and MMT/Clio.  Integral field spectrographs 
on \textit{Gemini Planet Imager} \citep[GPI][]{MacIntosh2008} 
and SPHERE \citep{Beuzit2008}
will sample the 1--2.5 $\mu m$ SED region with exceptional
 sensitivity and thus provide a detailed comparison between observed 
and predicted atmospheric properties of all planets.

Finally, ongoing collaborations such as the IDPS survey (Marois et al., in progress) 
and Gemini/NICI \citep{Liu2010} will 
better probe the frequency of wide, massive ($\sim$ 5--13 M$_{J}$, $>$ 30 AU) 
around nearby stars.  GPI and SPHERE will probe 1--5 M$_{J}$ planets 
at even smaller separations (e.g. 5--30 AU).
These surveys will produce a far more complete census of Jupiter-mass planets 
to better determine their ubiquity and constrain how the 
formation of planets like HR 8799's compare to that 
expected for lower-mass planets at smaller separations and wide-separation, 
low-mass brown dwarfs.

\acknowledgements We thank the referee, Jonathan Fortney, for a rapid report and 
suggestions which greatly improved our manuscript.  Roman Rafikov, Scott Kenyon, and Ed Thommes 
provided very detailed, highly informative discussions regarding planet formation by core 
accretion and planetary dynamics.  We also thank Kaitlin Kratter and Ruth Murray-Clay
for valuable discussions on the formation of wide-separation planets and brown dwarf companions and 
for aid in producing our Figure 15.  Stanimir Metchev and Marshall Perrin provided useful advice during 
the beginning stages of this project.  Finally, we thank David Lafreniere for numerous conversations regarding 
the technical details of the ADI/LOCI reduction procedures and for supplying us with some subroutines.
TC is supported by a NASA Postdoctoral Fellowship.
AB would like to acknowledge support in part under NASA ATP grant 
NNX07AG80G, HST grant HST-GO-12181.04-A, and JPL/Spitzer Agreements 
1417122, 1348668, and 1371432.  This work is based in part on 
on observations made with ESO Telescopes at the Paranal Observatory 
under programme ID 084.C-656.  We acknowledge the significant cultural role and reverence
that the Mauna Kea summit has always had within the
indigenous Hawaiian community. We are grateful to be able to conduct 
observations from this mountain. 

{}

\input{tab_obs.tex}
\input{tab_astrom.tex}
\input{tab_photall.tex}
\input{tab_photcomp.tex}
\input{tab_modelstandard.tex}
\input{tab_modelthick.tex}
\clearpage

\clearpage


\end{document}

%% file: tab_obs.tex
\begin{deluxetable}{llllllllllllll}
 \tiny
\setlength{\tabcolsep}{0pt}
\tabletypesize{\small}
\tablecolumns{11}
\tablecaption{Observations}
\tiny
\tablehead{{Telescope/Instrument}&{Filter}&{Date}&{Exposure Time (s)}&{Field Rotation (degrees)} & {Detections}}
\startdata
MMT/CLIO & L'& November 21, 2008 & 5460 & 83.88 & b, c, d\\
 & M & November 21, 2008 &9600 & 31.8 & --\\
 & [3.3] & September 12, 2009 & 6780 & 128.53 & b$^{1}$, c\\
Subaru/IRCS & z & August 15, 2009 & 4200 & 172 & b$^{1}$ \\
 & J & August 15, 2009 & 1080 & 7.4 & b\\
VLT/NaCo & K$_{s}$ & October 8, 2009 & 6185 & 63 & b, c, d, e\\
\enddata
\tablecomments{Note (1) -- Companion has a low signal-to-noise detection because it is intrinsically faint (photon-noise dominated 
region).}
\label{obstable}
\end{deluxetable}

%% file: tab_astrom.tex
\begin{deluxetable}{llllllllllllll}
 \tiny
\tabletypesize{\tiny}
\tablecolumns{11}
\tablecaption{Astrometry (E["], N["] Position)}
\tablehead{{Date}&{11-21-2008}&{8-15-2009}&{9-12-2009}&{10-8-2009}}
\startdata
Planet\\
HR 8799b & 1.532 $\pm$ 0.02, 0.796 $\pm$ 0.02 & 1.536 $\pm$ 0.01, 0.785 $\pm$ 0.01 & 1.538 $\pm$ 0.03, 0.777 $\pm$ 0.03&1.532 $\pm$ 0.007, 0.783 $\pm$ 0.007\\
HR 8799c & -0.654 $\pm$ 0.02, 0.700 $\pm$ 0.02& -&-0.634 $\pm$ 0.03, 0.697 $\pm$ 0.03&-0.627 $\pm$ 0.007, 0.716 $\pm$ 0.007\\
HR 8799d & -0.217 $\pm$ 0.02, -0.608 $\pm$ 0.02&-&-&-0.241 $\pm$ 0.007, -0.586 $\pm$ 0.007\\
HR 8799e & -&-&-&-0.306 $\pm$ 0.007, -0.217 $\pm$ 0.007\\ 
\enddata
\tablecomments{The 8-15-2009 astrometry listed for HR 8799b comes from the J band data because this data yields 
a higher signal-to-noise detection.}
\label{astromtable}
\end{deluxetable}

%% file: tab_photall.tex
\begin{deluxetable}{llllllllllllll}
 \tiny
\setlength{\tabcolsep}{0.0001in}
\tabletypesize{\scriptsize}
\tablecolumns{10}
\tablecaption{Photometry}
\tablehead{{Filter}&{z}&{J}&{H}&{CH4$_{S}$}&{CH4$_{L}$}&{K$_{s}$}&{[3.3]}&{L'}&{M}\\
{$\lambda$ ($\mu m$)}&{1.03}&{1.248}&{1.633}&{1.592}&{1.681}&{2.146}&{3.3}&{3.776}&{4.8}}
\startdata
Planet\\
b&18.24 $\pm$ 0.29&16.52 $\pm$ 0.14& 14.87 $\pm$ 0.17 & 15.18 $\pm$ 0.17 & 14.89 $\pm$ 0.18 & 14.05 $\pm$ 0.08 & 13.96 $\pm$ 0.28&12.68 $\pm$ 0.12 & $>$ 11.37\\
c&$>$ 16.48&14.65 $\pm$ 0.17 & 13.93 $\pm$ 0.17 & 14.25 $\pm$ 0.19 & 13.90 $\pm$ 0.19 & 13.13 $\pm$ 0.08 & 12.64 $\pm$ 0.20&11.83 $\pm$ 0.07 & $>$ 11.22\\
d&$>$ 15.03&15.26 $\pm$ 0.43 & 13.86 $\pm$ 0.22 & 14.03 $\pm$ 0.30 & 14.57 $\pm$ 0.23 & 13.11 $\pm$ 0.12 & $>$ 11.63&11.50 $\pm$ 0.12 & $>$ 11.15\\
e& &  & & & &12.89 $\pm$ 0.26&&11.61 $\pm$ 0.12\\
\enddata
\tablecomments{Magnitudes listed are the \textit{absolute magnitude} of the companions, assuming a distance of 39.4 pc.  
(1) H, CH4S, CH4L, and K band photometry for HR 8799bcd taken from \citep{Marois2008}.  
J band photometry for HR 8799c and d also taken from \citet{Marois2008}.  L' band (3.8 $\mu m$) photometry for 
HR 8799e taken from \citet{Marois2011}.  Photometry/upper limits at 3.3 $\mu m$, L' band and 
M band (4.8 $\mu m$) for HR 8799bcd are taken from this work.} 
\end{deluxetable}

%% file: tab_photcomp.tex
\begin{deluxetable}{llllllllllllll}
 \tiny
\tabletypesize{\small}
\tablecolumns{11}
\tablecaption{Adopted Photometry for Other Planet-Mass Objects and Low-Mass Brown Dwarfs}
\tiny
\tablehead{{Companion}&{D (pc)}&{J}&{H} & {K} & {L'} & {References}}
\startdata
2M 1207b & 52.4 & 16.40 $\pm$ 0.2 & 14.49 $\pm$ 0.21 & 13.33 $\pm$ 0.11 & 11.68 $\pm$ 0.14 & 1,2\\
1RXJ1609.1-210524 & 140 & 12.17 $\pm$ 0.12 & 11.139 $\pm$ 0.07 & 10.44 $\pm$ 0.18 & 9.14 $\pm$ 0.3 & 3,4\\ 
AB Pic b & 47.3 & 12.80 $\pm$ 0.10 & 11.31 $\pm$ 0.10 & 10.76 $\pm$ 0.08 & -99 & 5\\
HD 203030b & 40.8 & 15.08 $\pm$ 0.55 & 13.80 $\pm$ 0.12 & 13.16 $\pm$ 0.10 & -99 & 6\\ 
\enddata
\tablecomments{All magnitudes listed are absolute magnitudes.  References -- 1) \citet{Chauvin2004}, 
2) \citet{Mohanty2007}, 3) \citet{Lafreniere2008a}, 4) \citet{Ireland2010}, 5) \citet{Chauvin2005}, 
6) \citet{Metchev2006}.
}
\label{photcomptable}
\end{deluxetable}

%% file: tab_modelstandard.tex
\begin{deluxetable}{llllllllllllll}
 \tiny
\tabletypesize{\scriptsize}
\tablecolumns{12}
\tablecaption{Standard Model (Model E) Fitting Results}
\tablehead{{Model Run}&{$\chi^{2}_{min}$}&{log(g), T$_{eff}$ (for $\chi^{2}_{min}$)}&{C$_{k}$}&
{$\Delta \chi^{2}$ }& {log(g), log(T$_{eff}$) ($\chi^{2}$ $<$ $\Delta \chi^{2}$)}}
\startdata
\textbf{HR 8799b}\\
Model E solar, 3x & 279.0 & 4.5, 900K & 1 & 291.85& 4.5--5, 900--1000K\\
Model E solar, 3x & 264.5& 4.5, 900K & 0.91 & 286.4 & 4.5, 900K; 5, 1000K\\
(C$_{k}$ = 0.9--1.1)\\
Model E solar, 3x & 36.6 & 4.5, 1400 & 0.34 & 58.5 & 4, 1400K; 4.5, 1300--1500K; 5, 1400K\\
(C$_{k}$=0.2--2)\\
\\
\textbf{HR 8799c}\\
Model E solar, 3x & 120.8 & 5, 1200K & 1 & 142.7& 4.5, 1100K; 5, 1200K--1300K\\
Model E solar, 3x & 71.0 &5.0, 1300K & 0.9 & 92.5 & 5, 1300K\\
(C$_{k}$=0.9--1.1)\\
Model E solar, 3x & 17.6 &4.5, 1400&0.54 & 39.5 & 4, 1300-1700K; 4.5, 1300--1700K; 5, 1400--1700K\\
(C$_{k}$=0.2--2)\\
\\
\textbf{HR 8799d}\\
Model E solar, 3x & 17.0 & 4.5, 1100K & 1 & 38.5& 4.5, 1100K; 5, 1200K\\
Model E solar, 3x & 17.0& 4.5, 1100K & 0.9 & 38.5 & 4.5, 1100-1200K; 5, 1200--1300K\\
(C$_{k}$=0.9--1.1)\\
Model E solar, 3x & 10.91 & 4.5, 1300K & 0.64 & 32.8 & 4, 1300--1600K; 4.5, 900---1700K; 5, 1200--1700K\\
(C$_{k}$=0.2--2)\\
\enddata
\tablecomments{Where the metal rich models are considered (first three rows for each planet), they always 
provide the smallest $\chi^{2}$ value.}
\label{standardfittable}
\end{deluxetable}

%% file: tab_modelthick.tex
\begin{deluxetable}{llllllllllllll}
 \tiny
\tabletypesize{\tiny}
\tablecolumns{11}
\tablecaption{Thick Cloud Model (Model A) and ''Patchy" Cloud Approximation Fitting Results}
\tablehead{{Model Run}&{$\chi^{2}_{min}$}&{log(g), T$_{eff}$ (for $\chi^{2}_{min}$)}&{C$_{k}$}&
{$\Delta \chi^{2}$ }& {log(g), log(T$_{eff}$) ($\chi^{2}$ $<$ $\Delta \chi^{2}$)}}
\startdata
\textbf{HR 8799b}\\
Model A solar & 48.9 & 4.25, 900K & 1 & 70.8& 4, 900K; 4.25, 900K; 4.5, 900-1000K\\
Model A solar & 27.6& 4.5, 1000K & 0.9 & 47.7 & 4, 900K; 4.25, 900K; 4.5, 900-1000K\\
(C$_{k}$=0.9--1.1)\\
\\
Model A/E solar  & 91.5 & 4, 900K & 1 & 111.6 & 4, 800K\\
(60\% thick clouds)\\
Model A/E solar  & 85.5 & 4, 900K & 1 & 105.6 & 4, 800--900K\\
(60\% thick clouds, C$_{k}$=0.9--1.1)\\
Model A/E solar  & 51.4 & 4, 900K & 1  & 71.5 & 4, 800-900K\\
(90\% thick clouds)\\
Model A/E solar  & 20.6 & 4, 900K & 1  & 40.7 & 4, 800--900K\\
(90\% thick clouds, C$_{k}$=0.9--1.1)\\\\
\textbf{HR 8799c}\\
Model A solar & 60.7 & 4.5, 1100K & 1 & 82.6& 4.25, 1100K; 4.25, 1100K\\ 
& & & & & 4.5, 1100K--1200K\\
Model A solar & 43.5& 4.5 1200K & 0.9 & 63.6 & 4.25, 1100K; 4.5, 1100-1200K\\
(C$_{k}$=0.9--1.1)\\
\\
Model A solar& 14.1 &4, 1000K& 1 & 34.5 & 4, 1000K\\
(60\% thick clouds)\\
Model A solar& 14.0 &4, 1000K& 1 & 34.1 & 4, 1000-1100K\\
(60\% thick clouds, C$_{k}$=0.9--1.1)\\
Model A solar & 51.6 &4, 1000K&1 & 71.7 & 4, 1000K\\
(90\% thick clouds)\\
Model A solar & 48.0 &4, 1100K&1 & 68.1 & 4, 1000-1100K\\
(90\% thick clouds, C$_{k}$=0.9--1.1)\\\\
\textbf{HR 8799d}\\
Model A solar & 5.7& 4.25, 1100K & 1 & 25.8& 3.75, 1000K; 4, 1000-1100K\\
& & & & &4.25, 1100K; 4.5, 1100-1200K\\
Model A solar & 5.3& 4.5, 1200K & 0.98 & 27.2 & 3.75--4, 1000-1100K; 4.25, 1000-1200K\\ 
(C$_{k}$=0.9--1.1)& & & & &4.5, 1100-1200K\\
\\
Model A solar& 7.4 &4, 1000K&1 & 27.5 & 4, 1000-1100K\\
(60\% thick clouds)\\
Model A solar& 2.8 &4, 1000K&1 & 22.9 & 4, 1000-1100K\\
(60\% thick clouds, C$_{k}$=0.9--1.1)\\
Model A solar& 8.8 &4, 1000K&1 & 28.9 & 4, 1000-1100K\\
(90\% thick clouds)\\
Model A solar& 3.1&4, 1000K&1 & 23.2 & 4, 1000-1100K\\
(90\% thick clouds, C$_{k}$=0.9--1.1)\\
\enddata
\label{thickfittable}
\end{deluxetable}